\documentclass[11pt]{article}
\usepackage{graphicx}
\usepackage{amssymb}
\usepackage{amsmath}
\usepackage{epsfig} 
\usepackage[]{caption}
\usepackage{verbatim}
\usepackage{amsfonts}
\usepackage{amsthm}
\usepackage{enumerate}
\usepackage{float}
\usepackage{morefloats}
\usepackage{authblk}
\usepackage{algorithm}
\usepackage{algpseudocode}
\usepackage{authblk}

\usepackage{silence}
\usepackage{url}
\newcommand{\argmin}{\mathop{\rm arg~min}\limits}

\title{Analysis and modelling of macroscopic and microscopic dynamics of a pedestrian cross-flow}

\author{\scriptsize Francesco Zanlungo}
\affil{\tiny International Professional University of Technology in Osaka, 3-3-1 Umeda, Kita-ku, Osaka, 530-0001 Japan}
\author{\scriptsize Claudio Feliciani}
\affil{\tiny Research Center for Advanced Science and Technology, The University of Tokyo, 4-6-1 Komaba, Meguro-ku, Tokyo 153-8904, Japan}
\author{\scriptsize Zeynep Y\"ucel}
\affil{\tiny Graduate School of Natural Science and Technology, Okayama University, 3-1-1 Tsushima-naka Kita-ku, Okayama 700-8530 Japan}
\author{\scriptsize Katsuhiro Nishinari}
\affil{\tiny Research Center for Advanced Science and Technology, The University of Tokyo, 4-6-1 Komaba, Meguro-ku, Tokyo 153-8904, Japan}
\author{\scriptsize Takayuki Kanda}
\affil{\tiny Graduate School of Informatics, Kyoto University, Yoshidahonmachi, Sakyo-ku, Kyoto 606-8317, Japan}
\begin{document}
\maketitle
{\it \small
\section*{Abstract}
In this work we investigate the behaviour of a human crowd in a cross-flow. In the first part of our work we analyse the results of a set of controlled experiments in which subjects were divided into two groups,
organised in such a way to explore different density settings, and asked to walk through the crossing area. We study the results of the experiment by defining and analysing a
few macroscopic and microscopic observables. Along with analysing traditional indicators such as density and velocity, whose dynamics was, to the extent of our knowledge,
poorly understood for this setting, we pay particular attention to walking and body orientation, studying how these microscopic observables are influenced by density.
Furthermore, we report a preliminary but quantitative analysis on the emergence of self-organising patterns (stripes) in the crossing area, a phenomenon that had been previously qualitatively
reported for human crowds, and reproduced in models, but whose quantitative analysis with respect to density conditions is, again according to our knowledge, a novel contribution.

In the second part of our work, we try to reproduce the empirical results using a hierarchy of models, which differ in the details of the body shape (using a disk-shaped body vs a more realistic
elliptical shape) and in how collision avoiding is performed (using only information regarding ``centre of mass'' distance and velocity, or actually introducing body shape information). We verified
that the most detailed model (i.e., using body shape information and an elliptical body) outperforms in a significant way the simplest one (using only centre of mass distance and velocity, and
disk-shaped bodies). Furthermore, we observed that if elliptical bodies are introduced without introducing such information in collision avoidance, the performance of the model is relatively
poor. Nevertheless, the difference between the different models is relevant only in describing the ``tails'' of the observable distributions, suggesting that the more complex models
could be of practical use only in the description of high density settings.
Although we did not calibrate our model in order to reproduce  the ``stripe formation'' self-organising pattern, we verified that it naturally emerges in all models.}
\newpage
\tableofcontents
\newpage
\section{Introduction}
Numerical simulation has become a common practice in the design of pedestrian facilities. Several commercial and non-commercial crowd simulation software are available and used in the construction of train/metro stations, shopping malls, stadia, etc. or to study past accidents \cite{Lovreglio2020}. Crowd simulators are used in a variety of contexts, including urban planning \cite{Arisona2012}, performance assessment of transportation facilities \cite{Hoy2016} and, more recently, evaluation of policies preventing the spread of infectious diseases \cite{Comai2020}. However, due to their importance in terms of safety, evacuation scenarios are still the ones on which crowd simulations are typically targeted. Due to the widespread use of crowd simulators and the legal implications in case of accidents, a standard for the validation of evacuation codes has been recently proposed (ISO20414 \cite{ISO20414}). This shows that simulation models have gradually moved from a purely theoretical topic discussed among researchers to an instrument employed by facility managers, whose validation criteria are defined in international standards. As such, accuracy of simulation models plays an utmost role now than ever before. In this context, we should note that although the increasing use in realistic scenarios partially proves that simulation software have reached a sufficient degree of quality, they are still far from being perfect. The Social Force Model \cite{Helbing1995}, which is commonly used in commercial applications, has still many limitations and in this work we will discuss them and propose an approach which aims at highlighting behavioural features which are most needed in a pedestrian model.

To allow a critical assessment of the model proposed here, we will study the behaviour of a pedestrian crowd in a relatively simple geometry, the crossing area of two orthogonal corridors, each one characterised by a uni-directional flow. This is obviously an extremely idealised condition, since in the real world we do not often have perfectly orthogonal corridors and the ones existing are often created in large areas where people arrive from several directions heading to different destinations. Nevertheless, the cross-flow scenario represents a good trade-off between a simple experimental setup and realistic observed behaviour. Considering the complex interactions among pedestrians, which require both anticipatory and collision avoidance abilities, we believe the fundamental behaviour observed in crowds in such a setup is not much different than a real scenario, especially when people are heading toward a specific destination such as in evacuations. 

However, despite its remarkable properties, the cross-flow has not been explored in detail in the literature and the few studies considering this geometry have mostly analysed only macroscopic quantities, often as part of a comparison with other geometries, without discussing its properties on the microscopic level \cite{Cao2017,Cao2018}. On the other hand, most studies in pedestrian dynamics focus either on the behaviour of pedestrians in a simpler geometry, a single corridor (either in the uni- or bi-directional \cite{Zhang2012,Feliciani2016,Murakami2021} case), or on possible variations of the (mostly uni-directional) bottle-neck/evacuation from a room scenario \cite{Seyfried2009,Adrian2020,Feliciani2020B}.

All the above cases are obviously very interesting and still present open questions. Even the behaviour in a simple corridor is not completely understood; see for example many works discussing conditions causing the observed variations in fundamental diagrams (density/velocity relations) \cite{Chattaraj2009,Fujita2019,Ye2021}. Furthermore, when a second, opposite, flow is allowed, the dynamics becomes considerably more complex, and possibly not completely feasible to be analysed in controlled experiments (due to social norms on walking sides \cite{Zanlungo2012}, stronger influence of social groups \cite{Zanlungo2020}, etc.).

When variables accounting for conditions close to reality are further added, things get even more complex, since route choice \cite{Crociani2016} or exit selection \cite{Bode2015}, walking impairment or disabilities \cite{Geoerg2019}, collaborative vs competitive behaviours \cite{VonKruchten2017}, level of competitiveness \cite{Feliciani2020B}, social bonds \cite{Ye2021,zanlungo2014potential,zanlungo2015spatial,zanlungo2015mesoscopic,zanlungo2017intrinsic,zanlungo2019intrinsic}, available information \cite{Feliciani2020A}, etc. have to be taken into account; all aspects that may even be impossible to realistically reproduce as a whole in a controlled experiment setting.

Nevertheless, all the uni-directional settings share a strongly simplifying feature, namely that all pedestrians have roughly the same goal, and as a result have very similar velocities. While in general a collision avoidance model needs information concerning relative distance {\it and} velocity, as soon as all the pedestrians have similar velocities, the relative velocity may be (at least in a first, rough zeroth-order approximation) ignored. Although in the bi-directional corridor setting pedestrians are split in two streams with opposite goals and thus roughly opposite velocities, self-organisation (either induced by collision avoidance or, on a much faster scale, by social norm \cite{Zanlungo2012}) causes the streams to spatially separate. Although the time scale of this separation is a problem that still needs to be fully addressed, the ``stationary state'' is still (in a first approximation) fundamentally uni-directional \cite{Feliciani2018A}.

The above reasons explain why many research works may still rely on the ``Circular Specification'' (CS)  \cite{Helbing1995} of the Social Force Model, although this model ignores relative velocity and may thus not be able to describe collision scenarios with a more complex dynamics \cite{Johansson2007,Zanlungo2011}.

On the other hand, the geometry studied in this work naturally induces interactions in which pedestrians in different flows have velocity vectors with relative angles $\approx \pi/2$, and since, differently from the single corridor bi-directional scenario, it does not present a trivial solution in which two continuously flowing streams are physically separated by a single
stationary boundary, it does not allow for the above simplification (i.e., ignoring relative velocity).

Previous, although preliminary, studies suggest anyway the emergence of a self-organising pattern that reduces the interactions between the streams while optimising the overall flow, namely the presence of ``diagonal stripes'' \cite{Naka,Ando1988,Helbing2005}. This behaviour, which is well-known to be used in ``centrally organised'' marching parades \cite{YoutubeMarchingParade}, deserves to be deeply studied in self-organising pedestrian crowds. Theoretical justification for the emergence of such a pattern can be found in \cite{cividini2013diagonal,cividini2013wake} (using a discrete lattice model) and \cite{cividini2013crossing} (using a mean field approach). Such a pattern has also been reported in a very recent experimental analysis, which focuses in particular on the relation between stripe orientation and the crossing angle \cite{mullick2021analysis}. On the other hand, this work focuses on density conditions while keeping the crossing angle fixed at $\pi/2$.

As the stripes in a real crowd are expected to be highly irregular (and in particular both in the very low and very high density regimes), this phenomenon is to be properly studied with the help of clustering algorithms, to allow for a more general definition of ``stripe'', and indeed we intend to perform such a study in a future work.

Nevertheless, the nature of the problem is still geometrically simple, and a first analysis, performed in this work, may be based just on relative angles between first neighbours. This is indeed one of the observables that we are going to analyse in this work, which is aimed to study in detail the microscopic and macroscopic dynamics of such a crossing flow.

In more detail, we are going to analyse six different experimental conditions of the crossing scenario. In each condition, we changed the pedestrian density in the flows (while keeping the density {\it equal between the flows}), and we repeated the experiment six times for each condition. The analysis is based on the statistical study of nine different macroscopic and microscopic observables (to be defined in a rigorous way below).
Some of them are quite standard in pedestrian studies (evolution of density in time, speed distribution), while others are studied less often, such as the probability distribution of distances between first neighbours (either in the same or in the crossing flow), and the distribution of the velocity direction. Other features have, at least to our knowledge, been explored even less often, such as the distribution of relative position angles between first neighbours (either in the same and in the crossing flow). Finally, as we believe that, at least in the high density regime, the detailed orientation and movement of the body may play a role in collision avoidance, we study the body orientation angle, and its deviation with respect to the velocity direction.

The study is not limited to a statistical analysis of the data, but includes an attempt to explain them through the use of computational models. In this respect, we would like to stress that it is not our intention to introduce a new ``better'' or ``best'' collision avoidance model, and in particular not a general purpose model, but a model that may allow us to explain the observed data, with particular attention to those data that have not been reported before (orientation and relative angles).

To this respect, we have decided to use as a starting point the ``Collision Prediction'' (CP) formulation of the Social Force Model \cite{Zanlungo2011}. We use this model for three reasons: 1) it is a conceptually simple (although not necessarily trivial from a computational viewpoint) model that predicts, based on a linear approximation using current {\it relative position and velocity}, the future possible occurrence of a collision, and generates a force ``opposite to the collision'', and it is thus a possible way of overcoming the limitations of the CS Social Force Model specification while preserving a similar conceptual framework; 2) it fairly reproduces the emergence of self-organising stripes in a crossing scenario \cite{Zanlungo2007A,Zanlungo2007B,YoutubeModelComparison}; 3) it is well-know to us, as we first developed it.

Nevertheless, as the model does not include a specific description of the shape of the human body, nor a description of its orientation dynamics, we modify and extend it to take these aspects in consideration. As stated above, the proposed model is more an ``explanation tool'' than a ``computational tool'' to be used for practical purposes. We believe, based on the discussion above and the analysis to come in the rest of this work, that a model of collision avoidance that may properly describe the studied behaviour has to: 1) rely on information concerning relative position, velocity body shape and orientation to understand the possibility of a collision, and 2) determine the variation in velocity and orientation that may avoid such a collision. Our model is based on a conceptually simple implementation of these principles, although not necessarily in the most efficient or practical way.

As mentioned above, our work may be divided into a first part performing an analysis of experimental data, and a second part focusing on computational models to explain such results. In detail, Sections \ref{experiments} (describing the experiments), \ref{obs} (defining the observables), \ref{exp_res} (describing the results of the experiments), along with Appendices \ref{obs_def} (detailed definition of observables), \ref{obs_dens} (definition of observable dependence on density), \ref{depon_rhoI} (reporting the dependence of observables on the initial density), focus on the analysis of the experimental data. On the other hand Sections \ref{comp} (description of computational models), \ref{full_cal} (model calibration and evaluation), \ref{modcomp} (comparison between models), along with Appendices \ref{model_detailed} (detailed model description), \ref{statan} (description of statistical tools), \ref{evalmet} (description of evaluation metrics), \ref{standeval} (evaluation using the ``standard metric''), \ref{appcomp} (comparison of models on low and medium density ranges), \ref{graph_comp} (comparison of observable pdfs between different models), focus on computational models. In Appendix \ref{stripe} we provide a qualitative description of stripe formation in experiments and models.

\section{Experiments}
\label{experiments}
\subsection{Experimental setup}
At the scope of collecting experimental data on the behaviour of crowds in a cross-flow scenario, an experimental campaign was organised recruiting students as participants. Two waiting/starting lanes having an equal width $w=$ 3.0 m were prepared, each creating a similar unidirectional flow. These two flows were directed in such a way to have them crossing perpendicularly in a specific area where the trajectories of participants were collected. Experiments were conducted outdoors on the Tokyo University campus on December 7th, 2019. To protect participants from the rain roof-covered parts of the building were used and, due to space limitations, a starting lane had to be bent. The experimental setup is schematically presented in Fig. \ref{f1} (a) and in addition a top-view image of the experimental setup is also provided in Fig. \ref{f1} (b).

To allow an accurate setup of crowd density in each starting area and yet get a uniform distribution of participants, starting lines were drawn on the ground and participants were asked to freely take position between the external perimeter of the crossing area (whose borders delimited by pink lines are clearly visible in the image of Fig. \ref{f1} (b)) and the line determining the end of the waiting area. Staff members checked that subjects disposed themselves in a uniform way. The control of density in the crossing area was consequently achieved by setting different lengths $l$ in the starting areas (see Fig. \ref{f1} (a)). Namely, six different lengths have been considered: $l=$ 36.0 m, $l=$ 18.0 m, $l=$ 9.0 m, $l=$ 6.0 m, $l=$ 4.5 m and $l=$ 3.6 m.

This geometrical configuration was planned expecting a total number of 54 participants, which would lead to starting area densities $\rho_I=0.25$, $\rho_I=0.5$, $\rho_I=1$, $\rho_I=1.5$, $\rho_I=2$ and $\rho_I=2.5$ ped/m$^2$ (following the same order given for the lengths $l$ above). However, a slightly higher number of participants showed up on the day of the experiments: 56, 2 more than originally expected. As a consequence, real densities were slightly higher (3.7\%  higher to be precise). Nonetheless, to keep presentation of the results simple, we kept the original densities to label each experiment, also considering that those are simply initial conditions and densities really observed in the crossing area are measured in detail from the collected data.

For each initial density condition, 6 repetitions were performed, although this number was increased to 8 for the configuration having the lowest density ($\rho_I=0.25$ ped/m$^2$). To limit the capability of participants to develop efficient strategies through a learning process, configurations were not performed in increasing/decreasing order but following a shuffled schedule.

Considering that participants simply walked straight in the starting areas, marking was only performed delimiting each region with coloured signs on the ground. However, due to the strong interactions observed in the crossing area and its importance in the frame of data collection and analysis, this area has been delimited using chain partitions (see Fig. \ref{f1} (b)). It should be nonetheless remarked that chain partitions were low in height, and thus some participants were able to enter the ``forbidden region'' outside the crossing area by slightly bending their body. This detail was taken into consideration both in data analysis (definition of the density observable) and in simulation settings, as explained below.

\subsection{Experimental procedure}
An overall number of 56 participants voluntarily applied for the experiments. Only males were recruited since other experiments planned during the day had male-only conditions (and partially to avoid possible issues arising from body contact at high densities). Instructions to the participants were simple and yet clear. They were asked to line up in front of the starting lane uniformly occupying the space available until the line marking the end of the ``initial density condition area'' (see Fig. \ref{f1} (a)), and walk after the ``start'' signal would be given. Both groups had to walk straight, pass through the crossing area and keep walking for a short distance after leaving it (to avoid congestion due to stopping participants). Staff on-site helped to ensure a uniform distribution and to avoid participants taking repeatedly similar positions (front/back positions are typically preferred). In addition, staff at the exit encouraged participants to keep walking past the crossing area.

Experimental procedures were approved by the Ethical Commission of The University of Tokyo (approval number 19-274) and conform with the Declaration of Helsinki. Participants received clear information on the nature of the research, methods employed and disclosed data. After a briefing, they gave written permission for participation and data acquisition and received a remuneration in accordance to the university's policies.

\subsection{Data acquisition and processing}
A camera was placed approximately 6 m right above the centre of the crossing area to record the dynamics observed during the experiments. Resolution was set at 3000 $\times$ 4000 pixel and frame rate at 30 fps. Trajectories were extracted from the recorded videos using the PeTrack software \cite{Boltes2010,Boltes2013} employing hats with different colours worn by participants as markers. It is important to remark that the software employed allows to take into account both camera distortion (due to the wide-lens used) and perspective. This means that obtained trajectories are representative of planar motion and relative to the projection of people's head to the ground.

Since body orientation is an important part of the model presented hereafter, the following approach was taken to gain precise information on this quantity. 10 participants (5 in each starting lane) were given a tablet (Nexus 7, 2013) which was fixed to their body using a bib (see also \cite{Nagao2018} for more details). The tablet thus recorded the movements in the upper body part and more specifically the pedestrian chest orientation, which, except in cases of strong torsion, is expected to provide a reliable estimate of shoulder orientation (often regarded as the proxy for body orientation \cite{Willems2020}). Tablet (body) orientation was obtained by making use of the inbuilt gyroscope sensor. Previous research \cite{Feliciani2018B} showed that errors in terms of precision and accuracy of the gyroscope sensor are below 1\%. This allows obtaining an angle given the initial orientation and integrating the measured angular velocity over time. Due to the small error, dead-reckoning is thus not an issue for short time measurements such as in the case of this experiment. 

From the details presented above, it should be clear that to properly record body orientation it is necessary to know its initial value. For this reason, participants wearing a tablet were asked to orientate towards the crossing area (i.e., in such a way that the normal to their chest was aligned with the corridor's axis) before the start of the experiments so that angles could be computed accurately. However, as it could be possible that some participants did not follow the instructions properly, to verify the reliability of body orientation information, we overlapped the video recording of the experiments with ``animations'' that used ellipse shaped ``virtual pedestrians'' whose positions and orientations were given according to our tracking process, and proceeded to eliminate the few instances in which body orientation appeared to be unreliable.

\begin{figure}[htb]
\begin{center}
  \includegraphics[width=0.9\columnwidth]{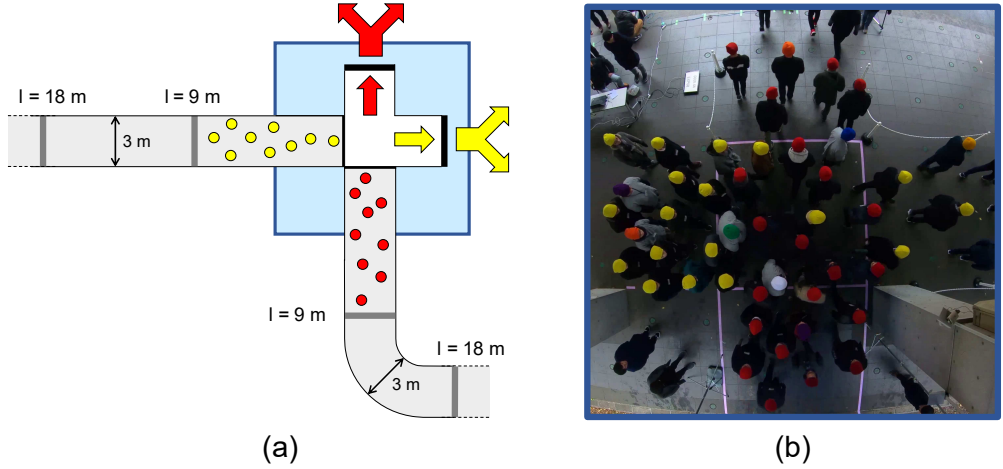}
\caption{\label{f1}(a): schematic representation of the experimental area. For the sake of simplicity, only a small number of pedestrians is represented and measures are given only for reference. (b): frame relative to a specific experiment with $\rho_I=2$ ped/m$^2$ (i.e., length of the starting area was 4.5 m). The image corresponds approximately to the blue area given in the schematic representation. The participants wearing caps different from red or yellow had been equipped with tablets to measure their body orientation.}
 \end{center}   
\end{figure}

Participants equipped with a tablet were given a cap with a different colour, allowing the coupling of their trajectory (obtained from the camera) with body orientation (from the gyroscope sensor). For example, it is possible to know that the participant wearing a green cap and a grey T-shirt and coming from the left (as seen in Fig. \ref{f1}) was using tablet no. 2 and thus it is possible to study its position and body orientation separately.

\section{Observables}
\label{obs}
In our work we use 9 observables to quantify cross-flow dynamics. In this section, we introduce them providing a qualitative description of their definition and meaning,
while a detailed, operative and quantitative definition is provided in Appendix \ref{obs_def}.

The {\it density in the crossing area}  $\rho(t)$ (1) is the time dependence of the number of pedestrians tracked at each time in the crossing region divided by the area of such region, and it is measured in ped/m$^2$. This is the only {\it macroscopic} (i.e., defined at the crowd-level, and not at the individual pedestrian-level) observable that we use. Nevertheless, it is strongly related to
a microscopic observable, the {\it exit time} (from the tracking area) $E_t$, and we will sometimes refer also to the latter observable because it allows for a more straightforward definition
of a probability distribution.

By $v$ we denote the {\it pedestrian speed} (2), whose probability distribution is denoted with $P(v)$, a notation used also for all the other pedestrian-level (microscopic) observables defined
below.

To investigate also the direction of the pedestrian velocity, we
study the {\it velocity direction angle} $\theta^v$ (3). This angle, defined in detail in Appendix \ref{obs_def}, assumes values in $[-\pi,\pi)$, $\theta^v=0$ denoting movement along the
  corridor axis. The angle is defined in such a way that $\theta^v>0$ denotes angles in the direction of the incoming cross-flow.

  A subset of ten subjects was carrying a sensor that allowed us to know their body orientation. This is denoted with $\theta$ (4), assuming values in   $[-\pi,\pi)$. Here $\theta=0$ corresponds to the
    state in which  the normal to the pedestrian chest is aligned with the corridor's axis. Again, $\theta>0$ denotes angles in the direction of the incoming cross-flow.

The difference between $\theta$ and $\theta^v$ is defined as $\delta \theta$ (5) in such a way that, again, $-\pi \leq \delta \theta <\pi$.

For all pedestrians we also measure, as a vector, the distance to their first neighbours, distinguishing between neighbours in the same flow and in the crossing flow.
In order to identify the presence of a self-organising pattern, neighbours are defined to be other pedestrians {\it on the front} of the pedestrian under consideration, i.e., located in the direction
of motion. In such a way, if a stripe pattern emerges, there will be a peak in the relative angle distribution corresponding to the stripe axis (if also neighbours on the back were present such
peak would be duplicated by symmetry).

The {\it magnitude of the relative distance to the first neighbour in the same flow} is denoted as $\delta^s$ (6), while the
{\it magnitude of the relative distance to the first neighbour in the crossing flow} is denoted as $\delta^o$ (7). The angles that these distance vectors form with the corridor axis are, respectively,
the {\it same flow first neighbour relative angle} $\phi^s$ (8) and {\it crossing flow first neighbour relative angle} $\phi^o$ (9). These angles are defined to assume
values in $[-\pi/2,\pi/2)$, $\phi=0$ identifying the corridor axis and $\phi>0$ the direction of the incoming cross-flow.

  \section{Experimental results}
    \label{exp_res}
  \subsection{Dependence of other observables on $\rho$}
  The most straightforward way to analyse the results of our experiments is to study the dependence of all observables on the initial density condition $\rho_I$, and such analysis is indeed reported in Appendix \ref{depon_rhoI}. We nevertheless noticed that results are more clearly
  interpreted if the remaining observables are studied as a function of the crossing area density $\rho$ (denoted  for each observable $O$ as $\langle O \rangle^{\rho}$, refer to Appendix \ref{obs_dens_cross} for the detailed operative definition).
Indeed, it is not surprising that the dynamics of the crowd is more directly determined by
  the actual density than by the initial density conditions, and thus we restrict ourselves to such analysis in the main text.

\begin{figure}
\begin{center}    
  \includegraphics[width=0.8\textwidth]{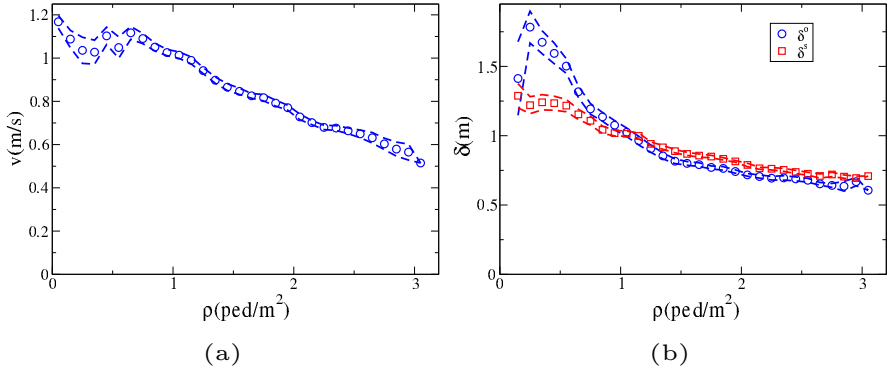}
  \caption{\label{f_v_delta_rho}(a): $\langle v \rangle^{\rho}$.  (b): $\langle \delta^o\rangle^{\rho}$ (blue circles) and $\langle \delta^s\rangle^{\rho}$ (red squares).
    Dashed lines provide standard error intervals.}
 \end{center}   
\end{figure}

\begin{figure}
  \begin{center}
  \includegraphics[width=0.8\textwidth]{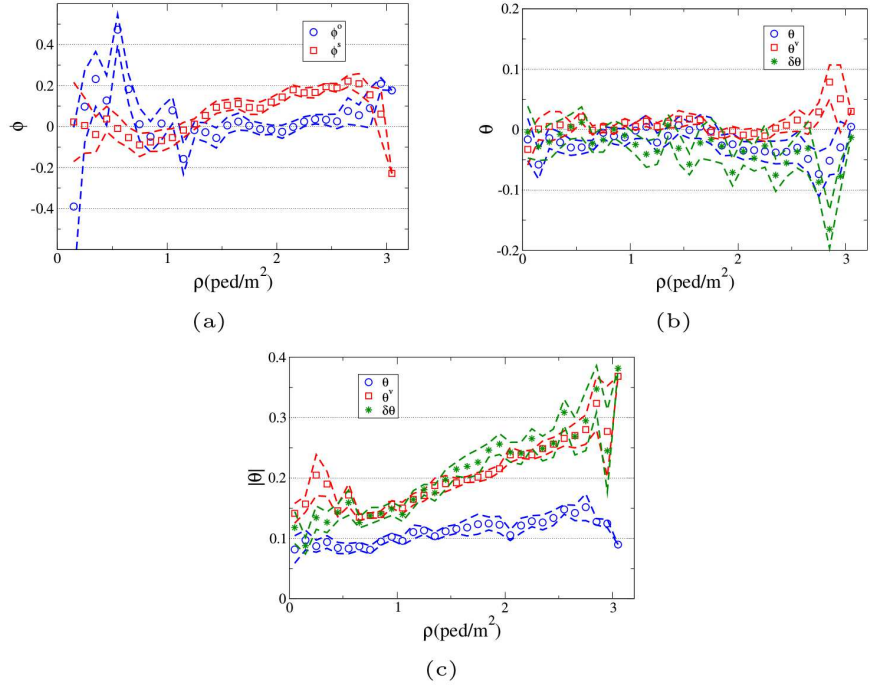}
  \caption{\label{f_phi_theta_rho}(a): $\langle \phi^o\rangle^{\rho}$ (blue circles) and $\langle \phi^s\rangle^{\rho}$ (red squares).
    (b): $\langle \theta\rangle^{\rho}$ (blue circles) and $\langle \theta^v\rangle^{\rho}$ (red squares) and $\langle \delta \theta\rangle^{\rho}$ (green stars).
    (c): $\langle |\theta|\rangle^{\rho}$ (blue circles) and $\langle |\theta^v|\rangle^{\rho}$ (red squares) and $\langle |\delta \theta|\rangle^{\rho}$ (green stars).
    Dashed lines provide standard error intervals.}
 \end{center} 
\end{figure}

  The dependence on $\rho$ of $v$ is shown in Fig.  \ref{f_v_delta_rho} (a),
  that of the $\delta$ observables in Fig.  \ref{f_v_delta_rho} (b), that of the $\phi$ observables in Fig.  \ref{f_phi_theta_rho} (a), and that of $\theta$
  observables in  Fig.  \ref{f_phi_theta_rho} (b).
  We also report in  Fig.  \ref{f_phi_theta_rho} (c) the absolute value of the $\theta$ observables (as $\theta$ observables are expected to be almost symmetric,
their average provides information on their asymmetry, while the average of the absolute value provides information on their spread).

In Fig. \ref{f_v_delta_rho} (b) observable $v$  presents an initial plateau for low density, but from $\rho\approx 0.7$ it turns into a decreasing function of $\rho$, a trend that does not change up to 3 ped/m$^2$. Such results are
in line with the usual ``fundamental diagram'' behaviours
(compare to the less clear behaviour when $v$ is studied as a function of $\rho_I$
in Fig. \ref{f_rho_v_rhoI} of Appendix \ref{depon_rhoI}).
Also $\delta^o$ and $\delta^s$ are decreasing functions of $\rho$ up to 3 ped/m$^2$, although they seem to be converging to a stable value (compare again to Fig. \ref{f_rho_v_rhoI} of Appendix \ref{depon_rhoI}, in which a plateau is reached at $\rho_I\approx 1.5$  ped/m$^2$).

In Fig. \ref{f_phi_theta_rho} (a) a bias towards positive values of $\phi^s$ emerges in a stable way for $\rho>1.2$ ped/m$^2$.
This result is related to the ``diagonal stripe formation'' and it is better discussed studying overall pdfs (see Section \ref{opdfs}). A weaker bias towards positive values is shown also by $\phi^o$ for
high $\rho$ values (and more strongly but also less regularly for low $\rho$ values).

In Fig. \ref{f_phi_theta_rho} (b) observables $\theta$ and $\delta \theta$ are weakly biased towards negative values,
while $\theta^v$ is weakly biased towards positive ones, such biases becoming stronger
as $\rho$ increases.

In Fig. \ref{f_phi_theta_rho} (c) observables $|\theta^v|$ and $|\delta \theta|$ are clearly increasing functions of  $\rho$.
Also $|\theta|$ appears to grow with $\rho$, although not as strongly as $|\theta^v|$ and $|\delta \theta|$. Interestingly, such a pattern for $|\theta|$ does not clearly emerge  when this observable is
studied as a function of $\rho_I$ in Appendix \ref{depon_rhoI} (see Fig. \ref{f_theta_rhoI}).

\subsection{Overall pdfs}
\label{opdfs}
\begin{figure}
\begin{center}
  \includegraphics[width=0.8\textwidth]{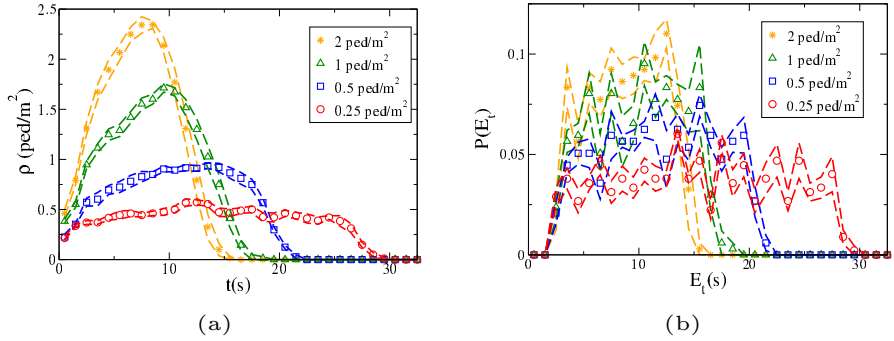}
  \caption{\label{frho_exit}(a): $\rho(t)$ for different initial conditions.  (b): $P(E_t)$ for different initial conditions.
    Dashed lines provide standard error intervals (computed over independent repetitions).}
 \end{center}   
\end{figure}

\begin{figure}
\begin{center}
  \includegraphics[width=0.8\textwidth]{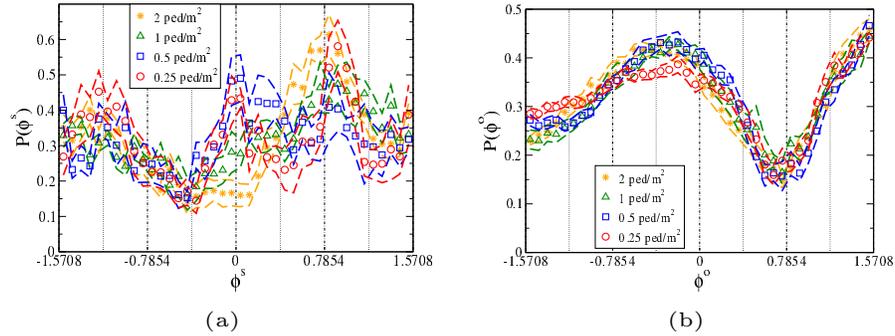}
  \caption{\label{fphi}(a): $P(\phi^s)$ for different initial conditions.
(b): $P(\phi^o)$ for different initial conditions.
    Dashed lines provide standard error intervals (computed over independent repetitions).}
 \end{center}   
\end{figure}

In Figs. \ref{frho_exit}$\sim$\ref{fphi} we also show the results concerning all observables for the $\rho_I=$ 0.25, 0.5, 1 and 2 ped/m$^2$ initial conditions (those concerning $\rho_I=$ 1.5 ped/m$^2$ are presented in the model calibration Section \ref{cal_sec}, while 
those concerning the highest $\rho_I=$ 2.5 ped/m$^2$ are presented in the model
evaluation Section \ref{ev_sec}.).

We may see in Fig. \ref{frho_exit} (a) that the maximum density in the crossing area grows with the initial density. Nevertheless, while it attains a value doubling the initial
condition for $\rho_I=$ 0.25 ped/m$^2$, it is just $\approx 1.25$ fold the initial condition at $\rho_I=$ 2 ped/m$^2$.

Fig. \ref{frho_exit} (b)
shows that the total time for the crowd to pass the crossing area decreases with time, i.e., the flow is increased. Nevertheless, an 8 fold increase in the initial condition
  density (corresponding to a 5 fold increase in peak density) results in half ``passing time''.

\begin{figure}
\begin{center}
  \includegraphics[width=0.8\textwidth]{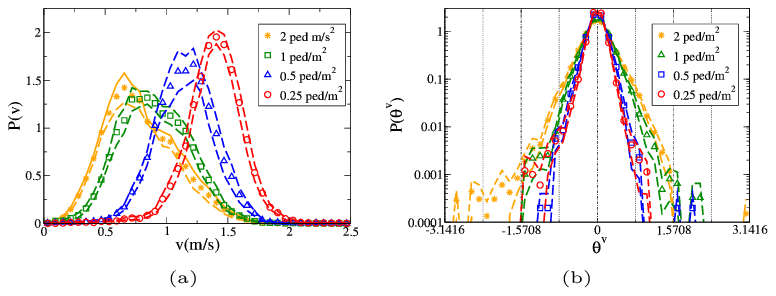}
  \caption{\label{fv}(a): $P(v)$ for different initial conditions.  (b): $P(\theta^v)$ for different initial conditions.
    Dashed lines provide standard error intervals (computed over independent repetitions).}
 \end{center}   
\end{figure}

This finding is obviously related to the decrease in velocity with increasing density, shown in Fig. \ref{fv} (a), as the velocity peak decreases from $\approx 1.4$ m/s
for the $\rho_I=$ 0.25 ped/m$^2$ initial condition to  $\approx 0.6$ m/s for $\rho_I=$ 2 ped/m$^2$.

As shown in Fig. \ref{fv} (b), motion becomes less ordered as density increases.

\begin{figure}
\begin{center}
  \includegraphics[width=0.8\textwidth]{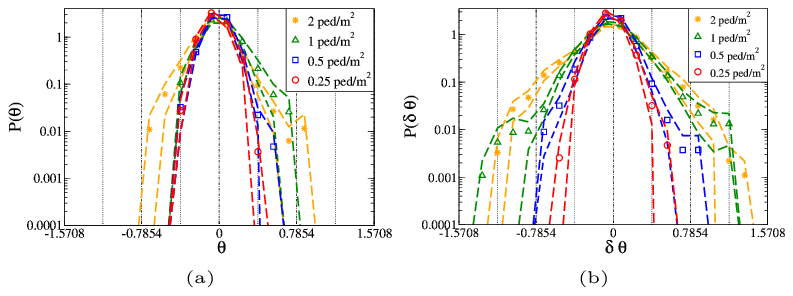}
  \caption{\label{ft}(a): $P(\theta)$ for different initial conditions.  (b): $P(\delta \theta)$ for different initial conditions.
    Dashed lines provide standard error intervals (computed over independent repetitions).}
 \end{center}   
\end{figure}

In an equivalent way, at high density, it is more probable to have large deviations between the body orientation and the corridor axis (Fig. \ref{ft} (a)), as well as between
the body orientation and the velocity one (Fig. \ref{ft} (b)). It is nevertheless to be noticed that $|\theta|$ seems to be strongly limited by $\pi/4$, and  $|\delta \theta|$ by $\pi/2$.

\begin{figure} 
\begin{center}
  \includegraphics[width=0.8\textwidth]{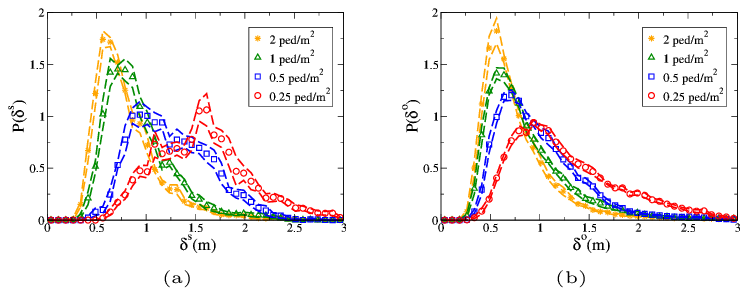}
  \caption{\label{fr}(a): $P(\delta^s)$ for different initial conditions.  (b): $P(\delta^o)$ for different initial conditions. 
    Dashed lines provide standard error intervals (computed over independent repetitions).}
 \end{center}   
\end{figure}

The distance between pedestrians in the same flow (Fig. \ref{fr} (a)) decreases with density, having a peak at $\approx 1.5$ m
for the $\rho_I=$ 0.25 ped/m$^2$ initial condition, and  $\approx 0.5$ m for $\rho_I=$ 2 ped/m$^2$.
At the highest initial density condition, the peak attains a similar value also for the distance to pedestrians in the other flow,
while at low density pedestrians get closer to people in the other flow than they do to people in their own flow
(peak at $\approx$ 1 m, Fig. \ref{fr} (b)).

Finally, we may see (Fig. \ref{fphi} (a)) that the angle to the first neighbour in the same flow is skewed to $\phi^s>0$ for all initial conditions (maximum
at $\approx \pi/4$). The distribution of this observables depends weakly on density, with the exception of the behaviour around $\phi^s=0$, since in the transition from low to high densities there is a strong decrease in the probability of pedestrians following each other (i.e., having $\phi^s\approx0$).

The $\phi^o$ distribution, shown in Fig. \ref{fphi} (b), is, compared to the $\phi^s$ one, less dependent on $\rho_I$, less noisy, and clearly shows the presence of a geometrical structure (stripes), with minima at
$\approx \pi/4$ (corresponding to the position of a pedestrian  from the same flow in case of a diagonal lane \cite{Naka,mullick2021analysis}).

\subsection{Comments}
Most of these results are qualitatively intuitive, although they had not, to the limits of our knowledge, been investigated before in a quantitative way. It was expected that the two flows would exhibit
some kind of organisation that would regulate the density in the crossing area (Figs. \ref{frho_exit} in the main text and \ref{f_rho_v_rhoI} (a) in Appendix \ref{depon_rhoI}), although the quantitative behaviour was not known in advance. Similarly, the velocity/density relation (see in particular Fig. \ref{f_v_delta_rho} (a))  is qualitatively
similar to those observed in uni-directional flows. As velocity decreases and collision avoidance increases, we naturally expect more variation in the velocity direction $|\theta^v|$
(Figs. \ref{f_phi_theta_rho} (c), and \ref{fv} (b)  in the main text and \ref{f_theta_rhoI} (b) in Appendix \ref{depon_rhoI}).
Also, the dependence on density (either in the crossing area or as initial conditions) of the distance between pedestrians, follows the expected qualitative behaviour
(Figs. \ref{f_v_delta_rho} (b), \ref{fr}  in the main text and \ref{f_delta_phi_rhoI} (a) in Appendix \ref{depon_rhoI}).

The other observables deserve more attention. The $\phi$ observables are related to the formation of ``stripes'' in the crossing area. The most intuitive and geometrically straightforward way to
identify the stripes is to notice the $\approx \pi/4$ peak in $\phi^s$ (Fig. \ref{fphi} (a)),
which shows that there is a tendency to have the forward first neighbour in the same flow walking at such an angle
in the direction from which the crossing flow is coming (see also the bias towards positive values in Figs. \ref{f_phi_theta_rho} (a)  in the main text and \ref{f_delta_phi_rhoI} (b) in Appendix \ref{depon_rhoI}).
This phenomenon has been shown to happen in many models, and had been reported in the literature \cite{Naka,Ando1988,Helbing2005},
but without quantitative data support (although a very recent result has investigated the relation between stripe orientation and crossing angle \cite{mullick2021analysis}). We believe that this aspect of the dynamics deserves to be studied more in detail using clustering algorithms and the like, but since that approach is quite
different from the one of this work (relying on straightforward geometrical observables) we leave it for a future work. As discussed below in the model section, we believe that this kind of pattern
formation is somehow necessary to reproduce a natural crossing of the flows (e.g., the density/time relations) and thus should be reproduced by any kind of ``good'' model (we report a qualitative
visualisation of stripes in experiments and models in Appendix \ref{stripe}).

It was also expected that, as density grows, $\delta \theta$ would increase. Nevertheless, as many models do not include the possibility of a discrepancy between velocity and body orientation (if they
include body orientation at all), it is very important to quantify this behaviour. We may see that at most densities, $|\theta^v| \approx |\delta \theta|$ (Figs. \ref{f_phi_theta_rho} (c)  in the main text and \ref{f_theta_rhoI} (b) in Appendix \ref{depon_rhoI}).
Obviously, this does not imply that $|\theta|\approx 0$,
but we have indeed verified that $|\theta|$ depends very weakly on initial density conditions (Fig. \ref{f_theta_rhoI} (b)  in Appendix \ref{depon_rhoI}).
The situation is more subtle when the
relation between $|\theta|$ and $\rho$ is analysed, as we may see that $|\theta|$ increases at high densities (Fig. \ref{f_phi_theta_rho} (c)).
Similarly, when pdfs are investigated (Fig. \ref{ft}), we see that at high (initial condition)
densities, rare events with large $|\theta|$ are more probable. In the computational model sections, we will try to understand how important $\theta$ is in reproducing the cross-flow dynamics.

Finally, the asymmetry in the $\theta$ observables (Figs.
\ref{f_phi_theta_rho} (b) and \ref{f_theta_rhoI} (a) in Appendix \ref{depon_rhoI}) is definitely an interesting, although weak, phenomenon.
The cross-flow setting obviously is asymmetrical, although to understand whether this is enough to
cause the asymmetry in the $\theta$ observables, or whether this is due to other aspects of the experimental setting, a comparison with (future) other similar experiments may be necessary.

\section{Computational models}
\label{comp}
Here we provide a short description of the main features of the models we use to reproduce and hopefully understand cross-flow dynamics. For the operational details
of the model implementation, we refer to Appendix \ref{model_detailed}.

The CP model, introduced in its preliminary form by \cite{Zanlungo2007A,Zanlungo2007B}, and developed in \cite{Zanlungo2011}, brings the concept of collision avoidance as a way to modify the velocity to avoid a ``future predicted
collision'' \cite{Reynolds1987,VanDenBerg2011,Curtis2014} in the SFM framework. Using relative velocity information, it improves with respect to the original SFM specification in describing non-trivial situations \cite{Zanlungo2011} (for a more straightforward modification of SFM to include velocity see \cite{Johansson2007}). It qualitatively reproduces the cross-flow stripe pattern \cite{Zanlungo2007A,Zanlungo2007B}, in a way that, at least from a
qualitative viewpoint appears to be ``more smooth'' than in other collision avoidance models \cite{YoutubeMarchingParade}.

The CP model intentionally does not consider the explicit shape or dimension of the pedestrians' bodies in the prediction of the collision. Actually, the predicted ``collision'' refers to a
point of future maximum approach based on a linear velocity and point like body approximation. The ``repulsive'' forces of SFM are then applied using such future ``approach distance''.
As discussed in \cite{Zanlungo2007A,Zanlungo2007B,Zanlungo2011}, this is done to introduce in the SFM framework the ``anticipation'' of a collision, and it is intended to reproduce human behaviour in low-medium density range.

Nevertheless, at higher density, when collisions may happen in short time, it has to be expected that the body shape and orientation have to be taken into consideration.
For example, in the above data analysis, we have seen that the deviation between the body orientation and the velocity, which is measured as $\delta \theta$, increases with density,
and we have also seen that the
deviation between the body orientation and the ``direction to the goal'' (corridor axis), which is measured as $\theta$, increases with density, although less strongly. One of the purposes of this work is to investigate the importance
of this degree of freedom in crowd dynamics.

There are a few subtle points to be considered when we analyse the importance of body orientation on crowd dynamics. It is clear that, as long as we know that body orientation and velocity
orientation are considerably not aligned (Fig. \ref{ft} (b)), a model according to which the two are identified is not completely satisfying from a scientific viewpoint \cite{Yamamoto2019}.
Nevertheless, this does not
imply that such a model may not be adequate for most practical purposes. A (non-obese, healthy, normally walking) person, normally occupies a larger ``cross section'' orthogonal to motion
if body orientation and velocity are aligned. For this reason, a model in which bodies are circular and orientation is simply ignored may, in some geometrical and density settings,
perform better than a model with more realistic body dynamics; and may even better fit real data. Nevertheless, this could lead to an over-fitting problem, i.e., the model could fail in describing
more challenging conditions in which body orientation may not be ignored anymore. Furthermore, in such conditions, pedestrians may take advantage of their body asymmetry by rotating to decrease their
``cross section'' beyond the one of a circular body, creating further differences between the two models \cite{Yamamoto2019}.

The discussion above shows clearly that both factors, namely the deviation between velocity and body orientation, denoted by us with $\delta \theta$, and the ``absolute''
(i.e., with respect to the goal, or main
walking direction) deviation $\theta$, play an important role. Nevertheless, it is not clear, just by analysing the data above, to understand if we are studying conditions in which proper introduction
of the body direction degree of freedom and of its dynamics are necessary to reproduce crowd behaviour. For this reason, we are going to introduce a class of models of increasing complexity,
to try to understand which elements are necessary to describe the observed behaviour.

To simplify, we may say that we want to compare our old model CP \cite{Zanlungo2011}, that did not take into consideration body orientation and its dynamics, with a model that includes such
degrees of freedom and their dynamics. Anyway, the distinction is a little bit more subtle, as we may consider whether body shape and the related degree of freedom may be introduced somehow passively,
i.e., by introducing the new physical degrees of freedom without modifying the decision dynamics, or more actively, i.e., by performing a detailed ``prediction of collisions'' between non-circular bodies. As a result, we will be actually comparing four models, summarised below and described in detail in Appendix
\ref{model_detailed}.
\subsection{Long range (circular)}
Except a few substantial differences described in Appendix \ref{model_detailed_cp}, this model is fundamentally an updated version of the one introduced in \cite{Zanlungo2011}.
It is a velocity-dependent implementation of a (Social) Force Model, i.e., a second order differential pedestrian model \cite{Adrian2019}, known to reproduce stripe formation in cross-flows
\cite{Zanlungo2007A,Zanlungo2007B,YoutubeMarchingParade}.

Following the Social Force framework, the ``social force'', or, more properly, the acceleration due to decisional dynamics, of a pedestrian is given by
\begin{equation}
  \label{sfmacc_i}
\ddot{\mathbf{r}}= -k_{v_p} (\dot{\mathbf{r}}-\mathbf{v}_p) + \mathbf{F}^{long}. 
\end{equation}
Here, the first term on the right refers to the tendency of pedestrians to walk with velocity $\mathbf{v}_p$, determined by their preferred speed and wanted walking direction (the parameter
$k_{v_p}$ having dimension $t^{-1}$),
while
the second term $\mathbf{F}^{long}$ introduces the interaction with the environment related to collision avoidance. The specifications of this force can be re-conducted to \cite{Reynolds1987}. Fundamentally, through first order approximation (i.e.
assuming constant velocity), the time and position of maximum approach to an obstacle (possibly a moving one, such as another pedestrian) is computed, and the interaction force (acceleration) is
determined based on such ``future condition''. We somehow improperly call this model ``long range'' because the moment of future maximum approach to the obstacle is computed based only on the
geometry of trajectories (relative positions and velocity of pedestrian centre of mass), with no regard to actual body shape. Thus, although it describes local interactions with pedestrians and other obstacles, it somehow does it at a ``longer range'' than the model using detailed body shape and size introduced in Section \ref{shortr}.

Although the model may in principle be applied to different body shapes (and indeed we will use it also for elliptical bodies), by not taking into consideration the shape of the body its natural
applications is to symmetrical ones (disks).
\subsection{Short range (ellipse)}
\label{shortr}
The model of \cite{Zanlungo2011} was designed for moderate densities and did not take into account the shape of the human body. In order to describe
the motion of pedestrians at high density, it is necessary to consider at least the fact that the 2D projection of the human body is not rotationally
symmetric \cite{Chraibi2010} (this asymmetry may be first
approximated by using ellipses instead of circles\footnote{Given the theoretical approach of this work,
  we decided to follow this path since an ellipse is the most natural geometrical generalisation of a circle.
  Nevertheless, this choice is not to be recommended when the priority is given to issues concerning implementation and computational efficiency \cite{Donev2005}.}).
When such an asymmetry is introduced, even if we still limit ourselves to the motion of pedestrians on a 2D plane, we need to introduce a new degree of freedom,
body orientation angle $\theta$. Assuming body orientation to be equal to velocity orientation would be a too strong limitation, since it would not allow pedestrians to rotate
their torso while avoiding a collision without changing considerably their motion direction \cite{Yamamoto2019}. We thus consider the pedestrian to be characterised by 3 degrees of freedom, 2D position
$\mathbf{r}=(x,y)$ and angle $\theta$. The latter variable identifies the orientation of an ellipse with axes $A \geq B$.

As we are operating in the Social Force Model paradigm, we deal with second order differential equations\footnote{Or better, with the corresponding
  difference equation obtained by applying a Euler integrator.} for the pedestrian linear and angular acceleration $\dot{\mathbf{v}\equiv \ddot{\mathbf{r}}}$ and $\dot{\omega}\equiv \ddot{\theta}$, as
functions of $\mathbf{r}$, $\dot{\mathbf{r}}\equiv \mathbf{v}$, $\theta$ and $\omega\equiv \dot{\theta}$.
Assuming the $\mathbf{r}$ and $\theta$ equations to be decoupled is highly unrealistic, since, while pedestrians are indeed
able to walk in a direction different from their body orientation, they prefer moving in the direction of their body orientation, \cite{Willems2020}. By ignoring for the moment body oscillations due to gait, we propose
the following equations

\begin{equation}
  \label{eqmr2}
\dot{\mathbf{v}}= -k_{v_p} (\mathbf{v}-\mathbf{v}_p)  - k^v_{\theta} (\mathbf{v}-v \mathbf{n}) + (1-\beta(t^c)) \mathbf{F}^{long} + \beta(t^c) \mathbf{F}^{short},
\end{equation}
and
\begin{equation}
   \label{eqmt} 
\dot{\omega}= -k_{\omega} \omega -k^{\omega}_{\theta} \delta \theta + \beta(t^c) T^{short}.
\end{equation}
%inserisci SIG_DRIFT se usato, ma meglio correggere k^v_{\theta}

Many terms deserve to be explained. $k_{\omega}$ accounts for the tendency of reducing body oscillations, while $k^v_{\theta}$
and $k^{\omega}_{\theta}$ for the
tendency of walking in the direction of body orientation, through the angle difference between velocity and body orientation given either by $\delta \theta$ (eq. \ref{def_deltath}
in Appendix \ref{obs_def})
and the vectorial difference between the pedestrian velocity $\mathbf{v}$ and the velocity they would have if walking with equal speed in the direction given by their chest
normal $\mathbf{n}$ (see eq. \ref{chest} in Appendix \ref{obs_def}).
$t^c$ is the time at which the first collision
will happen between the ellipse representing the pedestrian body and an obstacle (such as a wall or another pedestrian), computed using an event driven algorithm \cite{Foulaadvand2013} under the assumption
of no acceleration. The function $\beta$ introduces the idea that far away collisions are managed just using the model of \cite{Zanlungo2011}, while close ones use forces taking into account body shape
($\mathbf{F}^{short}$, defined below). In detail, $\beta$ introduces two time scales $\tau_1$, $\tau_2$ such that 
\begin{equation}
  \label{ebeta}
\beta(t)= \begin{cases} 
1 \text{ if } t \leq \tau_1,\\
\frac{\tau_2-t}{\tau_2-\tau_1} \text{ if } \tau_1 < t \leq \tau_2.\\
0 \text{ if } t>\tau_2.\\  
\end{cases}
\end{equation}
As stated above, we use an event driven algorithm to compute the time of the next collision between ellipses and obstacles (wall polygons or other ellipses). This algorithm can be used to reproduce
the dynamics of hard ellipses undergoing elastic collisions, i.e., it provides also the force and torque that the ellipse undergoes at the moment of collision, under the assumption of
impenetrability and conservation of energy and momentum\footnote{These are actually impulsive forces and torques, i.e., expressed as an instantaneous change in linear and angular momentum.}. These
forces and momentum can be used as the basis of a collision avoidance method. To understand the logic behind the model, let us first for simplicity consider the case of a 2D disc (i.e., ignoring $\theta$) colliding frontally with a wall
at velocity $\mathbf{v}$ in time $t^c$ ($t^c$ is the time from now at which the collision will happen if the pedestrian velocity is not modified). At the moment of collision,
the pedestrian will undergo a change in velocity $\mathbf{\Delta v}=-2\mathbf{v}$. If a force $\mathbf{\Delta v}/2t^c$ is applied on the pedestrian, collision will be avoided, with the pedestrian just
stopping short of the collision\footnote{Obviously, if the equations of motion are exactly integrated in continuous time it can be shown that the pedestrian reaches zero velocity
  half the way to the obstacle. Nevertheless, the approximation in the text still provides the correct order of magnitude.}. By generalising to the elliptical case, we define, with respect to pedestrian $i$, for all static obstacles (i.e., walls, represented as polygons)  and moving obstacles (i.e., pedestrians) $j$ in the environment the predicted time of collision with
$i$ as $t^c_{i,j}$, and the predicted change of momentum and angular momentum of $i$ in the collision as, respectively\footnote{Impulsive forces are assumed as orthogonal to the contact surface. As a results, they are central forces for disks, and in the Circular case we have $\mathbf{\Delta L}_{i,j}=\mathbf{0}$, consistent with our choice to ignore the variable $\theta$ for disks.}, $\mathbf{\Delta p}_{i,j}$ and $\mathbf{\Delta L}_{i,j}$.
Assuming $t^c_{i,j}= + \infty$ if no collision is predicted, and defining
\begin{equation}
t_i^c=\min_j t^c_{i,j}\;\; \overline{j}=\argmin_j{t^c_{i,j}},
\end{equation}
we finally have
\begin{equation}
\mathbf{F}_i^{short}= \frac{\gamma}{t_i^c} \mathbf{\Delta p}_{i\overline{j}},
  \end{equation}
\begin{equation}
\mathbf{T}_i^{short}= \frac{\gamma}{t_i^c} \mathbf{\Delta L}_{i\overline{j}},
\end{equation}
$\gamma$ being a model parameter.
\subsection{Physical dynamics}
\label{phys_dyn}
Equations \ref{eqmr2} and \ref{eqmt} realise the ``social'' interaction of pedestrians, i.e., their decisional process, that is, for simplicity's sake, realised at a fixed time step $\Delta t=0.05$
s (i.e., the decisional dynamics is solved by using an Euler integrator\footnote{As discussed in our previous works on the CP model \cite{Zanlungo2011},
  it is important that $\Delta t$ is comparable
to the time scale of human decisions, i.e., to the reaction times of humans to external events. In this work, we did not study model dependence on $\Delta t$.}). For the physical dynamics of the system, i.e., in order to deal with {\it actual} (as opposed to predicted) collisions
between pedestrians, we use again an event driven algorithm for elastic collisions between ellipses. Such an algorithm has the advantages of forcing absence of overlapping,
and thus constraints on space occupation, and providing a quantitative measure for the amount of collisions (kinetic energy, both linear and angular, exchanged in collisions\footnote{This information was not used in this work, since we were missing empirical data on possible collisions between pedestrians in the experiments.} \cite{Zanlungo2020}).
The negative aspect is that elastic collisions are not very realistic between pedestrians, but this problem should not be a serious one for a functional collision avoidance method, i.e.
a method in which collisions are extremely reduced\footnote{The implemented algorithm presented a convergence problem in extremely packed settings. Namely, events were predicted consistently
  at time steps close to machine precision, ``stalling'' the numerical integrator. Since such packed settings are not expected to occur in proper pedestrian behaviour,
  these convergence problems occur only in the first steps of the optimisation process. We thus introduced in our Genetic Algorithm optimisation procedure an exception to deal with such problems, namely halting the integration and
  giving a very low fitness to the solution that caused the problem.}.

\subsection{Possible model specifications}
As stated above, the main purpose of the modelling portion of this work is not to propose the best or most efficient model to reproduce the data, but to {\it understand which fundamental
  behaviours or physical features have to be introduced in the model to reproduce the crowd dynamics}. By this, we mean that although most of the details
(e.g., the specific form of interaction force function, or even something more fundamental such as using a second order model that specifies accelerations as opposed to using a first order approach
that acts directly on velocities) of the proposed model could be replaced by possibly better alternatives obtaining a {\it quantitative} change in the model's ability to reproduce crowd behaviour,
we are here interested in more fundamental and arguably {\it qualitative} differences.

We believe these differences to be the following:
\begin{enumerate}
\item Introducing or not an asymmetry in the shape of the human body, and the corresponding angle dynamics.
  \item Introducing or not a {\it short range} behaviour, in which pedestrians react to close-time collisions by taking explicitly into account the shape of their body.
\end{enumerate}

The asymmetry in the human body is introduced by using an ellipse with semi-axes $A \neq B$, and by using non-zero values of $k^v_{\theta}$, $k_{\omega}$, $k^{\omega}_{\theta}$.
We will call such a choice of parameters an {\it elliptical} specification of the proposed model\footnote{This should not be confused with the Elliptical Specification of SFM proposed by
  \cite{Johansson2007}, that corresponds to modify the SFM force potential creating an asymmetry due to the velocity direction, but unrelated to body shape.}. In this work, for simplicity's sake,
in elliptical specifications we use $A_E=0.225$ m, $B_E=0.1$ m, which may be considered to provide a rough approximation of the 2D projection of the body of Japanese
young adults that took part in the experiment \cite{AIST1992}.

On the other hand, using a circle with $R=A_c=B_c=\sqrt{A_E B_E}$ (so that the same 2D area is occupied by the pedestrian's body) and zero values for $k^v_{\theta}$, $k_{\omega}$, $k^{\omega}_{\theta}$,
corresponds to a {\it circular} specification of the proposed model\footnote{Again, this should not be confused with the original SFM model of \cite{Helbing1995}, which is referred to as a Circular
  Specification by  \cite{Johansson2007} and \cite{Zanlungo2011} due to the lack of velocity direction dependent asymmetry in its interaction force.}.

The short range behaviour is introduced by having non-zero values for parameters $\gamma$, $\tau_1$ and $\tau_2$. We refer to such a model specification as {\it full} (meaning it includes both
short and long range behaviours), while a model with $\gamma=\tau_1=\tau_2=0$ is referred to as {\it no-short} (range behaviour)\footnote{Since we consider the \cite{Zanlungo2011} model as a starting
  point, due to its ability to reproduce qualitatively the stripe formation in a cross-flow \cite{Zanlungo2007A,Zanlungo2007B,YoutubeMarchingParade}, all models include the long range module.}.

We will thus consider, and explicitly calibrate, four possible qualitatively different alternatives of the proposed model: 1) {\it Full Elliptical} (often just referred to as {\it Full})
2)  {\it Full Circular}, 3) {\it No-short Elliptical} and 4) {\it No-short Circular} (often referred to as CP).
In our analysis of the results, we will need to take into account that different specifications
of the model correspond to different dimensions of the parameter space, and thus different optimisation problems. Specifically, 8 parameters need to be calibrated for a {\it No-short Circular} specification,
11 for the {\it Full Circular} and {\it No-short Elliptical} ones, and 14 for a {\it Full Elliptical} specification (see Appendix \ref{model_detailed_par}).

\section{Full model calibration and evaluation}
\label{full_cal}
\subsection{Calibration}
\label{cal_sec}
We used Genetic Algorithms (24 runs, each one using 30 generations, 50 genomes and 8 different sets of mutation, crossover and selection hyper-parameters) to calibrate the {\it Full Elliptical} model (the quantitative comparison
to the other models is discussed in Section \ref{modcomp}).
The fitness function was based on an EMD (Earth Mover Distance, Appendix \ref{evalmet}) comparison between
the observables at initial densities of 0.25 and 1.5 ped/m$^2$, and the ones produced by the model. Initial conditions (entry points in the crossing area and corresponding velocity)
were taken from the data, and each experimental run was reproduced. The used observables are $\rho$ (through the $E_t$ pdf), $v$, $\theta_v$, $\theta$, $\delta \theta$, $\delta^s$ and $\delta^o$.
Through an exploratory phase, we observed that after calibration with the above observables, the model was able to reproduce qualitatively the stripe pattern,
and thus we decided not to use $\phi^s$ and $\phi^o$ in the calibration fitness function. The results concerning these observables are anyway shown
for evaluation. The fitness was the sum (of the negative of) the EMD distances between the experimental and model observable pdfs (sum over density initial conditions and observables; the used
pdfs for a given initial condition and observable are the average over repetitions, normalised to have 1 as the sum over all bin probabilities).

\begin{figure}
\begin{center}
  \includegraphics[width=0.8\textwidth]{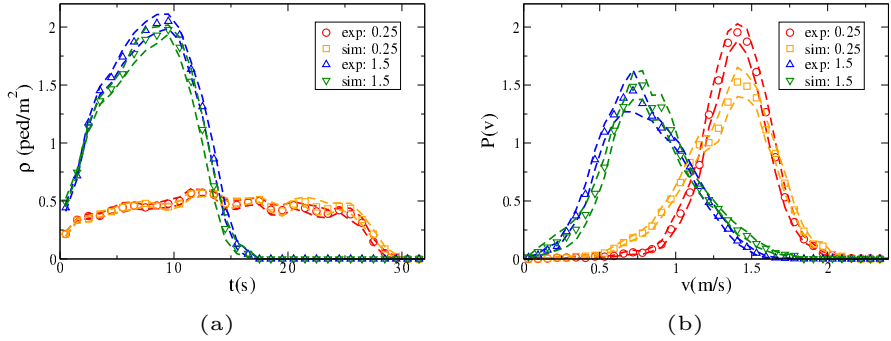}
  \caption{\label{f_rho_v_cal}(a): $\rho(t)$ for the experimental $\rho_I=$ 0.25 ped/m$^2$ initial condition (exp: 0.25, red circles), the calibrated
    $\rho_I=$ 0.25 ped/m$^2$ initial condition (sim: 0.25, orange squares), the experimental $\rho_I=$ 1.5 ped/m$^2$ initial condition (exp: 1.5, blue up triangles), the calibrated
    $\rho_I=$ 1.5 ped/m$^2$ initial condition (sim: 1.5, green down triangles).  (b): $P(v)$ for the experimental $\rho=$ 0.25 ped/m$^2$ initial condition (exp: 0.25, red circles), the calibrated
    $\rho=$ 0.25 ped/m$^2$ initial condition (sim: 0.25, orange squares), the experimental $\rho=$ 1.5 ped/m$^2$ initial condition (exp: 1.5, blue up triangles), the calibrated
    $\rho=$ 1.5 ped/m$^2$ initial condition (sim: 1.5, green down triangles).
    Dashed lines provide standard error intervals (computed over independent repetitions).}
 \end{center}   
\end{figure}

\begin{figure}
\begin{center}
  \includegraphics[width=0.8\textwidth]{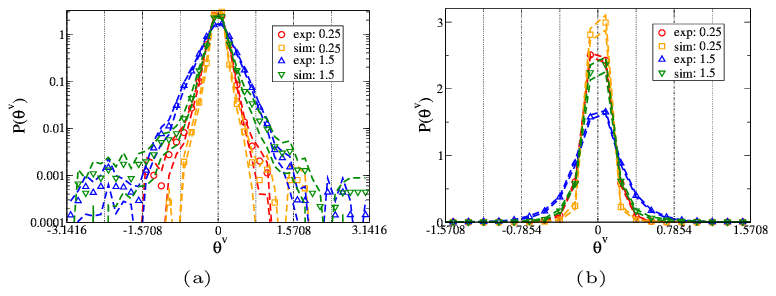}
  \caption{\label{f_thetav_cal}$P(\theta^v)$ for the experimental $\rho_I=$ 0.25 ped/m$^2$ initial condition (exp: 0.25, red circles), the calibrated
    $\rho_I=$ 0.25 ped/m$^2$ initial condition (sim: 0.25, orange squares), the experimental $\rho_I=$ 1.5 ped/m$^2$ initial condition (exp: 1.5, blue up triangles), the calibrated
    $\rho_I=$ 1.5 ped/m$^2$ initial condition (sim: 1.5, green down triangles). (a): logarithmic plot; (b): linear plot. Dashed lines provide standard error intervals (computed over independent repetitions).}
 \end{center}   
\end{figure}

\begin{figure}
\begin{center}
  \includegraphics[width=0.8\textwidth]{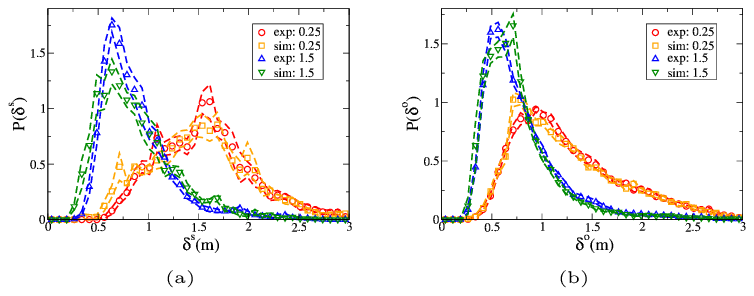}
  \caption{\label{f_r_cal}(a): $P(\delta^s)$ for the experimental $\rho_I=$ 0.25 ped/m$^2$ initial condition (exp: 0.25, red circles), the calibrated
    $\rho_I=$ 0.25 ped/m$^2$ initial condition (sim: 0.25, orange squares), the experimental $\rho_I=$ 1.5 ped/m$^2$ initial condition (exp: 1.5, blue up triangles), the calibrated
    $\rho_I=$ 1.5 ped/m$^2$ initial condition (sim: 1.5, green down triangles).  (b): $P(\delta^o)$ for the experimental $\rho_I=$ 0.25 ped/m$^2$ initial condition (exp: 0.25, red circles), the calibrated
    $\rho_I=$ 0.25 ped/m$^2$ initial condition (sim: 0.25, orange squares), the experimental $\rho_I=$ 1.5 ped/m$^2$ initial condition (exp: 1.5, blue up triangles), the calibrated
    $\rho_I=$ 1.5 ped/m$^2$ initial condition (sim: 1.5, green down triangles). Dashed lines provide standard error intervals (computed over independent repetitions).}
 \end{center}   
\end{figure}

\begin{figure}
\begin{center}
  \includegraphics[width=0.8\textwidth]{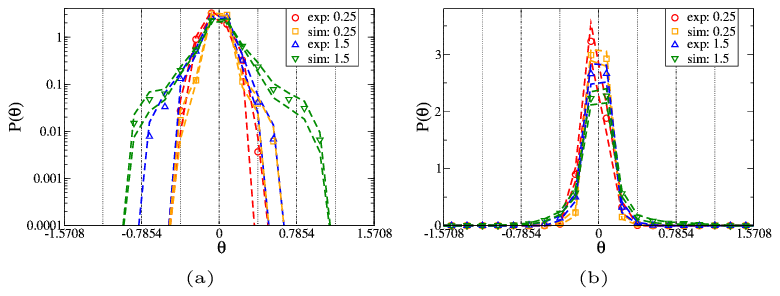}
  \caption{\label{f_theta_cal}$P(\theta)$ for the experimental $\rho_I=$ 0.25 ped/m$^2$ initial condition (exp: 0.25, red circles), the calibrated
    $\rho_I=$ 0.25 ped/m$^2$ initial condition (sim: 0.25, orange squares), the experimental $\rho_I=$ 1.5 ped/m$^2$ initial condition (exp: 1.5, blue up triangles), the calibrated
    $\rho_I=$ 1.5 ped/m$^2$ initial condition (sim: 1.5, green down triangles). (a): logarithmic plot.  (b): linear plot.
    Dashed lines provide standard error intervals (computed over independent repetitions).}
 \end{center}   
\end{figure}

\begin{figure}
\begin{center}
  \includegraphics[width=0.8\textwidth]{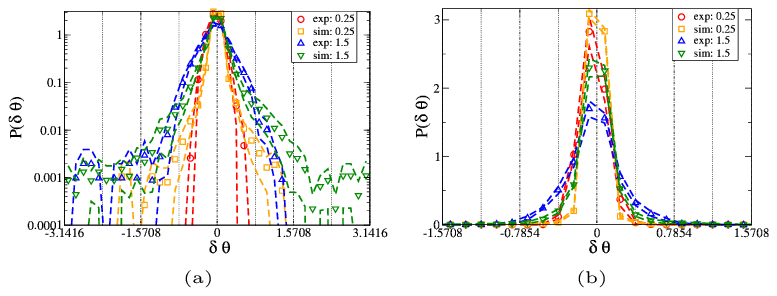}
  \caption{\label{f_delta_theta_cal}$P(\delta \theta)$ for the experimental $\rho_I=$ 0.25 ped/m$^2$ initial condition (exp: 0.25, red circles), the calibrated
    $\rho_I=$ 0.25 ped/m$^2$ initial condition (sim: 0.25, orange squares), the experimental $\rho_I=$ 1.5 ped/m$^2$ initial condition (exp: 1.5, blue up triangles), the calibrated
    $\rho_I=$ 1.5 ped/m$^2$ initial condition (sim: 1.5, green down triangles). (a): logarithmic plot.  (b): linear plot.
    Dashed lines provide standard error intervals (computed over independent repetitions).}
 \end{center}   
\end{figure}

\begin{figure}
\begin{center}
  \includegraphics[width=0.8\textwidth]{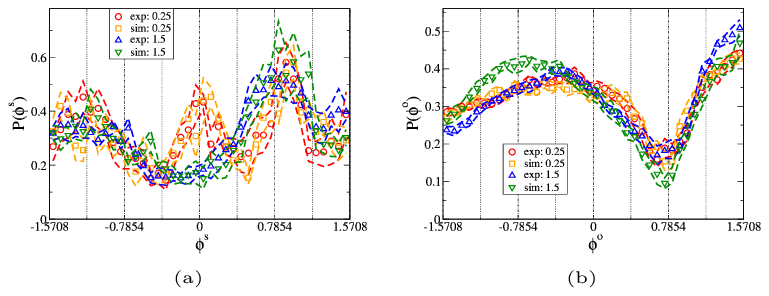}
  \caption{\label{f_phi_cal}(a): $P(\phi^s)$ for the experimental $\rho_I=$ 0.25 ped/m$^2$ initial condition (exp: 0.25, red circles), the calibrated
    $\rho_I=$ 0.25 ped/m$^2$ initial condition (sim: 0.25, orange squares), the experimental $\rho_I=$ 1.5 ped/m$^2$ initial condition (exp: 1.5, blue up triangles), the calibrated
    $\rho_I=$ 1.5 ped/m$^2$ initial condition (sim: 1.5, green down triangles).
(b): $P(\phi^o)$ for the experimental $\rho_I=$ 0.25 ped/m$^2$ initial condition (exp: 0.25, red circles), the calibrated
    $\rho_I=$ 0.25 ped/m$^2$ initial condition (sim: 0.25, orange squares), the experimental $\rho_I=$ 1.5 ped/m$^2$ initial condition (exp: 1.5, blue up triangles), the calibrated
    $\rho_I=$ 1.5 ped/m$^2$ initial condition (sim: 1.5, green down triangles). 
    Dashed lines provide standard error intervals (computed over independent repetitions).}
 \end{center}   
\end{figure}

Figs. \ref{f_rho_v_cal}-\ref{f_phi_cal} show the results for one of the best performing solutions.
%({\bf note for myself: solution 72}).


\subsection{Evaluation}
\label{ev_sec}
\begin{figure}
\begin{center}
  \includegraphics[width=0.8\textwidth]{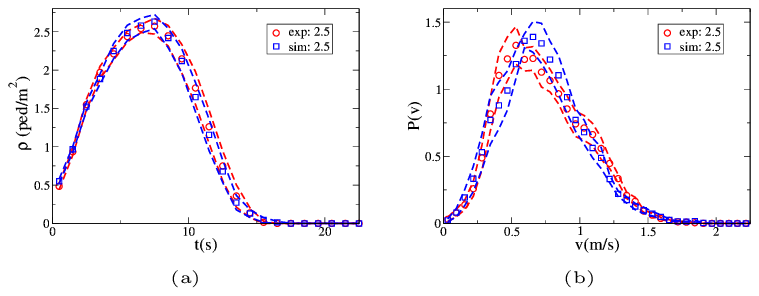}
  \caption{\label{f_rho_v_ev}(a): $\rho(t)$ for the experimental $\rho_I=$ 2.5 ped/m$^2$ initial condition (exp: 2.5, red circles) and the calibrated
    $\rho_I=$ 2.5 ped/m$^2$ initial condition (sim: 2.5, blue squares).
(b): $P(v)$ for the experimental $\rho_I=$ 2.5 ped/m$^2$ initial condition (exp: 2.5, red circles) and the calibrated
    $\rho_I=$ 2.5 ped/m$^2$ initial condition (sim: 2.5, blue squares).
    Dashed lines provide standard error intervals (computed over independent repetitions).}
 \end{center}   
\end{figure}

\begin{figure}
\begin{center}
  \includegraphics[width=0.8\textwidth]{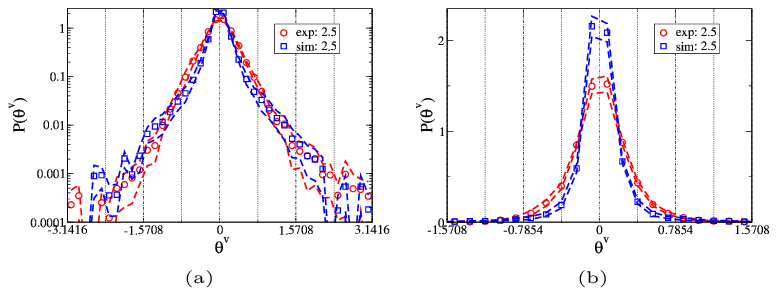}
  \caption{\label{f_thetav_ev}$P(\theta^v)$ for the experimental $\rho_I=$ 2.5 ped/m$^2$ initial condition (exp: 2.5, red circles) and the calibrated
    $\rho_I=$ 2.5 ped/m$^2$ initial condition (sim: 2.5, blue squares).  (a): logarithmic plot. (b): linear plot.
    Dashed lines provide standard error intervals (computed over independent repetitions).}
 \end{center}   
\end{figure}

\begin{figure}
\begin{center}
  \includegraphics[width=0.8\textwidth]{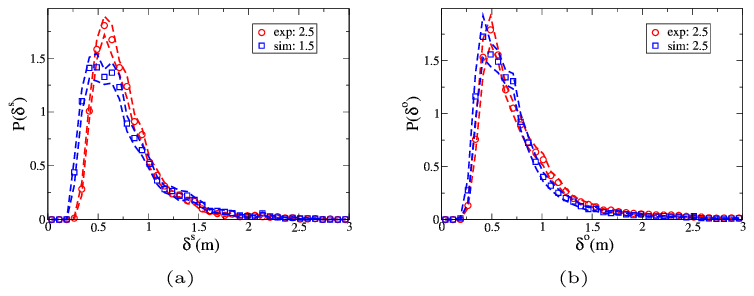}
  \caption{\label{f_r_ev}(a): $P(\delta^s)$ for the experimental $\rho_I=$ 2.5 ped/m$^2$ initial condition (exp: 2.5, red circles) and the calibrated
    $\rho_I=$ 2.5 ped/m$^2$ initial condition (sim: 2.5, blue squares). (b): $P(\delta^o)$ for the experimental $\rho_I=$ 2.5 ped/m$^2$ initial condition (exp: 2.5, red circles) and the calibrated
    $\rho_I=$ 2.5 ped/m$^2$ initial condition (sim: 2.5, blue squares).
    Dashed lines provide standard error intervals (computed over independent repetitions).}
 \end{center}   
\end{figure}

\begin{figure}
\begin{center}
  \includegraphics[width=0.8\textwidth]{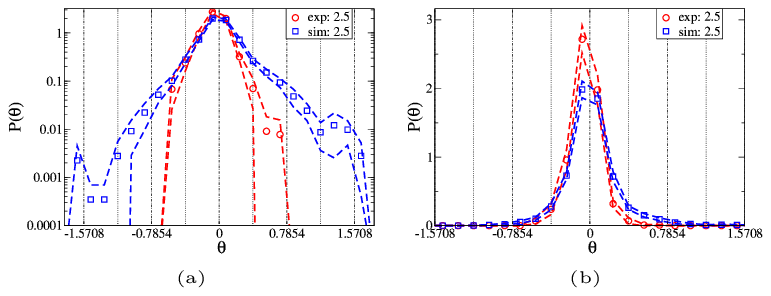}
  \caption{\label{f_theta_ev}$P(\theta)$ for the experimental $\rho_I=$ 2.5 ped/m$^2$ initial condition (exp: 2.5, red circles) and the calibrated
    $\rho_I=$ 2.5 ped/m$^2$ initial condition (sim: 2.5, blue squares).
 (a): logarithmic plot. (b): linear plot.
    Dashed lines provide standard error intervals (computed over independent repetitions).}
 \end{center}   
\end{figure}

\begin{figure}
\begin{center}
  \includegraphics[width=0.8\textwidth]{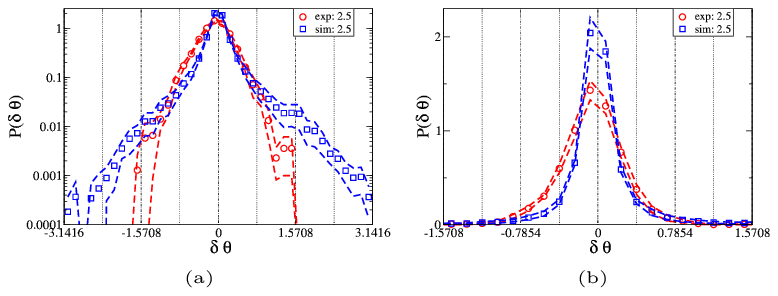} 
  \caption{\label{f_delta_theta_ev}$P(\delta \theta)$ for the experimental $\rho_I=$ 2.5 ped/m$^2$ initial condition (exp: 2.5, red circles) and the calibrated
    $\rho_I=$ 2.5 ped/m$^2$ initial condition (sim: 2.5, blue squares).
 (a): logarithmic plot. (b): linear plot.
    Dashed lines provide standard error intervals (computed over independent repetitions).}
 \end{center}   
\end{figure}

\begin{figure}
\begin{center}
  \includegraphics[width=0.8\textwidth]{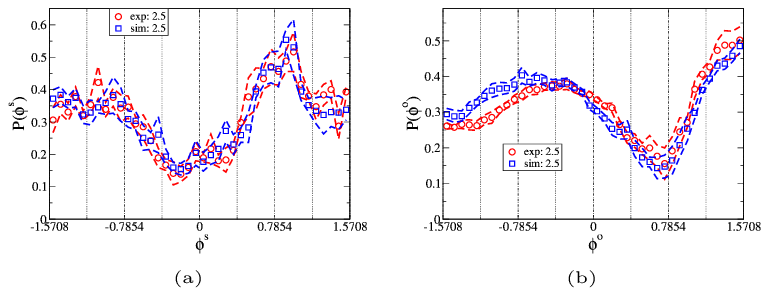}
  \caption{\label{f_phi_ev}(a): $P(\phi^s)$ for the experimental $\rho_I=$ 2.5 ped/m$^2$ initial condition (exp: 2.5, red circles) and the calibrated
    $\rho_I=$ 2.5 ped/m$^2$ initial condition (sim: 2.5, blue squares).
(b): $P(\phi^o)$ for the experimental $\rho_I=$ 2.5 ped/m$^2$ initial condition (exp: 2.5, red circles) and the calibrated
    $\rho_I=$ 2.5 ped/m$^2$ initial condition (sim: 2.5, blue squares).
    Dashed lines provide standard error intervals (computed over independent repetitions).}
 \end{center}   
\end{figure}

We run the model over the $\rho=2.5$ ped/m$^2$ initial condition and compare to experimental results in Figs. \ref{f_rho_v_ev}-\ref{f_phi_ev} (same solution as in calibration).

\subsection{Comments}
The examined solution appears to reproduce almost perfectly the time evolution of density (Figs. \ref{f_rho_v_cal} (a) and \ref{f_rho_v_ev} (a)) for all the three initial conditions,
both in calibration and evaluation. This is obviously one of the most important features of pedestrian crowd models. Not all GA solutions are equally good at reproducing the empirical data,
and some present a longer tail, due to few pedestrians being stuck for a long time in one corner (at the crossing point of the exiting flows).
According to a qualitative visual analysis of the simulations, this behaviour is probably mainly caused by the triviality of the path planning mechanism (Appendix \ref{pathfind}).
Nevertheless, as it happens for
high density scenarios, it is clearly accentuated by poor collision avoiding, and it does not appear in good performing models and solutions. The longer tail
is better analysed in the logarithmic plots of Fig. \ref{rho_comp} (b) in Appendix \ref{graph_comp}, which includes for all models the average over all solutions, including a few sub-optimal ones.

Concerning the speed distribution (Figs. \ref{f_rho_v_cal} (b) and \ref{f_rho_v_ev} (b)), the analysed solution  is able to reproduce
in a satisfying way calibration and evaluation distributions. On the other hand, some of the  solutions reproduced pdfs skewed towards higher velocities at low densities.
This tendency could be probably due to the fact that the model assumes a density-independent preferred speed, while the actual preferred
speed may be different between free walking and packed settings.

Concerning $\theta^v$, the chosen solution appears to describe better the tails (logarithmic plots in Figs. \ref{f_thetav_cal} (a) and \ref{f_thetav_ev} (a)) than the bulk (linear plots in Figs.
\ref{f_thetav_cal} (b) and \ref{f_thetav_ev} (b)) of the distributions. This behaviour is not typical, since, as shown in Fig. \ref{theta_v_comp} in Appendix \ref{graph_comp}, most solutions
have (for any density)
a tendency to overestimate the values of $P(\theta)$ at $\theta\approx 0$ and $|\theta|\approx \pi$, while intermediate values are underestimated.

Excluding a slight underestimation of the minimum value in the high-density support (non-zero domain) of $P(\delta^s)$, the reproduction of the $\delta$ observables is almost perfect at
low and high densities, in calibration and evaluation (Figs. \ref{f_r_cal} (a), \ref{f_r_cal} (b), \ref{f_r_ev} (a) and \ref{f_r_ev} (b)).

Reproducing the $\theta$ observables is probably the main task of the proposed model. The chosen solution is, compared to other solutions (see Appendix \ref{graph_comp}), relatively
good in reproducing the tails of the distributions
(logarithmic plots in Figs. \ref{f_theta_cal} (a), \ref{f_delta_theta_cal} (a), \ref{f_theta_ev} (a)
and \ref{f_delta_theta_ev} (a)), while still describing in a satisfactory way the $\theta$ bulks, although it underestimates the $\delta \theta$ ones
(linear plots in Figs. \ref{f_theta_cal} (b), \ref{f_delta_theta_cal} (b), \ref{f_theta_ev} (b) and \ref{f_delta_theta_ev} (b)).
Nevertheless, even this solution fails in reproducing the fact that in empirical distributions $|\delta \theta|$ is limited to assume values smaller than $\pi/2$, while
$|\theta|$ is limited to assume values smaller than $\pi/4$.
This tendency is stronger in less  performing  GA (Genetic Algorithm) solutions (Figs. \ref{theta_comp}, \ref{delta_theta_comp} in Appendix \ref{graph_comp}), that usually
overestimate the probability of having very large deviations from $0$, the latter effect being stronger for $\theta$ as compared to $\delta \theta$.

There are probably two reasons for this phenomenon. One is the choice
of the calibration method, since the EMD metric does not give a high weight to areas in which $P \ll 1$. The second one is related to the model itself. While our model
includes a linear recall force in $\delta \theta$ (Eq. \ref{eqmt}), Fig. \ref{f_theta_cal} (a) and  \ref{f_theta_ev} (a) suggest that pedestrians have a strong tendency, almost a constraint, to
avoid $|\theta|>\pi/4$. 
For this reason, non-linearity may be introduced in future models.

Finally, concerning $\phi$ observables, we may recall that these were not calibrated, so all the results in Figs. \ref{f_phi_cal} (a), \ref{f_phi_cal} (b), \ref{f_phi_ev} (a) and \ref{f_phi_ev} (b)
have to be considered as evaluation of a dynamical property of the system that has not been calibrated. It is our opinion that any model that is able to reproduce fairly the other observables
(and maybe even just the time-density relation), should exhibit the stripe pattern, and thus reproduce at least qualitatively the maxima and minima in the figures (see Appendix \ref{stripe} for a qualitative visualisation of stripe formation in experiments and models).
\section{Model comparison}
\label{modcomp}
To compare models, we evaluate them on the $\rho=2.5$ ped/m$^2$ initial condition. More in detail, we obtain for each model 24 independent GA solutions (calibrating on
$\rho=0.25$ ped/m$^2$ and $\rho=1.5$ ped/m$^2$ initial conditions), and evaluate them on the $\rho=2.5$ ped/m$^2$ initial condition. For each observable, we compute the EMD, and average
it over the different GA solutions. As a way to compare the overall performance of a model, we also provide a ``Total evaluation measure''  as an average over all the observables on which
the model was calibrated. Nevertheless, as observables $\theta$ and $\delta \theta$ are not defined for circular models, all comparisons involving a circular model are performed computing a
``No body (angles) evaluation measure''
(i.e., not considering $\theta$ and $\delta \theta$). Observables $\phi^s$ and $\phi^o$ are not included in the ``Total'' and ``No body'' evaluation measures, since
relative angles were not used in calibration. Nevertheless, the EMD values for these observables are reported in tables.

For each model we performed 24 calibration runs, using 8 different sets of hyper-parameters (i.e., each set is used 3 times).

In Appendix \ref{standapp}, we also use another metric to compare observables.
For each bin, we compare the simulation value to the experimental value, and divide the difference by the experimental standard error. Finally, we
compute the squared quadratic average of such difference for all observables. We call this measure ``Standard Metric''. While EMD verifies that the bulk of a distribution is located in the correct place,
the Standard Metric verifies that an observable assumes nowhere a value quite higher in the simulation observable distribution with respect to the experimental one.
The Standard Metric may be used directly on $\rho$, and may assume a high value if, for example, the simulated ``crossing time'' gets longer than the experimental one.
This difference may concern only few pedestrians, and then not have a strong impact on EMD. But it may be relevant in model evaluation, as we may expect that if a model calibrated
on $\rho=0.25$ ped/m$^2$ and $\rho=1.5$ ped/m$^2$ initial conditions presents a behaviour different from the experimental one on the evaluation $\rho=2.5$ ped/m$^2$ initial condition, this
behaviour needs to be identified properly (as it could become relevant at even higher densities).

As tools to quantify differences and guide our discussion (without any claim to ``prove'' or ``disprove'' anything), we compute for each metric an ANOVA $p$ value and effect size (see Appendix \ref{statan} for definitions). A comparison on how the model performs on the low (0.25 ped/m$^2$) and medium (1.5 ped/m$^2$)
initial density conditions on which it was calibrated is reported in Appendix \ref{appcomp}.

\subsection{Importance of short range}
We first verify what happens if we remove the short range behaviour, i.e., if we compare the {\it Full} model to a {\it No-short} Elliptical specification (see Table \ref{EMD} in the main text and Table \ref{Standard} in Appendix \ref{standeval}).

\begin{table}[h!]
\begin{center}
\caption{EMD metric comparison between Full and No-short elliptical models, evaluation at $\rho_I=$ 2.5 ped/m$^2$.}
\label{EMD}
\begin{tabular}{|c|c|c|c|c|c|}
\hline
& Full $\langle \; \rangle \pm \varepsilon (\sigma)$ & No-short $\langle \;  \rangle \pm \varepsilon (\sigma)$ & $p$ & effect size\\
\hline
Total &  $0.0185\pm 0.00084(0.004)$ & $0.0227\pm 0.00062(0.003)$ & 0.00023 &1.2\\
\hline
$\rho$ &  $0.00967\pm 0.0021(0.01)$ & $0.0142\pm 0.0017(0.0082)$ & 0.1 &0.49\\
\hline
$v$ &  $0.0194\pm 0.0034(0.016)$ & $0.0291\pm 0.003(0.015)$ & 0.037 &0.63\\
\hline
$\theta^v$ &  $0.00887\pm 0.00053(0.0025)$ & $0.0109\pm 0.00091(0.0044)$ & 0.063 &0.56\\
\hline
$\delta^s$ &  $0.0189\pm 0.0011(0.0054)$ & $0.0257\pm 0.00092(0.0044)$ & 2.5e-05 &1.4\\
\hline
$\delta^o$ &  $0.0164\pm 0.0015(0.0074)$ & $0.0198\pm 0.0024(0.011)$ & 0.24 &0.35\\
\hline
$\phi^s$ &  $0.0207\pm 0.0017(0.0083)$ & $0.0314\pm 0.0021(0.0099)$ & 0.00025 &1.2\\
\hline
$\phi^o$ &  $0.0424\pm 0.0032(0.015)$ & $0.0384\pm 0.0033(0.016)$ & 0.39 &0.26\\
\hline
$\theta$ &  $0.0156\pm 0.0008(0.0039)$ & $0.0164\pm 0.0019(0.0093)$ & 0.69 &0.12\\
\hline
$\delta \theta$ &  $0.0144\pm 0.00066(0.0032)$ & $0.018\pm 0.00099(0.0047)$ & 0.0049 &0.87\\
\hline
\end{tabular}
\end{center}
\end{table}

\subsection{Full models: Circular body versus elliptical body}
We then compare  (see Table \ref{EMD2}  in the main text and Table \ref{Standard2} in Appendix \ref{standeval}) the Full Elliptical body model to the Full Circular one.

\begin{table}[h!]
\begin{center}
\caption{EMD metric comparison between  Full Elliptical and Full Circular models, evaluation at $\rho_I=$ 2.5 ped/m$^2$.}
\label{EMD2}
\begin{tabular}{|c|c|c|c|c|c|}
\hline
& Elliptical $\langle \;  \rangle \pm \varepsilon (\sigma)$ & Circular $\langle \;  \rangle \pm \varepsilon (\sigma)$ & $p$ & effect size\\
\hline
No body &  $0.0146\pm 0.0014(0.0067)$ & $0.0155\pm 0.001(0.0049)$ & 0.61 &0.15\\
\hline
$\rho$ &  $0.00967\pm 0.0021(0.01)$ & $0.0136\pm 0.0018(0.0085)$ & 0.17 &0.41\\
\hline
$v$ &  $0.0194\pm 0.0034(0.016)$ & $0.0206\pm 0.0022(0.01)$ & 0.75 &0.095\\
\hline
$\theta^v$ &  $0.00887\pm 0.00053(0.0025)$ & $0.00818\pm 0.00082(0.0039)$ & 0.48 &0.21\\
\hline
$\delta^s$ &  $0.0189\pm 0.0011(0.0054)$ & $0.0192\pm 0.0012(0.0057)$ & 0.82 &0.069\\
\hline
$\delta^o$ &  $0.0164\pm 0.0015(0.0074)$ & $0.016\pm 0.0017(0.0081)$ & 0.87 &0.049\\
\hline
$\phi^s$ &  $0.0207\pm 0.0017(0.0083)$ & $0.0198\pm 0.0025(0.012)$ & 0.76 &0.09\\
\hline
$\phi^o$ &  $0.0424\pm 0.0032(0.015)$ & $0.0367\pm 0.0027(0.013)$ & 0.17 &0.41\\
\hline
\end{tabular}
\end{center}
\end{table}

\subsection{Full Elliptical versus Circular No-short (CP) model}
We also compare  (see Table \ref{EMD3}  in the main text and Table \ref{Standard3} in Appendix \ref{standeval}) the Full Elliptical model to the No-Short Circular (CP) model.
\begin{table}[h!]
\begin{center}
\caption{EMD metric comparison between Full Elliptical  and No-Short Circular  (CP) models, evaluation at $\rho_I=$ 2.5 ped/m$^2$.}
\label{EMD3}
\begin{tabular}{|c|c|c|c|c|c|}
\hline
& Full $\langle \;  \rangle \pm \varepsilon (\sigma)$ & CP $\langle \;  \rangle \pm \varepsilon (\sigma)$ & $p$ &  effect size\\
\hline
No body &  $0.0146\pm 0.0014(0.0067)$ & $0.0192\pm 0.0017(0.0081)$ & 0.041 &0.62\\
\hline
$\rho$ &  $0.00967\pm 0.0021(0.01)$ & $0.0139\pm 0.0025(0.012)$ & 0.21 &0.37\\
\hline
$v$ &  $0.0194\pm 0.0034(0.016)$ & $0.0289\pm 0.004(0.019)$ & 0.072 &0.54\\
\hline
$\theta^v$ &  $0.00887\pm 0.00053(0.0025)$ & $0.00937\pm 0.0007(0.0033)$ & 0.57 &0.17\\
\hline
$\delta^s$ &  $0.0189\pm 0.0011(0.0054)$ & $0.0239\pm 0.0012(0.0056)$ & 0.003 &0.93\\
\hline
$\delta^o$ &  $0.0164\pm 0.0015(0.0074)$ & $0.0201\pm 0.0022(0.011)$ & 0.17 &0.41\\
\hline
$\phi^s$ &  $0.0207\pm 0.0017(0.0083)$ & $0.0283\pm 0.0021(0.01)$ & 0.0079 &0.82\\
\hline
$\phi^o$ &  $0.0424\pm 0.0032(0.015)$ & $0.0291\pm 0.0036(0.017)$ & 0.0081 &0.82\\
\hline
\end{tabular}
\end{center}
\end{table}

\subsection{Circular Full versus circular No-short (CP) model}
We finally compare  (see Table \ref{EMD4}  in the main text and Table \ref{Standard4} in Appendix \ref{standeval}) how taking into consideration short range collisions impacts a circular model.

\begin{table}[h!]
\begin{center}
\caption{EMD metric comparison between Full Circular and No-short Circular models, evaluation at $\rho_I=$ 2.5 ped/m$^2$.}
\label{EMD4}
\begin{tabular}{|c|c|c|c|c|c|}
\hline
& Full $\langle \;  \rangle \pm \varepsilon (\sigma)$ & No-short $\langle \;  \rangle \pm \varepsilon (\sigma)$ & $p$ &  effect size\\
\hline
No body &  $0.0155\pm 0.001(0.0049)$ & $0.0192\pm 0.0017(0.0081)$ & 0.067 &0.55\\
\hline
$\rho$ &  $0.0136\pm 0.0018(0.0085)$ & $0.0139\pm 0.0025(0.012)$ & 0.93 &0.028\\
\hline
$v$ &  $0.0206\pm 0.0022(0.01)$ & $0.0289\pm 0.004(0.019)$ & 0.073 &0.54\\
\hline
$\theta^v$ &  $0.00818\pm 0.00082(0.0039)$ & $0.00937\pm 0.0007(0.0033)$ & 0.27 &0.33\\
\hline
$\delta^s$ &  $0.0192\pm 0.0012(0.0057)$ & $0.0239\pm 0.0012(0.0056)$ & 0.0071 &0.83\\
\hline
$\delta^o$ &  $0.016\pm 0.0017(0.0081)$ & $0.0201\pm 0.0022(0.011)$ & 0.15 &0.44\\
\hline
$\phi^s$ &  $0.0198\pm 0.0025(0.012)$ & $0.0283\pm 0.0021(0.01)$ & 0.012 &0.77\\
\hline
$\phi^o$ &  $0.0367\pm 0.0027(0.013)$ & $0.0291\pm 0.0036(0.017)$ & 0.098 &0.5\\
\hline
\end{tabular}
\end{center}
\end{table}

\subsection{Discussion}
\subsubsection{Long vs. Short range}
When comparing the two models that are able to describe the body orientation degree of freedom (the Full and No-short Elliptical models), we notice
a very strong tendency of the Full model to outperform the No-short concerning almost all observables for both metrics (see Table \ref{EMD}  in the main text and Table \ref{Standard} in Appendix \ref{standeval}).
The only observable on which the No-short model appears to perform better is $\phi^o$, on which the models are actually not calibrated. The overall evaluation fitness on the observables
for which the models have been calibrated is different in a significant way, with $p$ values smaller than $10^{-3}$ for both the EMD and Standard metrics, and  effect sizes larger than 1.
Quite low $p$ values and large effect sizes are attained concerning almost all the observables, excluding the body orientation $\theta$ (although there is a significant difference in $\delta \theta$,
i.e., in the deviation between body and velocity orientation), and the aforementioned $\phi^o$. The difference in $\rho$ is less strong than for other observables, not attaining statistical significance,
but the No-short model appears to describe poorly the distance between pedestrians, in particular $\delta^s$. As shown in Appendix \ref{appcomp}
(see Tables \ref{EMD_1_low}, \ref{Standard_1_low}, \ref{EMD_1_medium}, \ref{Standard_1_medium}), similar results are observed also in the calibration at
$\rho_I=1.5$ ped/m$^2$ (the behaviour at $\rho_I=0.25$ ped/m$^2$ is reproduced in a more similar way by both models, with the simpler model often slightly outperforming the more complex one).
There is not a significant change in the difference between the models when calibrated at $\rho_I=1.5$ ped/m$^2$ or evaluated at $\rho_I=2.5$ ped/m$^2$,
the difference being actually a little more pronounced in evaluation, showing that the more complex model does not suffer from over-fitting issues with respect to the simpler one.

It is quite clear, at least from these data, that when an elliptical body is considered, the actual collisions between such elliptical bodies should be predicted by the model.

Similar considerations can be made when we compare the two models that do not describe body orientation (the Full and No-short Circular models), although the difference in this case is less clear,
in particular when we use the Standard Metric (see Table \ref{EMD4}  in the main text and Table \ref{Standard4} in Appendix \ref{standeval}).
We indeed have, for the overall evaluation fitness,  $p=0.067$ and a 0.55 effect size for the EMD metric,
but $p=0.61$ and a 0.15 effect size for the Standard Metric. The differences between the two circular models are actually stronger when the calibration phase is considered
(see Tables \ref{EMD_4_low}, \ref{Standard_4_low}, \ref{EMD_4_medium}, \ref{Standard_4_medium} in Appendix \ref{appcomp}),
with significant differences in particular at $\rho_I=1.5$ ped/m$^2$. These results suggest
that over-fitting issues may be present when comparing the Full and No-short Circular models.

The, somehow intuitive but nevertheless important, conclusion is that predicting the details of short-time collisions is more important when a non-symmetrical body shape is considered.
\subsubsection{Elliptical vs. Circular}
A straightforward comparison between the newly proposed Full Elliptical model and our previous CP (Circular no-short) model (see Table \ref{EMD3}  in the main text and Table \ref{Standard3} in Appendix \ref{standeval}),
shows that the new model significantly outperforms the old one
in overall EMD fitness ($p=0.041$, effect size 0.62) and in general performs better on most observables, although a comparison with calibration results
(see Tables \ref{EMD_3_low}, \ref{Standard_3_low}, \ref{EMD_3_medium}, \ref{Standard_3_medium} in Appendix \ref{appcomp}), in which the difference between the two models is more pronounced,
suggests that the more complex model (i.e., the model with more parameters) may suffer more (as expected) of over-fitting.

Nevertheless, the problem of comparing Circular models to Elliptical ones is more subtle than just referring to their calibration and evaluation performances.
Circular models are not able to reproduce orientation dynamics, so for any kind of situation in which such a dynamics is relevant, Elliptical ones (or other
models including body orientations) should be used. As we have seen that the deviation of $\delta \theta$ from 0 increases with density, this is probably necessary at very high density.
Furthermore, intuitive considerations and previous studies \cite{Yamamoto2019}, suggest this dynamics to be very important in narrow spaces.

A different issue is to understand the regimes in which using body orientation models is necessary. For the studied dynamics, we observe (see Table \ref{EMD2}  in the main text and Table \ref{Standard2} in Appendix \ref{standeval})
that when the results are analysed using the EMD metric, that concerns the position of the distribution bulks, very little difference is observed between the Full Elliptical and Full Circular
models (it should nevertheless be noted that the Elliptical model has been calibrated to reproduce also body orientation, and thus it is ``trying to solve'' a more complex dynamics). Nevertheless,
the situation is different when, by using the Standard Metric, a larger weight is assigned to the tails. In this case,  concerning the overall evaluation fitness on the observables
for which the Circular model has been calibrated, we have $p=0.27$ and an effect size of 0.33, which, although very far from providing compelling evidence, are stronger than those found using the
EMD metric. In particular, we have $p=0.11$ and an effect size of 0.48 for the $\rho$ observable (which is one of the observables on
which pedestrian models are traditionally evaluated).

Fig. \ref{rho_comp} (b) in Appendix \ref{graph_comp} shows a comparison between the empirical $\rho(t)$ at $\rho_I=2.5$ ped/m$^2$ and the evaluation results in all models (for each model the standard error intervals obtained
averaging over all solutions are shown). It can be noticed that all models exhibit a slight tendency to have a fatter tail than the empirical distribution. This is due to some pedestrians getting
stuck in an ``impasse'' near the top-right corner of the crossing area in Fig. \ref{f1}. This problem could probably be avoided by using a better path finding behaviour. Nevertheless, the occurrence of such situations are
probably the closest the virtual pedestrians ever get to an actual ``clogged'' situation in our environment. It appears that the Full Elliptical model is better than the Full Circular model
in dealing with such situations. The use of the Elliptical model could thus be necessary at higher densities/narrower environments than the ones analysed in this work.

Another interesting point can be raised referring to Fig. \ref{theta_v_comp} (b) in Appendix \ref{graph_comp}, showing a comparison for the pdf of the $\theta^v$ observable, namely the only directional observable
that may be reproduced by Circular models. We see again that all models present fatter tails than the empirical distribution, but this effect is much stronger in the Circular model(s).
Again, the Full Elliptical model seems better at describing rare events. These results are in agreement with the discussion following Fig. \ref{ft}, in which we suggested that
in describing rare, high local density events, body orientation could result to be important. Further comparisons between empirical data and model evaluations are reported in Appendix \ref{graph_comp}. 

As an overall consideration concerning model comparison, we may say that although the fundamental properties of the dynamics are already well-described by the CP model (refer to the qualitative graphs
of Appendix \ref{graph_comp}), including the short range interaction leads to a, although reduced, quantitative and qualitative improvement.
It is furthermore natural, also from a purely theoretical viewpoint,
to introduce such dynamics by taking explicitly into account the actual shape of the human body.
The proposed approximation, using 2D ellipses, indeed leads to an improvement in the description of rare events (tails), for which probably a detailed description of torso movements is needed.

Nevertheless, from a practical viewpoint, at least concerning the description of cross-flows up to 2-3 ped/m$^2$, the improvement in the description of the dynamics could be too marginal compared
to the increase in computational cost
(for a quantitative and qualitative improvement in the description of the cross-flow it would probably be more important to improve path finding or other forms of strategic choices
that have not been considered in this work.)
\subsubsection{Limitations and possible improvements}
Arguably, explicit body orientation dynamics will be needed also from a practical viewpoint at higher densities and in more complex (and narrower) environments.
To this end, it may be useful to analyse some of the shortcomings of the proposed model. As can be observed in Figs. \ref{theta_comp}, \ref{delta_theta_comp} in Appendix \ref{graph_comp},
the proposed model does not include the strong tendency of human pedestrians to ``not turn their back'' to
their goal. Although this is partially achieved by some solutions (Figs. \ref{f_theta_cal} (a), \ref{f_delta_theta_cal} (a), \ref{f_theta_ev} (a), \ref{f_delta_theta_ev} (a)),
this happens to the detriment of the description of the bulk distributions (Figs. \ref{f_theta_cal} (b), \ref{f_delta_theta_cal} (b), \ref{f_theta_ev} (b), \ref{f_delta_theta_ev} (b)).
The proposed model is linear in $\theta$ angles (i.e., given by a quadratic potential), and thus an improvement could be using a steeper potential for large $\theta$.

Another shortcoming is the description of low $\delta$ values. A qualitative analysis of Fig. \ref{ds_comp} in Appendix \ref{graph_comp}  shows that Elliptical models underestimate the minimum value
assumed by $\delta^s$, although they describe in a more realistic way the initial slope of the pdf. The Standard Metric penalises a method when describing a finite possibility for
an observable to assume values outside its empirical range, which leads to the low performance of the Full Elliptical model concerning $\delta^s$ in Table \ref{Standard2} in Appendix \ref{standeval} (on the other hand, the Full Elliptical model
appears to have a good qualitative description of low $\delta^o$ values, see Fig. \ref{do_comp} in Appendix \ref{graph_comp}). This is probably due to the limits of the 2D ellipse approximation.
Although, as described in
Appendix \ref{step_app}, we tried to include in the basic CP model a tendency ``not to overlap steps'' to account for the fact that our model completely ignores the complex 3D dynamics of leg
movement, a more detailed description of limb shape and movement could be needed in future (this effect is intuitively particularly important for elliptical models, since when people are walking in a
file and with their torso perpendicular to walking direction, if the movement of the limbs are ignored, they may get closer due to the smaller extension of their body in the walking direction).

Finally, the detailed description of $\phi$ observables could be improved (see Appendix \ref{graph_comp}). We nevertheless remind that we decided not to calibrate this behaviour in the current work,
and confined ourselves to observe how it was ``naturally'' emerging in the models.

It may thus be useful to better study models describing body orientation, and possibly develop more simple and efficient ones to decrease issues of calibration and computational cost. Nevertheless, at least for the problem
and the range of density studied in this work, it does not appear that body orientation is {\it crucial} to reproduce the observed dynamics. From a practical viewpoint, for example, it may be that
a quantitative and qualitative improvement in the description of the cross-flow may be attained by improving path finding (or other forms of strategic choices) aspects
that have not been considered in this work.
\section{Conclusions}
In this work, we analysed the results of a set of controlled experiments in which subjects were divided into two groups,
organised in such a way to explore different density settings, and asked to walk through a crossing area.
By defining a few macroscopic and microscopic observables, including traditional indicators such as density and velocity, but also observables relating relative distance between pedestrians in the
crowd, and observables relating walking and body orientation, we studied in a qualitative and quantitative way how the cross-flow dynamics is affected by density conditions.
We also report a preliminary but quantitative analysis on the emergence of self-organising patterns (stripes) in the crossing area, a phenomenon that had been previously qualitatively
reported for human crowds, and reproduced in models, but whose quantitative analysis (in particular with respect to density conditions) is, to the best of our knowledge, a novel contribution.

In the second part of our work, we tried to reproduce the empirical results using a hierarchy of models, which differ in the details of the body shape (using a disk-shaped body vs a more realistic
elliptical shape) and in how collision avoidance is performed (using only information regarding ``centre of mass'' distance and velocity, or actually introducing body shape information). We verified
that the most detailed model (i.e., using body shape information and an elliptical body) outperforms in a significant way the simplest one (using only centre of mass distance, velocity, and
disk-shaped bodies). Furthermore, we observed that if elliptical bodies are introduced without using such information in collision avoidance, the performance of the model is relatively
poor. Nevertheless, the difference between the different models is relevant only in describing the ``tails'' of the observable distributions, suggesting that the more complex models
could be of practical use only in the description of high density settings.
Although we did not calibrate our model in order to reproduce  the ``stripe formation'' self-organising pattern, we verified that it naturally emerges in all models.

\section{Acknowledgements}
This research work was in part supported by: JSPS KAKENHI Grant Number 18H04121, JSPS KAKENHI Grant Number 20K14992, JST-Mirai Program Grant Number JPMJMI20D1.
\clearpage
\appendix
\section{Definition of observables}
\label{obs_def}
We provide a detailed definition of the observables used in this work. We start with some general definitions that will prove to be useful.

The cardinality of a set $S$ is defined as $\#(S)$.
Vectors are denoted by boldface, such as $\mathbf{a}$. The standard Euclidean inner (scalar) product between vectors $\mathbf{a}$ and $\mathbf{b}$ is given by
\begin{equation}
(\mathbf{a},\mathbf{b}),
\end{equation}
and the Euclidean norm by
\begin{equation}
a=(\mathbf{a},\mathbf{a})^{\frac{1}{2}}.
\end{equation}

As we always deal with 2D vectors, we may consider their vector product as a scalar (given by its projection on the right-handed normal to the plane). Namely, given an arbitrary right handed frame, we define
\begin{equation}
\langle \mathbf{a},\mathbf{b}\rangle \equiv (\mathbf{a} \times \mathbf{b})_z=a_xb_y-a_yb_x.
\end{equation}

If we are only interested in the absolute value of $\langle \mathbf{a},\mathbf{b}\rangle$, we need not to worry about the right-handedness of the frame.

We often use the two-value arctan function, defined as
\begin{equation}
  \text{atan2}(y,x)=
  \begin{cases}
    \arctan\left(\frac{y}{x}\right), & \text{ if } x > 0\\
    \arctan\left(\frac{y}{x}+ \pi\right), & \text{ if } x < 0, y \geq 0\\
    \arctan\left(\frac{y}{x} - \pi\right), & \text{ if } x < 0, y <0\\
    \frac{\pi}{2}, & \text{ if } x = 0, y>0\\
    -\frac{\pi}{2}, & \text{ if } x = 0, y<0\\
\text{ undefined }, & \text{ if } x=0, y=0.
\end{cases}
\end{equation}

In our analysis we will consider only pedestrians that, at a given time $t$, are located inside a tracking area, which is defined as  an $L$  times $L$ square ($L=3.4$ m), adding a 0.2 m border on the sides of
the pink area in Fig. \ref{f1}, to take into account the aforementioned absence of hard walls, and as a consequence has an area $A=L^2$. For each pedestrian $i$ located inside the tracking area (at a given
tracking time $t$) we define the position vector $\mathbf{r}_i$ as the vectorial distance from the origin (e.g., located in the centre of the crossing area).
The distance between two pedestrians is defined as
\begin{equation}
  \mathbf{r}_{i,j}=\mathbf{r}_j-\mathbf{r}_i.
\end{equation}

At each time $t$, we also define $f^1(t)$ and $f^2(t)$ as the subset of (tracked) pedestrians belonging, respectively, to flow 1 and 2. Since these sets correspond to tracked pedestrians,
they formally depend on tracking time $t$, although such a dependence is shown only when needed. Their union is the set of tracked pedestrians
\begin{equation}
  T(t)=f^1(t) \cup f^2(t).
  \end{equation}
Each flow belongs to a corridor, whose
axis direction (oriented as the marching direction of the pedestrians, i.e., their goal) is defined through the normalised vector $\mathbf{j}^k$, $k=\{1,2\}$.
We also define the orthogonal normalised vectors as $\mathbf{i}^1=-\mathbf{j}^2$, $\mathbf{i}^2=-\mathbf{j}^1$. It should be noted that, through this definition,
$\{\mathbf{i}^k,\mathbf{j}^k\}$ are not necessarily right-handed. This decision is made because we want $\mathbf{i}^k$ to identify the direction of the incoming cross-flow, in order to
have, by symmetry, the same expected distributions for the relative angle observables to be defined below.

Furthermore, for each corridor $k$, we define a reference frame using
\begin{equation}
  \label{pedownframe}
a^k_x=(\mathbf{a},\mathbf{i}^k)\;\;\; a^k_y=(\mathbf{a},\mathbf{j}^k),
\end{equation}
($a^k_{x,y}$ being the components of an arbitrary vector $\mathbf{a}$ in frame $k$)
and for each pedestrian $i$, we define the function
\begin{equation}
F(i)=
\begin{cases}
1, & \text{ if } i \in f^1\\
2, & \text{ if } i \in f^2.
\end{cases}
\end{equation}

An {\it empirical probability distribution}\footnote{The term ``distribution'' is used since observables are theoretically continuous, although
  from an empirical point of view they are computed over discrete bins. In general, in figures we normalise them in such a way that their integral equals 1,
  although in EMD computations they are normalised in such a way that the sum over bins equals 1.} of an observable $O$ is computed by defining a bin size
\begin{equation}
\Delta O=\frac{O_{\text{max}}-O_{\text{min}}}{n_b},
\end{equation}
where $n_b$ is the number of bins, and $O_{\text{min}}$, $O_{\text{max}}$ define the interval in which we study the observable distribution.

The number of observations belonging to a bin $j$ is then defined as
\begin{equation}
\begin{split}
n^O_j=
\#(&\text{ observations } k \text{ with value } O_k
\\
&\text{ such that } j \Delta O +O_{\text{min}}\leq O_k < (j+1) \Delta O+O_{\text{min}})
\end{split}
\end{equation}
and the empirical probability distribution of $O$ is
\begin{equation}
  P(O)=P\left(\left(j+\frac{1}{2}\right) \Delta O+O_{\text{min}}\right)\equiv P_j =\frac{n^O_j}{\sum_{l} n^O_l}.
\end{equation}
All empirical probability distributions considered in this work are computed by using $n_b=40$ bins.

For each experimental condition we performed $n_e$ independent repetitions. In graphs, we compare averages and standard errors over all independent repetitions.

Namely, if $P^k_j$ is the value assumed by the $j$th bin of a given observable with an experimental condition in the $k$th independent repetition, graphs will
show average values and standard errors
\begin{equation}
    \langle P_j \rangle \pm\epsilon_j,
\end{equation}
defined according to
\begin{equation}
  \label{obavp}
    \langle P_j \rangle=\frac{\sum_{k=1}^{n_e} P^k_j}{n_e}, 
\end{equation}
and
\begin{equation}
  \label{obavp2}  
  \begin{split}
   &\langle (P_j)^2 \rangle=\frac{\sum_{k=1}^{n_e} (P^k_j)^2}{n_e},\\ 
    &\epsilon_j=\sqrt{\frac{\langle (P_j)^2 \rangle-\langle P_j \rangle^2}{n_e-1}}.
    \end{split}
\end{equation}

%The above definitions are used also in the computation of the EMD and Standard metrics defined in \ref{evalmet}.
\subsection{Density $\rho(t)$}
The density $\rho(t)$ is the density of pedestrians in the crossing area as a function of time, and it is measured in ped/m$^2$. The formal and computational definition is the following.
A bin size $\Delta T_{\rho}$ is defined (in the following, we use  $\Delta T_{\rho}=1$ s).
For each instant $t_k=k \delta t_{\text{track}}$ at which tracking was performed ($\delta t_{\text{track}}=1/30$ s being the tracking time interval,
and $k\in \mathbb{N}$), the number of pedestrians in the tracking area is defined as $n_k=\#(T(t))$. We define the set $M_j$ as
  consisting of all observation times $k$ such that $j \Delta T_{\rho}\leq k (\delta t_{\text{track}}) < (j+1) \Delta T_{\rho}$. The density in the crossing area is then defined as
  \begin{equation}
\rho(t_j)=\rho\left(\left(j+\frac{1}{2} \Delta T_{\rho}\right)\right)=\frac{\sum_{k \in M_j} n_k}{A \#(M_j)},
  \end{equation}
  $A=L^2$ being the size of the crossing area.
  \subsubsection{Exit time pdf $P(E_t)$}
  Between all the observables considered in this work, $\rho$ is the only one that is not defined as a probability density function (pdf).
  Since our model calibration process (to be described below)
  and part of our model evaluation method rely on the Earth Mover's Distance (EMD) \cite{Cohen1999}, which is defined for pdfs, we tried to define a corresponding observable which may be defined
  as a pdf (simply normalising the $\rho(t)$ interval is not a feasible solution, since the absolute value of $\rho$ is of capital importance in pedestrian dynamics).

  The chosen observable is the {\it exit time} pdf $P(E_t)$. Computationally, it is obviously defined using discrete bins. For each observation time
  $t_k= k \delta t_{\text{track}}$, we define $e_k$ as the number of pedestrians exiting the crossing area, i.e.
  \begin{equation}
e_k = \left[\cap_{t'>t_k} (i\not \in T(t'))\right]\cap (i\in T(t_k)).
  \end{equation}
  We then define $\overline{e}_j$ for any bin $t_j=j\Delta T_{\rho}$ as 
   \begin{equation}
\overline{e}_j=\sum_{k \in M_j} \#(e_k),
  \end{equation} 
  and finally define
  the empirical probability distribution of $E_t$ as
  \begin{equation}
P(E_t)\equiv P\left(\left(j+\frac{1}{2} \Delta T_{\rho}\right)\right) \equiv \frac{\overline{e}_j}{\sum_{l} \overline{e}_l}.
  \end{equation}
  As all experiment repetitions were finished in less than 40 s, we used 40 bins to define the probability distribution. The same number of bins is used for all the observables,
  to make the EMD comparison straightforward.
  \subsection{Speed pdf $P(v)$}
  Given the velocity $\mathbf{v}_i$ of pedestrian $i$, $v_i$ gives the pedestrian speed. We consider a possible maximum speed of 2.5 m/s, and in order to have 40 bins we define
  $\Delta v=0.0625$ m/s.
     \subsection{Velocity direction pdf $P(\theta^v)$}
     We define the velocity direction angle as
     \begin{equation}
       \label{tv}
       \theta^v_i=\text{atan2}((v_i)^{F(i)}_x,(v_i)^{F(i)}_y).
     \end{equation}
     We remind that $(v_i)^{F(i)}_{\{x,y\}}$ are the $\{x,y\}$ components of $i$'s velocity vector as measured in the reference frame corresponding to $i$'s corridor.
     The function $\text{atan2}(v_y,v_x)$ is chosen in such a way to have $-\pi \leq \theta^v < \pi$, with
     $\theta^v=0$ for velocities aligned with the corridor's axis, and $\theta^v>0$ for angles in the direction of the incoming cross-flow.

     The pdf $P(\theta^v)$ is empirically defined by taking bins of size $(2 \pi)/40$. Since the angle $\theta^v$ is defined with respect to a clear and physically meaningful axis
     (the corridor direction), and since we verified that the empirical probability distribution is clearly centred around zero and is $\approx 0$ for $\theta^v\approx \pm \pi$,
     all the statistical analysis concerning the observable are performed by treating $\theta^v$ (and similarly $\theta$ and $\delta \theta$ to be defined below) as linear (i.e., we are not using circular
     statistics).

      \subsection{Body direction pdf $P(\theta)$}
      As explained in Section \ref{experiments}, 10 subjects were carrying a tablet, fixed to their chest. Through a gyroscope, we could know the angular velocity of the tablet, from which (by time integration, and assuming
      the pedestrians as having their chest orthogonal to the corridor direction as an initial condition) we could obtain the normal unit vector to the chest $\mathbf{n}_i$, and define
      \begin{equation}
        \label{chest}
       \theta_i=\text{atan2}((n_i)^{F(i)}_x,(n_i)^{F(i)}_y).
     \end{equation}
      The pdf $P(\theta)$ is empirically defined by taking bins of size $(2 \pi)/40$.
      \subsection{Body direction deviation pdf $P(\delta \theta)$}
      We also measure the deviation between the body and velocity directions as
      \begin{equation}
        \label{def_deltath}
       \delta \theta_i=\text{Mod}_{2 \pi}(\theta_i-\theta^v_i+\pi)-\pi,
      \end{equation}
      and define again its empirical pdf $P(\delta \theta)$ by taking bins of size $(2 \pi)/40$.
    
      \subsection{Same flow first neighbour relative distance pdf $P(\delta^s)$}
For each pedestrian $i$, we define the {\it first (forward) neighbour in the same flow} as
\begin{equation}
N^s_i=\argmin_j{r_{i,j}, \text{ with } F(i)=F(j), \text{ and } (\mathbf{r}_{i,j},\mathbf{j}^{F(i)})\geq 0}.
  \end{equation}
We then define
the distance to the first neighbour in the same flow as
\begin{equation}
\delta^s_i=r_{i,N^s_i}.
  \end{equation}
Namely, $\delta^s_i$ is the distance to the closest neighbour in the tracking area belonging to the same flow,
{\it and located on the front} of $i$.

Considering the size of the crossing area, the empirical pdf $P(\delta_s)$ is defined using bins of size $3/40=0.075$ m.
      \subsection{Crossing flow first neighbour relative distance pdf $P(\delta^o)$}
      For each pedestrian $i$, we also define the {\it first (forward) neighbour in the crossing flow} as
\begin{equation}
N^o_i=\argmin_j{r_{i,j}, \text{ with } F(i) \neq F(j), \text{ and } (\mathbf{r}_{i,j},\mathbf{j}^{F(i)})\geq 0},
  \end{equation}
and the distance to the first neighbour in the crossing flow as
\begin{equation}
\delta^o_i=r_{i,N^o_i}.
  \end{equation}
Namely, $\delta^o_i$ is the distance to the closest neighbour in the tracking area belonging to the {\it crossing} flow,
{\it and located on the front} of $i$. Also $P(\delta^o)$ is defined using bins of size $3/40=0.075$ m.
\subsection{Same flow first neighbour relative angle pdf $P(\phi^s)$}
For each pedestrian $i$, we then define
\begin{equation}
\phi^s_i=\text{atan2} ((r_{i,N^s_i})^{F(i)}_x,(r_{i,N^s_i})^{F(i)}_y),
\end{equation}
as the relative angle (in $i$'s frame) to the first (forward) neighbour in the same flow. As we require the neighbour to be in the forward position,
we have $\phi^s_i \in [-\pi/2,\pi/2]$, and we treat again this observable as a linear one in our statistical analysis. $P(\phi^s)$ is defined using bins of size $\pi/40$.
The same considerations apply to $\phi^o$ defined below.
\subsection{Opposite flow first neighbour relative angle pdf $P(\phi^o)$}
Finally we define
\begin{equation}
\phi^o_i=\text{atan2} ((r_{i,N^o_i})^{F(i)}_x,(r_{i,N^o_i})^{F(i)}_y),
\end{equation}
as the relative angle (in $i$'s frame) to the first (forward) neighbour in the crossing flow.
\section{Definition of observable dependence on density}
\label{obs_dens}
\subsection{Dependence on experimental conditions (initial density $\rho_I$)}
\label{obs_dens_int}
For each initial condition $\rho_I$, repetition $j$ and observable $O$, we compute the average value
\begin{equation}
\langle O \rangle_j(\rho_I)\equiv O^{\rho_I}_{j} \equiv\frac{\sum_{k,l} O^{\rho_I}_{j,k,l}}{\#(\{\rho_I,j\})},
\end{equation}
where $O^{\rho_I}_{j,k,l}$ stands for the observation concerning pedestrian $l$ at time $k(\delta t_{\text{track}})$ during repetition $j$ with initial condition $\rho_I$, and 
$\{\rho_I,j\}$ stands for the whole set of observations concerning repetition $j$ with initial condition $\rho_I$.
Furthermore, we compute the average value for the initial condition as
\begin{equation}
\langle O \rangle^{\rho_I}=\frac{\sum_{j} O^{\rho_I}_{j}}{\#(\{\rho_I\})},
\end{equation}
and the corresponding standard error as
\begin{equation}
\varepsilon_{O^{\rho_I}}=\sqrt{\frac{\langle O^2 \rangle^{\rho_I}-{\langle O \rangle^{\rho_I}}^2}{\#(\{\rho_I\})-1}}.
\end{equation}
Here $\#(\{\rho_I\})$ stands for the number of repetitions with initial condition $\rho_I$.
\subsection{Dependence on crossing area density $\rho$}
\label{obs_dens_cross}
While the above discussion allows us to see how the observables depend on the initial condition, equally or even more interesting is to study their dependence on the crossing area density $\rho$.
We may define a density interval $\Delta \rho$ (that for practical purposes is fixed to $0.1$ ped/m$^2$), and denote $\{(m+1/2) \Delta \rho,j\}$ as the set of observations
concerning pedestrian $l$ during repetition $j$ {\it for all times $k$ such that} $m \Delta  \rho \leq \rho_k < (m+1) \Delta  \rho$. Then we can proceed as above to obtain
\begin{equation}
\langle O \rangle_j^{\rho}\equiv O^{\rho}_{j} \equiv\frac{\sum_{k,l} O^{\rho}_{j,k,l}}{\#(\{\rho,j\})},
\end{equation}
\begin{equation}
\langle O \rangle^{\rho}=\frac{\sum_{j} O^{\rho}_{j}}{\#(\{\rho\})},
\end{equation}
\begin{equation}
\varepsilon_{O^{\rho}}=\sqrt{\frac{\langle O^2 \rangle^{\rho}-{\langle O \rangle^{\rho_I}}^2}{\#(\{\rho\})-1}},
\end{equation}
where, to simplify the notation, the discrete value $(m+1/2) \Delta  \rho$ has been replaced by $\rho$.
\section{Dependence of observables on the initial density  $\rho_I$}
\label{depon_rhoI}
For each observable $O$, we compute its dependence on the initial condition density $\rho_I$, denoted as $\langle O  \rangle^{\rho_I}$ (for a detailed operational definition of these average values
and their standard errors and deviations, refer to Appendix \ref{obs_dens_int}).

\begin{figure}
\begin{center}
  \includegraphics[width=0.8\textwidth]{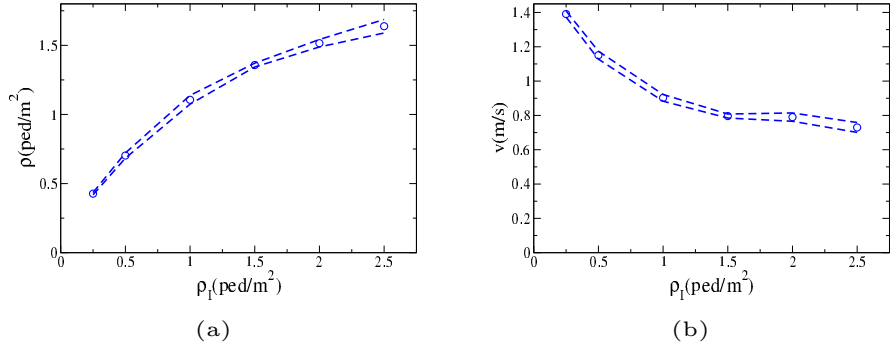}
  \caption{\label{f_rho_v_rhoI}(a): $\langle \rho\rangle^{\rho_I}$. (b): $\langle v \rangle^{\rho_I}$.
    Dashed lines provide standard error intervals.}
 \end{center}   
\end{figure}

\begin{figure}
\begin{center}
  \includegraphics[width=0.8\textwidth]{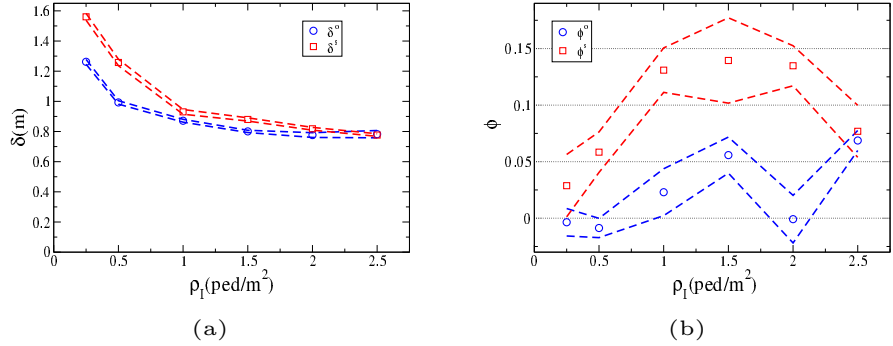}
  \caption{\label{f_delta_phi_rhoI}(a): $\langle \delta^o\rangle^{\rho_I}$ (blue circles) and $\langle \delta^s\rangle^{\rho_I}$ (red squares).
(b): $\langle \phi^o\rangle^{\rho_I}$ (blue circles) and $\langle \phi^s\rangle^{\rho_I}$ (red squares).
    Dashed lines provide standard error intervals.}
 \end{center}   
\end{figure}

\begin{figure}
\begin{center}
  \includegraphics[width=0.8\textwidth]{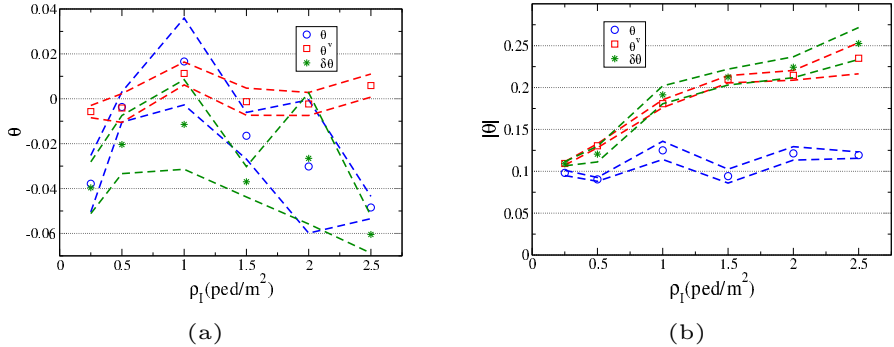}
  \caption{\label{f_theta_rhoI}(a): $\langle \theta\rangle^{\rho_I}$ (blue circles) and $\langle \theta^v\rangle^{\rho_I}$ (red squares) and $\langle \delta \theta\rangle^{\rho_I}$ (green stars).
    (b): $\langle |\theta|\rangle^{\rho_I}$ (blue circles) and $\langle |\theta^v|\rangle^{\rho_I}$ (red squares) and $\langle |\delta \theta|\rangle^{\rho_I}$ (green stars).
    Dashed lines provide standard error intervals.}
 \end{center}   
\end{figure}

$\langle \rho  \rangle^{\rho_I}$ is reported in Fig. \ref{f_rho_v_rhoI} (a). Furthermore, the dependence
on $\rho_I$ of $v$ is shown in Fig. \ref{f_rho_v_rhoI} (b), that of the $\delta$ observables in Fig. \ref{f_delta_phi_rhoI} (a),
that of the $\phi$ observables in Fig. \ref{f_delta_phi_rhoI} (b), and finally the one
of $\theta$
observables in Fig.   \ref{f_theta_rhoI} (a). We also report in Fig.  \ref{f_theta_rhoI} (b) the absolute value of the $\theta$ observables.

We may observe that, as expected, $\rho$ is an increasing function of $\rho_I$. Nevertheless, while at $\rho_I=0.25$ ped/m$^2$ we have $\rho\approx 2 \rho_I$,
at $\rho_I=2.5$ ped/m$^2$ we have $\rho\approx 3/5 \rho_I$ (this is the average value over the experiment; maxima, as shown in those sections, e.g., Section \ref{opdfs},
in which we report pdfs, are considerably higher, in particular
for high $\rho_I$, although the growth is still sub-linear). We observe that $v$, $\delta^o$ and $\delta^s$ are decreasing functions, although they reach a plateau around $\rho_I\approx 1.5$  ped/m$^2$.
We also observe that $\phi^s$ is clearly biased towards positive values. This result, which is related to the ``diagonal stripe formation'' and it is better discussed studying full pdfs (see again Section \ref{opdfs}),
is particularly strong in the 1-2 ped/m$^2$ range. A weaker bias towards positive values is shown also by $\phi^o$. $\theta$ and $\delta \theta$ are weakly biased towards negative values,
while $\theta^v$ is weakly biased towards positive ones. Finally, $|\theta|$ has no clear dependence on $\rho_I$, while $|\theta^v|$ and $|\delta \theta|$ are clearly increasing with  $\rho_I$.
\section{Stripe formation in experiments and models}
\label{stripe}
\begin{figure}[h!]
\begin{center}
  \includegraphics[width=0.8\textwidth]{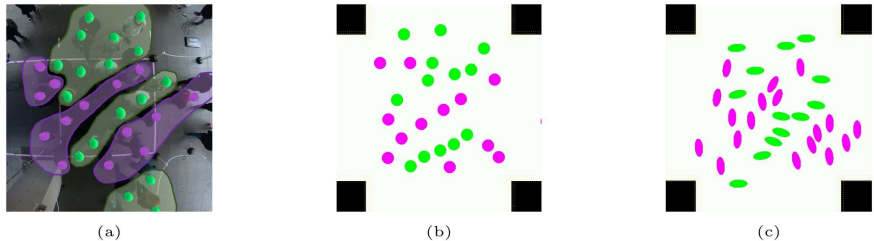}
  \caption{\label{fstripes}Frames showing in a qualitative and visual way stripe formation under the $\rho^I=1.5$ ped/m$^2$ initial condition. (a): experiments with subjects, (b) Full Circular model simulation, (c) Full Elliptical model simulation. Pedestrians in green move from top to bottom, those in purple from left to right.}
 \end{center}   
\end{figure}

In Fig. \ref{fstripes} we show some situations in which pedestrians are clearly self organised in stripes. Fig. \ref{fstripes} (a) shows a frame from the experiments (colours are edited to show clearly which pedestrians belong to which stream, and to suggest the presence of the ``clusters'', while Fig. \ref{fstripes} (b) and Fig. \ref{fstripes} (c) show, respectively, frames from simulations using the Full Circular and Full Elliptical models. Although all frames are obtained using the same initial density condition $\rho^I=1.5$ ped/m$^2$, they are extracted from different experiment repetitions. The full simulation video corresponding to the results of Fig. \ref{fstripes} (b) may be found at \cite{video_stripes_b} while that corresponding to the results of Fig. \ref{fstripes} (c) may be found at \cite{video_stripes_c}.
\section{Detailed description of computational models}
\label{model_detailed}
We start with a detailed description of the long range module, focusing in particular on the differences with the model as introduced in \cite{Zanlungo2011}. We recall that all the models used in this work
are based on the long range module, and thus the properties of the latter are relevant to all models.
\subsection{Details in the implementation of the long range module}
\label{model_detailed_cp}
The model has been formalised in  \cite{Zanlungo2011}, although it was introduced in a preliminary version in \cite{Zanlungo2007A,Zanlungo2007B}. Furthermore, while \cite{Zanlungo2011} dealt with pedestrian motion in a large environment (absence
of non-human obstacles), the model was updated to deal with obstacles in \cite{Zanlungo2017}. Moreover, a new term is added in this work to better describe the density/velocity relation in high density
crowds. Basically, we may consider the  model to consist of 3 terms, the {\it inter-pedestrian collision prediction term} $\mathbf{F}^{p}$, the {\it obstacle collision prediction term}
$\mathbf{F}^{o}$, and the {\it step overlap term} $\mathbf{F}^{s}$. The sum of these terms provides an ``interaction social force''
\begin{equation}
  \label{3terms}
\mathbf{F}^{long}=\mathbf{F}^{p}+\mathbf{F}^{o}+\mathbf{F}^{s},
\end{equation}
namely an acceleration that pedestrians perform in order to
change their velocity based on their perception of the environment\footnote{In \cite{Helbing1995} the Social Force Model was introduced as the sum of terms involving
  purely physical dynamics (body contact) and terms concerning the pedestrian decisions dynamics. Since the action of physical forces obviously depends on
  pedestrians' masses, the latter were explicitly present in the model. When, as in the current work, the SFM framework is used to describe only decisional dynamics,
  it is tacitly assumed that pedestrians will apply a force that determines their wanted acceleration, and the term ``force'' is interchangeable with ``acceleration''.
  This could be formalised by multiplying all acceleration terms by the pedestrian's mass, but since we never write equations mixing physical and decisional dynamics,
  such a formalisation is left as implicit. For further details on the relation between ``force'' and ``acceleration'' (second order) models, see \cite{Adrian2019}.}. We recall that according
to the Social Force Model paradigm, the pedestrians' decision process determines their acceleration as
\begin{equation}
  \label{sfmacc}
\ddot{\mathbf{r}}= -k_{v_p} (\dot{\mathbf{r}}-\mathbf{v}_p) + \mathbf{F}^{long}. 
\end{equation}
Here $\mathbf{v}_p$ is the preferred velocity of the pedestrians, given by
\begin{equation}
  \label{sfmacc_spef}
\mathbf{v}_p=v_p \hat{\mathbf{g}},
\end{equation}
where the preferred speed $v_p$ is assumed to be constant in time for each pedestrian (although it may obviously be different between pedestrians),
while the preferred goal direction $\hat{\mathbf{g}}$ may be a function of position and time. $k_{v_p}$ is a parameter of the
model and has the dimension of $t^{-1}$.
\subsubsection{Inter-pedestrian collision prediction term}
The basic idea of the CP model is the following: for each pedestrian $i$ with position $\mathbf{r}_i$ and velocity $\mathbf{v}_i\equiv \dot{\mathbf{r}}_i$, we define as usual the relative position
and velocity with respect to another pedestrian $j$ respectively as
\begin{equation}
\mathbf{r}_{i,j} =\mathbf{r}_j-\mathbf{r}_i,
\end{equation}
and
\begin{equation}
\mathbf{v}_{i,j} =\mathbf{v}_j-\mathbf{v}_i.
\end{equation}
We define $j$ to be {\it visible} to $i$ ($j \in V^p_i$) if
\begin{equation}
  \label{evis}
(\mathbf{v}_i,\mathbf{r}_{i,j})>0,
\end{equation}
and {\it approaching} ($j \in A^p_i$) if 
\begin{equation}
(\mathbf{r}_{i,j},\mathbf{v}_{i,j})<0.
\end{equation}
Finally, we define a pedestrian as {\it colliding}  ($j \in C^p_i$) if
\begin{equation}
\left|\left\langle \mathbf{r}_{i,j},\frac{\mathbf{v}_{i,j}}{v_{i,j}}\right\rangle\right| <d_{ext}.
\end{equation}
This latter condition means that, assuming linear motion, $i$ and $j$ will reach a distance smaller than a threshold $d_{ext}$, i.e., the maximum distance at which an interaction is expected
(not necessarily related to body size). These conditions are slightly different from those used in \cite{Zanlungo2011} and take advantage of the introduction of
hard-core potentials of \cite{Zanlungo2017} (see eq. \ref{step1}).

For all pedestrians $j$ satisfying all three conditions we then compute the time $t_{i,j}$ at which the minimum distance
between $i$ and $j$ will be reached, assuming both will keep a constant velocity,
namely
\begin{equation}
t_{i,j}=-\left(\mathbf{r}_{i,j},\frac{\mathbf{v}_{i,j}}{v_{i,j}^2}\right).
\end{equation}
We then define the {\it time of next closest approach to a pedestrian}
\begin{equation}
t_i^p=\min_{j\in(V^p_i \cap A^p_i \cap C^p_i)} t_{i,j},
\end{equation}
and, again assuming linear motion at constant velocity, we compute, for all {\it visible} pedestrians $j$,
\begin{equation}
\mathbf{r}^{min}_{i,j}=\mathbf{r}_j(t)-\mathbf{r}_i(t)+(t_i^p-t)\mathbf{v}_{i,j},
\end{equation}
i.e., the difference between the (predicted) position of the (moving, i.e., pedestrian) obstacle $\mathbf{r}_j$ and the one of the pedestrian $\mathbf{r}_i$ at $t_i^p$.

Such ``future distance'' $\mathbf{r}^{min}_{i,j}$ is then used to define the collision avoidance term of a velocity dependent specification of the social force model  as
\begin{equation}
  \label{fepl11}
\mathbf{F}^p_{i,j}=-\frac{v_i}{\max(t_i^p,\Delta t)} f(r^{min}_{i,j}) \frac{\mathbf{r}^{min}_{i,j}}{{r}^{min}_{i,j}}.
\end{equation}

The term $v_i/\max(t_i^p,\Delta t)$ is dimensionally an acceleration and it is introduced assuming that the desired acceleration must be roughly enough for the pedestrian to stop in a time $t_i^p$,
unless such time is smaller than the ``reaction time'' $\Delta t$ (the integration step of the behavioural model).
As a result, $f$ should be dimensionless, and of order 1.

While in \cite{Zanlungo2011}, for historical reasons, the choice of $f$ resembled the one of \cite{Helbing1995}, just replacing current
with future positions, following the analysis of \cite{Zanlungo2007A,Zanlungo2007B,Zanlungo2017} in the current work we use hard core potentials
\begin{equation}
  \label{step1}
f(r)= \begin{cases} 
C &\text{ if } r \leq d_{int},\\
C \frac{d_{ext}-r}{d_{ext}-d_{int}} &\text{ if } d_{int} < r \leq d_{ext},\\
0 &\text{ if } r>d_{ext}.\\  
\end{cases}
\end{equation}
$C$ being a dimensionless constant and $d_{int}$, $d_{ext}$ distance thresholds.

Finally, the interaction force in eq. \ref{3terms} due to the interaction with the other pedestrians is
\begin{equation}
\mathbf{F}_i^{p}=\sum_{j \in (V^p_i\cap C^p_i)} \mathbf{F}_{i,j}^p,
\end{equation}
where the sum is performed over all {\it visible} and {\it colliding} pedestrians $j$ (i.e., including
also non-approaching ones).

\subsubsection{Obstacle collision prediction term}

Obstacles (walls) are handled using the approach of \cite{Zanlungo2017}. In that work, intended for a robot application, obstacles are expressed as ``a cloud of points'',
and the force generated by such an obstacle is given by an average over all ``cloud points''.
This choice in \cite{Zanlungo2017} was motivated by the fact that obstacles are perceived by robots as sets of scan points from a sensor
(although obviously algorithms to extract higher level information are available).
Although here we are dealing with human behaviour, using the same kind of approach appears as a way to express the long range
behaviour with respect to obstacles (averaged interaction with the whole obstacle, and not with an explicit collision point).

In detail, we repeat the operations performed above for the interaction of pedestrian $j$ with every other pedestrian in the environment also on a set of ``obstacle points''\footnote{Specifically,
  we replace walls with a regular grid of points with a 0.05 m lattice step.}. Defining $\mathbf{r}_{i,j}$ as the relative distance between the pedestrian $i$ and the obstacle point $j$, and
assuming the velocity of the obstacle equal to zero, we define $j$ to be {\it near} ($j \in N^o_j$) if   
\begin{equation}
\mathbf{r}_{i,j} <R_{max},
\end{equation}
{\it on the front} ($j \in A^o_j$) if
\begin{equation}
\left(\frac{\mathbf{r}_{i,j}}{r_{i,j}},\frac{\mathbf{v}_{i}}{v_{i}}\right)>\frac{\sqrt{2}}{2},
\end{equation}
and {\it colliding}  ($j \in C^o_j$) if
\begin{equation}
\left|\left\langle \mathbf{r}_{i,j},\frac{\mathbf{v}_{i}}{v_{i}}\right\rangle\right| <d_{ext}.
\end{equation}

These conditions assure that the collision avoidance process is, although defined as ``long range'', still of a local nature, i.e., it accounts only for obstacles close in space and collision time.
The distance threshold $R_{max}$ was chosen to be equal to 2 m, so that, in general, a pedestrian is interacting with a single obstacle (or very few obstacles) and it is meaningful to use an average
of the grid points. The short range model, on the other hand, deals with obstacles as polygons, and computes explicit collision points. The {\it on the front} condition
is stricter than the {\it visible} one ($j \in V^o_j$), namely eq. \ref{evis},
to allow pedestrians to be able to walk close to a wall if needed.

For all {\it near}, {\it colliding} and {\it on the front} points, we can compute again
\begin{equation}
t_{i,j}=\left(\mathbf{r}_{i,j},\frac{\mathbf{v}_{i}}{v_{i}^2}\right).
\end{equation}
and define the {\it time of next closest approach to an obstacle}
\begin{equation}
t_i^o=\min_{j\in(N^o_j \cap A^o_j \cap C^o_j)} t_{i,j}.
\end{equation}
Then, for all {\it near} and {\it visible} obstacles (the set of visible obstacles $V^o_i$ is computed in analogy with eq. \ref{evis}) 
\begin{equation}
\mathbf{r}^{min}_{i,j}=\mathbf{r}_j(t)-\mathbf{r}_i(t)-(t_i^p-t)\mathbf{v}_{i},
\end{equation}
and
\begin{equation}
  \label{fepl11_2}
\mathbf{F}^o_{i,j}=-\frac{v_i}{\max(t_i^o,\Delta t)} f(r^{min}_{i,j}) \frac{\mathbf{r}^{min}_{i,j}}{{r}^{min}_{i,j}}.
\end{equation}
Finally, we compute
\begin{equation}
  \label{eqav}
\mathbf{F}^o_i=\frac{1}{\#(N^o_i \cap V^o_i)}\sum_{j \in (N^o_i \cap V^o_i)} \mathbf{F}^o_{i,j}.
\end{equation}
The presence of the threshold $R_{max}$ prevents this average from yielding a very low value due to the sum of many small contributions.

\subsubsection{Step overlap term}
\label{step_app}
It has been suggested \cite{Curtis2014} that models based on prediction of collisions are ``too efficient'', as they typically present too high speeds at high densities. Indeed, the CP model would allow,
in the noiseless limit, a crowd of pedestrians with the same goal direction to move with their preferred speed regardless of their densities, or at least up to the maximum density compatible
with physical constraints. This is due to the fact that in such a condition, since all pedestrians have the same velocity, no collision is
predicted\footnote{More precisely, $t_i^p \to \infty$.} and there is no need of changing velocity.

This is clearly unrealistic, because pedestrians are not 2D rigid bodies. When a uni-directional, single flow motion is considered, the most important correction to the ``2D rigid body''
assumption amounts to take into consideration
the swing of the legs. As a first order assumption, we proceeded as follows\footnote{The proposed term can hardly be regarded as ``long range'', but we include it in this section because
  it is considered as part of the ``basic'' model.}.

To each pedestrian we associate an elliptical\footnote{A rectangular shape could be used to decrease computational time.} ``leg swing space'', with semi-axes $A_l$ and $B_l$.
$A_l$ is given by half of the pedestrian maximum linear extension (for a circular body with radius $R$, $A_l=R$; for an elliptical body with semi-axes $A>B$, $A_l=A$),
while $B_l$ is a parameter of the model that represents half
of the leg swing length. Such an elliptical space is centred at the body centre of mass (elliptical body centre), and oriented in such a way that the $B_l$
axis overlaps with the velocity direction\footnote{This means that, for an elliptical body, the body minor axis $B$ is not necessarily aligned with $B_l$, since we allow body direction
  and velocity direction to differ.}.
For each pedestrian pair $\{i,j\}$, and time $t$, assuming $j \in V^p_i$,  the set of visible (i.e., located on $i$'s front) pedestrians, we verify if their ``leg swing space'' overlap, and in case they do, we evaluate ``how much they overlap'' in the following way. We scale
both ``leg swing spaces'' of a linear factor $s$ (i.e., we replace the semi-axes with $s A_l$ and $s B_l$), and define $0 \leq s^f_{i,j} < 1$ as the maximum $s$ such that the two ellipses do not overlap.

Finally, if we add the following extra force term to the interaction term $\mathbf{F}^{long}$ acting on pedestrian $i$
\begin{equation}
  \label{eqover}
\mathbf{F}_i^s=-\min\left(k_s \left(\sum_{j\in (V^p_i)'} \left(1-s^f_{i,j}\right)\right),\left(\Delta t\right)^{-1}\right) \mathbf{v},
\end{equation}
where $(V^p_i)'$ is the set of visible pedestrians for which the overlapping condition is satisfied.

This term is a velocity-dependent ``friction'' term (parameter $k_s$ having dimension $t^{-1}$) which is caused by the overlapping of the pedestrians' ``leg swing spaces''. The minimum with respect to the inverse of the time step
is taken to be sure that, after a decision step, the pedestrian is slowed down, but not pushed into the opposite direction (see Section \ref{phys_dyn}
for a discussion of the meaning of the time step $\Delta t$
in this model).
\subsection{``Path finding'' and randomness}
\label{pathfind}
All proposed models include the ``preferred velocity'' term $-k_{v_p} (\dot{\mathbf{r}}-\mathbf{v}_p)$, so it is important to specify how the preferred velocity direction is computed.

The geometry of the crossing corridor is very simple, so that, as long as pedestrians are located inside the borders of their own
corridor,
  their preferred velocity
  should be directed as their walking direction $\mathbf{j}^{F(i)}$, which in the following discussion we may assume to be given by $(0,1)$ (i.e., we describe it according to
  the simulated pedestrian's own frame of reference as described by eq. \ref{pedownframe}). In this model, we include
  the ``unpredictability'' of human behaviour by defining the goal vector as
  \begin{equation}
    \label{enoise}
    \hat{\mathbf{g}}=\frac{\mathbf{g}}{g},\;\;\;\mathbf{g}=(N^1(\sigma_g),1+N^2(\sigma_g)),
    \end{equation}
  $N^m(\sigma)$ being independent realisations of a Gaussian random variable with zero mean and $\sigma$ standard deviation.

  As a result of collision avoidance, simulated pedestrians may be temporally located in the other stream's corridor, and this situation has to be managed by goal assignment. We verified that using a
  ``minimum path'' approach caused an overcrowding around the corridors' corners, which did not appear to be qualitatively similar to human behaviour. We thus adopted the following ``ad hoc'' procedure.
  Let us identify the positions of the crossing areas' 4 corners in each simulated pedestrian's own frame of reference with coordinates
  \begin{equation}
    \begin{split}
      c_x&=\pm \frac{L}{2},\\
      c_y&=\pm \frac{L}{2}.
      \end{split}
    \end{equation}
    Here, $L$ is the width corridor, and the origin of the reference frame is the centre of the crossing area. Since in each pedestrian's frame the walking direction is identified with
the $y$ axis, we assume the pedestrians to be located out of their own corridor if the following applies for the pedestrian's position $\mathbf{r}=(x,y)$:
   \begin{equation}
|x|\geq\frac{L}{2}-0.5 \text{m},\;-\frac{L}{2} \leq y \leq \frac{L}{2}.
   \end{equation}
   The 0.5 m term concerning the $x$ position is introduced assuming that pedestrians prefer to keep some distance from the corridor's wall.

   When such a condition is attained, pedestrians are assumed to re-enter the corridor by walking towards the corner identified by the position
   \begin{equation}
\overline{\mathbf{c}}=\left(\text{sign}(x)\left(\frac{L}{2}-0.5\right),\frac{L}{2}\right)\equiv (\overline{c}_x,\overline{c}_y).
   \end{equation}
   Nevertheless, to avoid overcrowding around the corner, we assign a stronger weight to the $x$ direction (walking back into the correct corridor) with respect to the $y$ direction
   (walking towards the goal) defining the goal vector as
  \begin{equation}
    \label{enoise2}
    \hat{\mathbf{g}}=\frac{\mathbf{g}}{g},\;\;\;\mathbf{g}=(2(\overline{c}_x-x+N^1(\sigma_g)),\overline{c}_y-y+N^2(\sigma_g)).
    \end{equation}   

  As stated above, this is definitely an ``ad hoc'' solution to the problem, but we are not aware of any quantitative studies regarding pedestrian path choices under such conditions. Nevertheless, we believe that,
  given the purpose of this work, it was important to adopt a relatively simple approach common to all computational models. For practical applications and future works, it may be interesting to try a machine learning approach aimed also at reproducing pedestrian path choices.
  \subsection{Parameters}
  \label{model_detailed_par}
The ``long range'' model (with the addition of the step overlap term), is characterised by the following parameters (assuming the body size of the pedestrian as given by
ellipse semi-axes $A \geq B$, with $A=B$ representing the circular case): $C$ (dimensionless), $d_{int}$ (length), $d_{ext}$ (length)  (eq. \ref{step1}), the mean $\mu_v$ and standard
deviation $\sigma_v$ of the preferred velocity $v_p$ distribution (assumed to be normally distributed), $k_{v_p}$ (eq. \ref{sfmacc}, inverse of time), the threshold under which walls are considered in
collision avoidance $R_{max}$ (length),
the size of the leg swing space $B_l$ (length), $k_s$ (eq. \ref{eqover}, inverse of time), and the ``goal noise'' standard deviation $\sigma_g$ (eq. \ref{enoise}).
In this work we assumed, following \cite{Zanlungo2011} $k_{v_p}=1.52$ s$^{-1}$, and fixed $R_{max}=2$ m,
leaving the 8 parameters $C$, $d_{int}$, $d_{ext}$, $\mu_v$, $\sigma_v$, $B_l$, $k_s$ and $\sigma_g$ to be calibrated.

A further parameter is the integration step $\Delta t$, that was fixed to $0.05$ s. The equations governing the behavioural dynamics (collision avoidance) of pedestrians are written
as differential equations but implemented as difference equations, by discretising through an Euler scheme. The integration step is common to all behavioural modules (including the short range ones
described below), and assumes that pedestrians take decisions at regular time steps and act based on them. Using a Euler scheme with a time step comparable to human reaction times is thus not
an issue, as long as, as described in Section \ref{phys_dyn}, physical dynamics is resolved at a finer scale.

The ``short range'' module introduces 6 new parameters:  $k^v_{\theta}$ (inverse of time), $k_{\omega}$ (inverse of time) and $k^{\omega}_{\theta}$ (inverse of squared time),
$\gamma$ (dimensionless), $\tau_1$ (time) and $\tau_2$ (time) (eqs. \ref{eqmr2}, \ref{eqmt}, \ref{ebeta}).

\subsection{Constraints}
All terms explained until now are considered to be the results of pedestrian decisional processes, and as such they should provide velocities and accelerations that are compatible with
realistic pedestrian behaviour. For this reason, in case after computing the discrete equations, the values of $|\dot{\mathbf{v}}|$ and $\dot{\omega}$ are higher
than thresholds $a_{max}=5$ m s$^{-2}$ and $\dot{\omega}_{max}=2 \pi $ s$^{-2}$, the final acceleration and angular accelerations are scaled (i.e., preserving their direction and sign) to such maximum
values. In an equivalent way, thresholds $v_{max}=3$ m/s and $\omega_{max}=2 \pi $ s$^{-1}$ are applied to velocity and angular velocity.

\section{Statistical analysis of observables}
\label{statan}
In this appendix we provide a definition of some recurrent statistical terms used in this work.

We assume that $O$ is an observable whose value is measured over $n$ repetitions, assumed statistically independent, of an experiment. Here, both the term ``observable'' and ``experiment'' are used in a very broad sense. ``Observable'' may refer not only to those defined in Section \ref{obs}, but also to the values assumed by the evaluation metrics defined in Appendix \ref{evalmet}. By ``experiment'', we may refer to the actual experiments with human pedestrians, to the simulations using the same initial conditions of a given experiment repetition, or even to an ``experiment with pedestrians-simulation'' pair in which the simulation uses the same initial conditions as the experiment.

The ``$n$ independent repetitions'' are assumed to share some common experimental condition (e.g., the initial density $\rho_I$) or a set of common experimental conditions (e.g., initial density and model specification used in simulation). Given these assumptions, we use some statistical tools to understand what our experiments tell us about the expected behaviour under these conditions and
about its statistical variation. All these tools are quite basic and are intended to provide some quantitative tool for comparison, without any claim to ``prove'' or ``disprove'' anything (although
terms like ``statistically significant'' are sometimes used in our manuscript when the evidence appears to be compelling, or on the other hand, when we want to stress the lack of such evidence).

Given these premises, writing the value assumed by $O$ in the $k$th repetition as $O_k$ we define the mean of the observable as
\begin{equation}
<O>=\frac{\sum_{k=1}^{n} O_k}{n},
\end{equation}
its standard deviation as
\begin{equation}
\sigma=\sqrt{\frac{\sum_{k=1}^{n} O_k^2}{n}-<O>^2},
\end{equation}
and its standard error as
\begin{equation}
\label{standard}
\varepsilon=\frac{\sigma}{\sqrt{n-1}}.
\end{equation}

When comparing results between two different conditions (e.g., when comparing the evaluation metric values attained by different models), which we may call condition $A$ and $B$, a possible rule
of thumb is to say that $O$ assumes a different value under the two conditions provided that
\begin{equation}
\label{thumb}
|<O>^A-<O>^B| \gg \sqrt{2} \max(\varepsilon^A,\varepsilon^B).
\end{equation}

This rule of thumb is obviously related to the ANOVA $p$ values reported in the text. In general, when comparing results between 2 different conditions, corresponding to $n^A$ and $n^B$ repetitions, we may define the overall mean as
\begin{equation}
<O>=\frac{<O>^A n^A+<O>^B n^B}{n^A+n^B}
\end{equation}
and then the deviations of each condition from this average as
\begin{equation}
d_A=<O>-<O>^A,\; d_B=<O>-<O>^B.
\end{equation}
The degrees of freedom are 
\begin{equation}
\gamma_1=1,\qquad \gamma_2=n^A+n^B-2.
\end{equation}
and the $F$ value is then defined as
\begin{equation}
F=\frac{\left(d_A^2 n^A+d_B^2 n^B \right)}{\left((\sigma^A)^2 n^A+(\sigma^B)^2 n^B \right)}\frac{\gamma_2}{\gamma_1}.
\end{equation}
In our tables, we report  the celebrated
$p$-value, that provides the probability, under the hypothesis of independence of data, that the difference between the distributions is due to chance \cite{Ash2011}
\begin{equation}
p=1-\int_0^F f_{\gamma_1,\gamma_1}(x) dx.
\end{equation}
The $F$ distribution has to be computed numerically \cite{Press1988}, but a value $F \gg 1$ assures a small $p$-value.

Let us see how this relates to the rule of thumb for standard errors. Let us assume, as usual in our comparisons
\begin{equation}
n^A=n^B=n
\end{equation}
We have
\begin{equation}
<O>=\frac{<O>^A+<O>^B}{2},
\end{equation}
\begin{equation}
|d_A|=|d_B|=\frac{|<O>^A-<O>^B|}{2},
\end{equation}
and
\begin{equation}
F=\frac{|<O>^A-<O>^B|^2}{(\sigma^A)^2+(\sigma^B)^2} (n-1).
\end{equation}
Using
\begin{equation}
\frac{(\sigma^{A,B})^2}{n-1} = (\varepsilon^{A,B})^2,
\end{equation}
we get the expression
\begin{equation}
F\approx\frac{|<O>^A-<O>^B|^2}{(\varepsilon^A)^2+(\varepsilon^B)^2}>\frac{|<O>^A-<O>^B|^2}{2( \max(\varepsilon^A,\varepsilon^B))^2},
\end{equation}
so that the rule of thumb eq. \ref{thumb} corresponds to have a high $F$-value and thus a low $p$-value.

The $F$ value is high if the $\sigma$s are smaller than the $d$s, i.e., if the variations inside the categories are smaller than the ones outside the categories, 
and if the total number of observations is high. It may be useful to use also an estimator that does not depend on the number of observations. In this work, we also use effect size, defined as
\begin{equation}
\begin{split}
\delta&=\frac{<O>^A-<O>^B}{\overline{\sigma}}, 
\\
\overline{\sigma}&=\sqrt{\frac{(n^A-1)(\sigma^A)^2+(n^B-1)(\sigma^B)^2}{n^A+n^B-2}}.
\end{split}
\end{equation}

While a $p$-value tells us about the significance of the statistical difference between two distributions, the difference may often be so small that it can be verified only if a large amount of
data are collected. But if we have also $\delta\approx 1$, then the two distributions are different enough to be distinguished also using a relatively reduced amount of data.

\section{Evaluation metrics}
\label{evalmet}
\subsection{Earth Mover's Distance}
The Earth Mover's Distance (EMD, \cite{Cohen1999}) may be used to compare probability distributions.

The EMD process can be visualised as filling holes by moving piles of dirt. Assume that $P$ and $Q$ denote two pdfs, and that a proper metric, named \textit{ground distance}
is defined to measure the distance between the bins $i$ and $j$ \footnote{It is common practice to use the Euclidean distance from $i$ to $j$ as the ground distance. Namely, in the 1D case this reduces to $|i-j|$.}. Suppose also that a flow $f(i,j)$ is applied to morph $P$ to $Q$, namely a (signed) quantity is subtracted from $P(i)$ and added to $Q(j)$ in the process of making the $P$ and $Q$ distributions more similar.
  
In this contest, EMD can be formulated and solved essentially as a transportation problem.
Namely, EMD aims at finding the amount of flow $f$ that minimises the overall cost of morphing $P$ to $Q$. Explicitly, the work required to morph $P$ to $Q$ (or vice versa) \mbox{\it given an explicit flow} $f$ is,
\begin{equation}
\label{eq:emd_work}
 \sum_i \sum_j f(i,j)d(i,j),
\end{equation}
where $f(i,j)$ and $d(i,j)$ are respectively, the flow and the ground distance between $P(i)$ and $Q(j)$.
By solving for the optimal flow and normalising it with the total flow, EMD is described as,

\begin{equation}
EMD(P, Q) = \frac{\min_f \sum_{i}\sum_j f(i,j)d(i,j)}{\sum_{i}\sum_j f(i,j)}.
\end{equation}

  The normalisation operation in the above equation yields the average distance travelled by unit weight under the optimal flow.

When computing the EMD between two discrete histograms defined on the same array of bins $i=0,...,N-1$, the following algorithm may be used
\begin{align}
\label{eq:EMDalgo1}
\begin{split}
  \text{EMD}_0&=0,\\
  \text{EMD}_{i+1}&=Q_i-P_i+\text{EMD}_{i},\\
  \text{EMD}&=\frac{\sum_{i=1}^N |\text{EMD}_{i}|}{N}.
  \end{split}
\end{align}

In the last step, we normalise dividing by the number of bins $N$.

In this work, the comparison is performed between observables averaged over independent repetitions of experiments according to eq. \ref{obavp}.
\subsection{``Standard'' metric}
As shown by eq. \ref{eq:EMDalgo1}, when the absolute difference between the two histograms is small with respect to the sum over all bins, $|Q_i-P_i|\ll 1$, the contribution
of bin $i$ to the overall EMD will be $\ll 1$ (due to the final $1/N$ normalisation), {\it regardless of the ratio} $|Q_i|/|P_i|$. Nevertheless, such relative difference may be relevant
to our study.

We were particularly interested in identifying simulated observable distributions presenting relatively high values in regions in which the empirical distribution presents relatively high values,
and we defined the following evaluation metric, that we name ``Standard'' since it is based on a comparison using standard errors.

Let us assume that the empirical distribution of the observable under consideration can be expressed through $\langle E_i \rangle \pm \epsilon_i$, defined according to eqs. \ref{obavp}, \ref{obavp2}.
Furthermore, let us assume the corresponding simulated distribution as given by  $\langle S_i \rangle$.

Since we want to compare the difference between empirical and simulated distributions with the uncertainty in the empirical data
\begin{equation}
  \label{defdq}
\delta_i =\frac{S_i-E_i}{\epsilon_i},
\end{equation}
the simulated distribution standard error does not play a role in the definition.

Eq. \ref{defdq} is not defined when we have $E_i=\epsilon_i=0$. To deal with such situations, we compute
\begin{equation}
\overline{\epsilon}=\frac{\min_{\epsilon_i \neq 0} \epsilon_i}{2}
\end{equation}
and define $\delta_i$ for the $E_i=\epsilon_i=0$ case as
\begin{equation}
\delta_i =\frac{S_i}{\overline{\epsilon}}.
\end{equation}

Finally the Standard Metric is defined as
\begin{equation}
S=\sqrt{\frac{\sum_{i=1}^N \delta_i^2}{N}}.
\end{equation}

%%%%%%%%%%%%%%%%%%%%%%%%%%%%%%%%%%%%%%%%%%%%%%%%%%%%%%%%%%%%%%%%%%%%%%%%%%%%%%%%%%%%%%%%%%%%%%%%%%%%%%%%%%%
\section{Evaluation using the Standard Metric}
\label{standeval}
In Tables \ref{Standard}, \ref{Standard2}, \ref{Standard3}, \ref{Standard4} we evaluate the different models at $\rho_I=2.5$ ped/m$^2$ and compare them using the standard metric.
\label{standapp}
\begin{table}[h!]
\begin{center}
\caption{Standard Metric comparison between Full and No-short elliptical models, evaluation at $\rho_I=$ 2.5 ped/m$^2$}
\label{Standard}
\begin{tabular}{|c|c|c|c|c|c|}
\hline
& Full $\langle \; \rangle \pm \varepsilon (\sigma)$ & No-short $\langle \;  \rangle \pm \varepsilon (\sigma)$ & $p$ & effect size\\
\hline
Total &  $5.18\pm 0.55(2.6)$ & $10.8\pm 1.5(7)$ & 0.00084 &1.1\\
\hline
$\rho$ &  $3.35\pm 1.1(5.2)$ & $4.48\pm 0.88(4.2)$ & 0.43 &0.24\\
\hline
$v$ &  $3.71\pm 0.55(2.7)$ & $5.18\pm 0.65(3.1)$ & 0.092 &0.51\\
\hline
$\theta^v$ &  $7.49\pm 1(4.9)$ & $9.95\pm 1.5(7.2)$ & 0.18 &0.4\\
\hline
$\delta^s$ &  $7.52\pm 1.2(6)$ & $27.6\pm 5.2(25)$ & 0.00051 &1.1\\
\hline
$\delta^o$ &  $5.87\pm 2.1(9.9)$ & $28\pm 7.2(35)$ & 0.0049 &0.87\\
\hline
$\phi^s$ &  $2.27\pm 0.074(0.36)$ & $2.6\pm 0.09(0.43)$ & 0.0079 &0.82\\
\hline
$\phi^o$ &  $4.88\pm 0.3(1.4)$ & $5.16\pm 0.33(1.6)$ & 0.53 &0.18\\
\hline
$\theta$ &  $5.34\pm 0.33(1.6)$ & $5.19\pm 0.66(3.2)$ & 0.84 &0.061\\
\hline
$\delta \theta$ &  $6.16\pm 0.5(2.4)$ & $8.59\pm 1.2(5.6)$ & 0.061 &0.57\\
\hline
\end{tabular}
\end{center}
\end{table}

\begin{table}[h!]
\begin{center}
\caption{Standard Metric comparison between Full Elliptical and Full Circular models, evaluation at $\rho_I=$ 2.5 ped/m$^2$.}
\label{Standard2}
\begin{tabular}{|c|c|c|c|c|c|}
\hline
& Elliptical $\langle \; \rangle \pm \varepsilon (\sigma)$ & Circular $\langle \;  \rangle \pm \varepsilon (\sigma)$ & $p$ & effect size\\
\hline
No body &  $5.59\pm 0.91(4.4)$ & $7.05\pm 0.93(4.4)$ & 0.27 &0.33\\
\hline
$\rho$ &  $3.35\pm 1.1(5.2)$ & $5.65\pm 0.9(4.3)$ & 0.11 &0.48\\
\hline
$v$ &  $3.71\pm 0.55(2.7)$ & $3.69\pm 0.4(1.9)$ & 0.98 &0.0068\\
\hline
$\theta^v$ &  $7.49\pm 1(4.9)$ & $18.1\pm 3.1(15)$ & 0.0024 &0.95\\
\hline
$\delta^s$ &  $7.52\pm 1.2(6)$ & $5.46\pm 0.81(3.9)$ & 0.17 &0.41\\
\hline
$\delta^o$ &  $5.87\pm 2.1(9.9)$ & $2.41\pm 0.15(0.73)$ & 0.1 &0.49\\
\hline
$\phi^s$ &  $2.27\pm 0.074(0.36)$ & $2.19\pm 0.079(0.38)$ & 0.43 &0.24\\
\hline
$\phi^o$ &  $4.88\pm 0.3(1.4)$ & $4.67\pm 0.29(1.4)$ & 0.63 &0.14\\
\hline
\end{tabular}
\end{center}
\end{table}

\begin{table}[h!]
\begin{center}
\caption{Standard Metric comparison between Full Elliptical  and No-Short Circular  (CP) models, evaluation at $\rho_I=$ 2.5 ped/m$^2$.}
\label{Standard3}
\begin{tabular}{|c|c|c|c|c|c|}
\hline
& Full $\langle \; \rangle \pm \varepsilon (\sigma)$ & CP $\langle \;  \rangle \pm \varepsilon (\sigma)$ & $p$ & effect size\\
\hline
No body &  $5.59\pm 0.91(4.4)$ & $6.39\pm 0.89(4.3)$ & 0.53 &0.19\\
\hline
$\rho$ &  $3.35\pm 1.1(5.2)$ & $4.79\pm 1.8(8.4)$ & 0.49 &0.21\\
\hline
$v$ &  $3.71\pm 0.55(2.7)$ & $5.82\pm 0.78(3.7)$ & 0.032 &0.65\\
\hline
$\theta^v$ &  $7.49\pm 1(4.9)$ & $12.5\pm 2(9.8)$ & 0.032 &0.65\\
\hline
$\delta^s$ &  $7.52\pm 1.2(6)$ & $5.72\pm 0.21(1)$ & 0.16 &0.42\\
\hline
$\delta^o$ &  $5.87\pm 2.1(9.9)$ & $3.09\pm 0.18(0.86)$ & 0.19 &0.4\\
\hline
$\phi^s$ &  $2.27\pm 0.074(0.36)$ & $2.52\pm 0.062(0.3)$ & 0.013 &0.76\\
\hline
$\phi^o$ &  $4.88\pm 0.3(1.4)$ & $4.61\pm 0.35(1.7)$ & 0.56 &0.17\\
\hline
\end{tabular}
\end{center}
\end{table}

\begin{table}[h!]
\begin{center}
\caption{Standard  metric comparison between Full Circular and No-short Circular models, evaluation at $\rho_I=$ 2.5 ped/m$^2$.}
\label{Standard4}
\begin{tabular}{|c|c|c|c|c|c|}
\hline
& Full $\langle \; \rangle \pm \varepsilon (\sigma)$ & No-short $\langle \;  \rangle \pm \varepsilon (\sigma)$ & $p$  & effect size\\
\hline
No body &  $7.05\pm 0.93(4.4)$ & $6.39\pm 0.89(4.3)$ & 0.61 &0.15\\
\hline
$\rho$ &  $5.65\pm 0.9(4.3)$ & $4.79\pm 1.8(8.4)$ & 0.67 &0.13\\
\hline
$v$ &  $3.69\pm 0.4(1.9)$ & $5.82\pm 0.78(3.7)$ & 0.019 &0.72\\
\hline
$\theta^v$ &  $18.1\pm 3.1(15)$ & $12.5\pm 2(9.8)$ & 0.15 &0.44\\
\hline
$\delta^s$ &  $5.46\pm 0.81(3.9)$ & $5.72\pm 0.21(1)$ & 0.75 &0.094\\
\hline
$\delta^o$ &  $2.41\pm 0.15(0.73)$ & $3.09\pm 0.18(0.86)$ & 0.0061 &0.85\\
\hline
$\phi^s$ &  $2.19\pm 0.079(0.38)$ & $2.52\pm 0.062(0.3)$ & 0.0015 &0.99\\
\hline
$\phi^o$ &  $4.67\pm 0.29(1.4)$ & $4.61\pm 0.35(1.7)$ & 0.88 &0.044\\
\hline
\end{tabular}
\end{center}
\end{table}

\section{Comparison on low and medium density ranges}
\label{appcomp}
\subsection{Full Elliptical vs No-short Elliptical}
We compare the Full Elliptical and No-short Elliptical models on the $\rho_I=0.25$ ped/m$^2$  (see Tables \ref{EMD_1_low} and \ref{Standard_1_low}) and $\rho_I=1.5$ ped/m$^2$
(see Tables \ref{EMD_1_medium} and \ref{Standard_1_medium}) initial conditions
on which they were calibrated.
\begin{table}[h!]
\begin{center}
\caption{EMD metric comparison between Full and No-short elliptical models, $\rho_I=0.25$ ped/m$^2$ initial condition.}
\label{EMD_1_low}
\begin{tabular}{|c|c|c|c|c|c|}
\hline
& Full $\langle \; \rangle \pm \varepsilon (\sigma)$ & No-short $\langle \;  \rangle \pm \varepsilon (\sigma)$ & $p$ &  effect size\\
\hline
Total &  $0.00885\pm 0.00026(0.0013)$ & $0.00862\pm 0.0002(0.00098)$ & 0.49 &0.21\\
\hline
$\rho$ &  $0.00458\pm 0.00013(0.00062)$ & $0.00458\pm 0.00012(0.00056)$ & 1 &0\\
\hline
$v$ &  $0.0189\pm 0.0015(0.0073)$ & $0.0166\pm 0.00083(0.004)$ & 0.2 &0.39\\
\hline
$\theta^v$ &  $0.00459\pm 0.0003(0.0014)$ & $0.00474\pm 0.00023(0.0011)$ & 0.7 &0.12\\
\hline
$\delta^s$ &  $0.00969\pm 0.00081(0.0039)$ & $0.00761\pm 0.0007(0.0033)$ & 0.057 &0.58\\
\hline
$\delta^o$ &  $0.00957\pm 0.00089(0.0043)$ & $0.0113\pm 0.00068(0.0033)$ & 0.13 &0.45\\
\hline
$\phi^s$ &  $0.0103\pm 0.00082(0.0039)$ & $0.00896\pm 0.00096(0.0046)$ & 0.29 &0.31\\
\hline
$\phi^o$ &  $0.00845\pm 0.00072(0.0035)$ & $0.00862\pm 0.00063(0.003)$ & 0.86 &0.053\\
\hline
$\theta$ &  $0.00721\pm 0.00012(0.00058)$ & $0.00742\pm 0.0001(0.0005)$ & 0.2 &0.39\\
\hline
$\delta \theta$ &  $0.00743\pm 0.00025(0.0012)$ & $0.00808\pm 0.00021(0.00099)$ & 0.055 &0.58\\
\hline
\end{tabular}
\end{center}
\end{table}

\begin{table}[h!]
\begin{center}
\caption{Standard Metric comparison between Full and No-short elliptical models, $\rho_I=0.25$ ped/m$^2$ initial condition.}
\label{Standard_1_low}
\begin{tabular}{|c|c|c|c|c|c|}
\hline
& Full $\langle \; \rangle \pm \varepsilon (\sigma)$ & No-short $\langle \;  \rangle \pm \varepsilon (\sigma)$ & $p$ & effect size\\
\hline
Total &  $3.63\pm 0.12(0.57)$ & $3.79\pm 0.12(0.56)$ & 0.37 &0.27\\
\hline
$\rho$ &  $0.898\pm 0.034(0.16)$ & $0.901\pm 0.024(0.11)$ & 0.94 &0.023\\
\hline
$v$ &  $4.46\pm 0.33(1.6)$ & $3.67\pm 0.21(1)$ & 0.05 &0.59\\
\hline
$\theta^v$ &  $6.04\pm 0.45(2.2)$ & $6.65\pm 0.42(2)$ & 0.33 &0.29\\
\hline
$\delta^s$ &  $2.78\pm 0.22(1)$ & $3.34\pm 0.38(1.8)$ & 0.21 &0.37\\
\hline
$\delta^o$ &  $3.16\pm 0.26(1.2)$ & $3.72\pm 0.13(0.65)$ & 0.057 &0.58\\
\hline
$\phi^s$ &  $1.41\pm 0.043(0.2)$ & $1.33\pm 0.038(0.18)$ & 0.19 &0.39\\
\hline
$\phi^o$ &  $1.84\pm 0.06(0.29)$ & $1.93\pm 0.049(0.24)$ & 0.24 &0.35\\
\hline
$\theta$ &  $3.89\pm 0.18(0.87)$ & $3.99\pm 0.28(1.3)$ & 0.78 &0.085\\
\hline
$\delta \theta$ &  $4.21\pm 0.27(1.3)$ & $4.24\pm 0.17(0.81)$ & 0.92 &0.031\\
\hline
\end{tabular}
\end{center}
\end{table}

\begin{table}[h!]
\begin{center}
\caption{EMD metric comparison between Full and No-short elliptical models, $\rho_I=1.5$ ped/m$^2$ initial condition.}
\label{EMD_1_medium}
\begin{tabular}{|c|c|c|c|c|c|}
\hline
& Full $\langle \; \rangle \pm \varepsilon (\sigma)$ & No-short $\langle \;  \rangle \pm \varepsilon (\sigma)$ & $p$ & effect size\\
\hline
Total &  $0.0128\pm 0.00056(0.0027)$ & $0.0162\pm 0.001(0.005)$ & 0.0062 &0.85\\
\hline
$\rho$ &  $0.00975\pm 0.00094(0.0045)$ & $0.011\pm 0.001(0.0049)$ & 0.35 &0.28\\
\hline
$v$ &  $0.0211\pm 0.0025(0.012)$ & $0.0295\pm 0.0041(0.02)$ & 0.087 &0.52\\
\hline
$\theta^v$ &  $0.00813\pm 0.00056(0.0027)$ & $0.00892\pm 0.0006(0.0029)$ & 0.34 &0.28\\
\hline
$\delta^s$ &  $0.0202\pm 0.0009(0.0043)$ & $0.0274\pm 0.00095(0.0046)$ & 1.3e-06 &1.6\\
\hline
$\delta^o$ &  $0.0114\pm 0.0012(0.0058)$ & $0.0139\pm 0.003(0.014)$ & 0.44 &0.23\\
\hline
$\phi^s$ &  $0.0153\pm 0.0015(0.0072)$ & $0.0174\pm 0.0021(0.01)$ & 0.41 &0.25\\
\hline
$\phi^o$ &  $0.0312\pm 0.0026(0.012)$ & $0.0328\pm 0.0025(0.012)$ & 0.67 &0.13\\
\hline
$\theta$ &  $0.00978\pm 0.00076(0.0037)$ & $0.00969\pm 0.0019(0.0092)$ & 0.97 &0.012\\
\hline
$\delta \theta$ &  $0.00916\pm 0.00071(0.0034)$ & $0.0127\pm 0.00081(0.0039)$ & 0.0021 &0.96\\
\hline
\end{tabular}
\end{center}
\end{table}

\begin{table}[h!]
\begin{center}
\caption{Standard Metric comparison between Full and No-short elliptical models, $\rho_I=1.5$ ped/m$^2$ initial condition.}
\label{Standard_1_medium}
\begin{tabular}{|c|c|c|c|c|c|}
\hline
& Full $\langle \; \rangle \pm \varepsilon (\sigma)$ & No-short $\langle \;  \rangle \pm \varepsilon (\sigma)$ & $p$ & effect size\\
\hline
Total &  $5.85\pm 0.36(1.7)$ & $11.3\pm 2(9.7)$ & 0.01 &0.79\\
\hline
$\rho$ &  $2.35\pm 0.37(1.8)$ & $2.61\pm 0.29(1.4)$ & 0.58 &0.16\\
\hline
$v$ &  $8.73\pm 1(5)$ & $11\pm 1.4(6.6)$ & 0.19 &0.39\\
\hline
$\theta^v$ &  $6.29\pm 0.37(1.8)$ & $7.41\pm 0.72(3.4)$ & 0.17 &0.41\\
\hline
$\delta^s$ &  $9.65\pm 0.99(4.8)$ & $35.5\pm 8.9(43)$ & 0.006 &0.85\\
\hline
$\delta^o$ &  $3.55\pm 0.55(2.7)$ & $11.3\pm 3.2(15)$ & 0.022 &0.7\\
\hline
$\phi^s$ &  $1.73\pm 0.069(0.33)$ & $1.88\pm 0.075(0.36)$ & 0.15 &0.43\\
\hline
$\phi^o$ &  $3.98\pm 0.26(1.2)$ & $4.57\pm 0.24(1.2)$ & 0.099 &0.5\\
\hline
$\theta$ &  $5.17\pm 0.4(1.9)$ & $4.69\pm 0.81(3.9)$ & 0.6 &0.16\\
\hline
$\delta \theta$ &  $5.26\pm 0.56(2.7)$ & $6.83\pm 1.2(6)$ & 0.26 &0.34\\
\hline
\end{tabular}
\end{center}
\end{table}

\subsection{Full Elliptical vs Full Circular}
We compare the Full Elliptical and Full Circular models on the $\rho_I=0.25$ ped/m$^2$  (see Tables \ref{EMD_2_low} and \ref{Standard_2_low}) and $\rho_I=1.5$ ped/m$^2$
(see Tables \ref{EMD_2_medium} and \ref{Standard_2_medium}) initial conditions
on which they were calibrated.

\begin{table}[h!]
\begin{center}
\caption{EMD metric comparison between  Full Elliptical and Full Circular models,  $\rho_I=0.25$ ped/m$^2$ initial condition.}
\label{EMD_2_low}
\begin{tabular}{|c|c|c|c|c|c|}
\hline
& Elliptical $\langle \;  \rangle \pm \varepsilon (\sigma)$ & Circular $\langle \;  \rangle \pm \varepsilon (\sigma)$ & $p$ & effect size\\
\hline
No body &  $0.00947\pm 0.00036(0.0018)$ & $0.00996\pm 0.00037(0.0018)$ & 0.35 &0.28\\
\hline
$\rho$ &  $0.00458\pm 0.00013(0.00062)$ & $0.00515\pm 0.00016(0.00077)$ & 0.0082 &0.82\\
\hline
$v$ &  $0.0189\pm 0.0015(0.0073)$ & $0.02\pm 0.0013(0.006)$ & 0.58 &0.16\\
\hline
$\theta^v$ &  $0.00459\pm 0.0003(0.0014)$ & $0.00385\pm 0.00029(0.0014)$ & 0.079 &0.53\\
\hline
$\delta^s$ &  $0.00969\pm 0.00081(0.0039)$ & $0.00915\pm 0.00056(0.0027)$ & 0.59 &0.16\\
\hline
$\delta^o$ &  $0.00957\pm 0.00089(0.0043)$ & $0.0117\pm 0.00095(0.0046)$ & 0.12 &0.47\\
\hline
$\phi^s$ &  $0.0103\pm 0.00082(0.0039)$ & $0.0101\pm 0.00074(0.0036)$ & 0.89 &0.043\\
\hline
$\phi^o$ &  $0.00845\pm 0.00072(0.0035)$ & $0.0111\pm 0.00093(0.0045)$ & 0.032 &0.65\\
\hline
\end{tabular}
\end{center}
\end{table}

\begin{table}[h!]
\begin{center}
\caption{Standard Metric comparison between Full Elliptical and Full Circular models,  $\rho_I=0.25$ ped/m$^2$ initial condition.}
\label{Standard_2_low}
\begin{tabular}{|c|c|c|c|c|c|}
\hline
& Elliptical $\langle \; \rangle \pm \varepsilon (\sigma)$ & Circular $\langle \;  \rangle \pm \varepsilon (\sigma)$ & $p$ & effect size\\
\hline
No body &  $3.47\pm 0.14(0.67)$ & $3.51\pm 0.17(0.8)$ & 0.84 &0.058\\
\hline
$\rho$ &  $0.898\pm 0.034(0.16)$ & $0.949\pm 0.025(0.12)$ & 0.23 &0.36\\
\hline
$v$ &  $4.46\pm 0.33(1.6)$ & $4.98\pm 0.49(2.3)$ & 0.38 &0.26\\
\hline
$\theta^v$ &  $6.04\pm 0.45(2.2)$ & $4.92\pm 0.4(1.9)$ & 0.069 &0.55\\
\hline
$\delta^s$ &  $2.78\pm 0.22(1)$ & $3.13\pm 0.3(1.4)$ & 0.35 &0.28\\
\hline
$\delta^o$ &  $3.16\pm 0.26(1.2)$ & $3.57\pm 0.22(1.1)$ & 0.23 &0.36\\
\hline
$\phi^s$ &  $1.41\pm 0.043(0.2)$ & $1.36\pm 0.044(0.21)$ & 0.44 &0.23\\
\hline
$\phi^o$ &  $1.84\pm 0.06(0.29)$ & $2.14\pm 0.094(0.45)$ & 0.0091 &0.8\\
\hline
\end{tabular}
\end{center}
\end{table}

\begin{table}[h!]
\begin{center}
\caption{EMD metric comparison between  Full Elliptical and Full Circular models, $\rho_I=1.5$ ped/m$^2$ initial condition.}
\label{EMD_2_medium}
\begin{tabular}{|c|c|c|c|c|c|}
\hline
& Elliptical $\langle \;  \rangle \pm \varepsilon (\sigma)$ & Circular $\langle \;  \rangle \pm \varepsilon (\sigma)$ & $p$ & effect size\\
\hline
No body &  $0.0141\pm 0.00077(0.0037)$ & $0.015\pm 0.00067(0.0032)$ & 0.38 &0.26\\
\hline
$\rho$ &  $0.00975\pm 0.00094(0.0045)$ & $0.0101\pm 0.00069(0.0033)$ & 0.74 &0.097\\
\hline
$v$ &  $0.0211\pm 0.0025(0.012)$ & $0.0248\pm 0.0019(0.009)$ & 0.24 &0.35\\
\hline
$\theta^v$ &  $0.00813\pm 0.00056(0.0027)$ & $0.00694\pm 0.0005(0.0024)$ & 0.12 &0.47\\
\hline
$\delta^s$ &  $0.0202\pm 0.0009(0.0043)$ & $0.0224\pm 0.0012(0.0058)$ & 0.15 &0.43\\
\hline
$\delta^o$ &  $0.0114\pm 0.0012(0.0058)$ & $0.0108\pm 0.0018(0.0088)$ & 0.77 &0.086\\
\hline
$\phi^s$ &  $0.0153\pm 0.0015(0.0072)$ & $0.0192\pm 0.0023(0.011)$ & 0.16 &0.43\\
\hline
$\phi^o$ &  $0.0312\pm 0.0026(0.012)$ & $0.0259\pm 0.0024(0.012)$ & 0.14 &0.44\\
\hline
\end{tabular}
\end{center}
\end{table}

\begin{table}[h!]
\begin{center}
\caption{Standard Metric comparison between Full Elliptical and Full Circular models, $\rho_I=1.5$ ped/m$^2$ initial condition.}
\label{Standard_2_medium}
\begin{tabular}{|c|c|c|c|c|c|}
\hline
& Elliptical $\langle \; \rangle \pm \varepsilon (\sigma)$ & Circular $\langle \;  \rangle \pm \varepsilon (\sigma)$ & $p$ & effect size\\
\hline
No body &  $6.11\pm 0.43(2.1)$ & $6.28\pm 0.4(1.9)$ & 0.78 &0.083\\
\hline
$\rho$ &  $2.35\pm 0.37(1.8)$ & $2.62\pm 0.23(1.1)$ & 0.53 &0.19\\
\hline
$v$ &  $8.73\pm 1(5)$ & $6.69\pm 0.86(4.1)$ & 0.14 &0.44\\
\hline
$\theta^v$ &  $6.29\pm 0.37(1.8)$ & $8.96\pm 1.1(5.2)$ & 0.025 &0.68\\
\hline
$\delta^s$ &  $9.65\pm 0.99(4.8)$ & $10.1\pm 1.6(7.5)$ & 0.82 &0.069\\
\hline
$\delta^o$ &  $3.55\pm 0.55(2.7)$ & $3.03\pm 0.3(1.5)$ & 0.42 &0.24\\
\hline
$\phi^s$ &  $1.73\pm 0.069(0.33)$ & $1.91\pm 0.076(0.37)$ & 0.093 &0.51\\
\hline
$\phi^o$ &  $3.98\pm 0.26(1.2)$ & $3.84\pm 0.24(1.1)$ & 0.69 &0.12\\
\hline
\end{tabular}
\end{center}
\end{table}

\subsection{Full Elliptical vs No-short Circular (CP)}
We compare the Full Elliptical and No-short Circular models on the $\rho_I=0.25$ ped/m$^2$  (see Tables \ref{EMD_3_low} and \ref{Standard_3_low}) and $\rho_I=1.5$ ped/m$^2$
(see Tables \ref{EMD_3_medium} and \ref{Standard_3_medium}) initial conditions
on which they were calibrated.

\begin{table}[h!]
\begin{center}
\caption{EMD metric comparison between Full Elliptical  and No-Short Circular  (CP) models, $\rho_I=0.25$ ped/m$^2$ initial conditions.}
\label{EMD_3_low}
\begin{tabular}{|c|c|c|c|c|c|}
\hline
& Full $\langle \;  \rangle \pm \varepsilon (\sigma)$ & CP $\langle \;  \rangle \pm \varepsilon (\sigma)$ & $p$ & effect size\\
\hline
No body &  $0.00947\pm 0.00036(0.0018)$ & $0.0108\pm 0.00046(0.0022)$ & 0.031 &0.66\\
\hline
$\rho$ &  $0.00458\pm 0.00013(0.00062)$ & $0.00479\pm 0.00013(0.0006)$ & 0.24 &0.35\\
\hline
$v$ &  $0.0189\pm 0.0015(0.0073)$ & $0.0198\pm 0.0009(0.0043)$ & 0.62 &0.15\\
\hline
$\theta^v$ &  $0.00459\pm 0.0003(0.0014)$ & $0.00418\pm 0.00023(0.0011)$ & 0.28 &0.32\\
\hline
$\delta^s$ &  $0.00969\pm 0.00081(0.0039)$ & $0.0116\pm 0.001(0.005)$ & 0.15 &0.43\\
\hline
$\delta^o$ &  $0.00957\pm 0.00089(0.0043)$ & $0.0135\pm 0.0014(0.0066)$ & 0.02 &0.71\\
\hline
$\phi^s$ &  $0.0103\pm 0.00082(0.0039)$ & $0.00959\pm 0.00096(0.0046)$ & 0.57 &0.17\\
\hline
$\phi^o$ &  $0.00845\pm 0.00072(0.0035)$ & $0.00992\pm 0.00045(0.0021)$ & 0.09 &0.51\\
\hline
\end{tabular}
\end{center}
\end{table}

\begin{table}[h!]
\begin{center}
\caption{Standard Metric comparison between Full Elliptical  and No-Short Circular  (CP) models, $\rho_I=0.25$ ped/m$^2$ initial conditions.}
\label{Standard_3_low}
\begin{tabular}{|c|c|c|c|c|c|}
\hline
& Full $\langle \; \rangle \pm \varepsilon (\sigma)$ & CP $\langle \;  \rangle \pm \varepsilon (\sigma)$ & $p$  & effect size\\
\hline
No body &  $3.47\pm 0.14(0.67)$ & $4.41\pm 0.22(1.1)$ & 0.00079 &1.1\\
\hline
$\rho$ &  $0.898\pm 0.034(0.16)$ & $0.908\pm 0.022(0.11)$ & 0.8 &0.077\\
\hline
$v$ &  $4.46\pm 0.33(1.6)$ & $5.3\pm 0.43(2.1)$ & 0.13 &0.46\\
\hline
$\theta^v$ &  $6.04\pm 0.45(2.2)$ & $5.74\pm 0.37(1.8)$ & 0.61 &0.15\\
\hline
$\delta^s$ &  $2.78\pm 0.22(1)$ & $5.6\pm 0.57(2.7)$ & 3.1e-05 &1.4\\
\hline
$\delta^o$ &  $3.16\pm 0.26(1.2)$ & $4.47\pm 0.31(1.5)$ & 0.0021 &0.96\\
\hline
$\phi^s$ &  $1.41\pm 0.043(0.2)$ & $1.29\pm 0.035(0.17)$ & 0.045 &0.61\\
\hline
$\phi^o$ &  $1.84\pm 0.06(0.29)$ & $2.22\pm 0.058(0.28)$ & 4.2e-05 &1.3\\
\hline
\end{tabular}
\end{center}
\end{table}

\begin{table}[h!]
\begin{center}
\caption{EMD metric comparison between Full Elliptical  and No-Short Circular  (CP) models, $\rho_I=1.5$ ped/m$^2$ initial conditions.}
\label{EMD_3_medium}
\begin{tabular}{|c|c|c|c|c|c|}
\hline
& Full $\langle \;  \rangle \pm \varepsilon (\sigma)$ & CP $\langle \;  \rangle \pm \varepsilon (\sigma)$ & $p$ & effect size\\
\hline
No body &  $0.0141\pm 0.00077(0.0037)$ & $0.0199\pm 0.0012(0.0057)$ & 0.00019 &1.2\\
\hline
$\rho$ &  $0.00975\pm 0.00094(0.0045)$ & $0.0125\pm 0.0011(0.0052)$ & 0.062 &0.56\\
\hline
$v$ &  $0.0211\pm 0.0025(0.012)$ & $0.0377\pm 0.004(0.019)$ & 0.001 &1\\
\hline
$\theta^v$ &  $0.00813\pm 0.00056(0.0027)$ & $0.00765\pm 0.00045(0.0022)$ & 0.51 &0.2\\
\hline
$\delta^s$ &  $0.0202\pm 0.0009(0.0043)$ & $0.0268\pm 0.00096(0.0046)$ & 7.6e-06 &1.5\\
\hline
$\delta^o$ &  $0.0114\pm 0.0012(0.0058)$ & $0.0147\pm 0.0013(0.006)$ & 0.066 &0.55\\
\hline
$\phi^s$ &  $0.0153\pm 0.0015(0.0072)$ & $0.0177\pm 0.0014(0.0066)$ & 0.23 &0.36\\
\hline
$\phi^o$ &  $0.0312\pm 0.0026(0.012)$ & $0.0254\pm 0.0024(0.012)$ & 0.11 &0.48\\
\hline
\end{tabular}
\end{center}
\end{table}

\begin{table}[h!]
\begin{center}
\caption{Standard Metric comparison between Full Elliptical  and No-Short Circular  (CP) models, $\rho_I=1.5$ ped/m$^2$ initial conditions.}
\label{Standard_3_medium}
\begin{tabular}{|c|c|c|c|c|c|}
\hline
& Full $\langle \; \rangle \pm \varepsilon (\sigma)$ & CP $\langle \;  \rangle \pm \varepsilon (\sigma)$ & $p$ & effect size\\
\hline
No body &  $6.11\pm 0.43(2.1)$ & $8.28\pm 0.48(2.3)$ & 0.0015 &1\\
\hline
$\rho$ &  $2.35\pm 0.37(1.8)$ & $2.77\pm 0.28(1.3)$ & 0.36 &0.27\\
\hline
$v$ &  $8.73\pm 1(5)$ & $12.3\pm 1.7(8.2)$ & 0.084 &0.52\\
\hline
$\theta^v$ &  $6.29\pm 0.37(1.8)$ & $8.94\pm 0.88(4.2)$ & 0.0079 &0.82\\
\hline
$\delta^s$ &  $9.65\pm 0.99(4.8)$ & $12.5\pm 0.9(4.3)$ & 0.039 &0.63\\
\hline
$\delta^o$ &  $3.55\pm 0.55(2.7)$ & $4.95\pm 0.42(2)$ & 0.05 &0.59\\
\hline
$\phi^s$ &  $1.73\pm 0.069(0.33)$ & $1.92\pm 0.081(0.39)$ & 0.091 &0.51\\
\hline
$\phi^o$ &  $3.98\pm 0.26(1.2)$ & $4.03\pm 0.28(1.3)$ & 0.89 &0.041\\
\hline
\end{tabular}
\end{center}
\end{table}

\subsection{Full Circular vs No-short Circular (CP)}
We compare the Full Circular and No-short Circular models on the $\rho_I=0.25$ ped/m$^2$  (see Tables \ref{EMD_4_low} and \ref{Standard_4_low}) and $\rho_I=1.5$ ped/m$^2$
(see Tables \ref{EMD_4_medium} and \ref{Standard_4_medium}) initial conditions
on which they were calibrated.

\begin{table}[h!]
\begin{center}
\caption{EMD metric comparison between Full Circular and No-short Circular models, $\rho_I=0.25$ ped/m$^2$ initial conditions.}
\label{EMD_4_low}
\begin{tabular}{|c|c|c|c|c|c|}
\hline
& Full $\langle \;  \rangle \pm \varepsilon (\sigma)$ & No-short $\langle \;  \rangle \pm \varepsilon (\sigma)$ & $p$ & effect size\\
\hline
No body &  $0.00996\pm 0.00037(0.0018)$ & $0.0108\pm 0.00046(0.0022)$ & 0.17 &0.41\\
\hline
$\rho$ &  $0.00515\pm 0.00016(0.00077)$ & $0.00479\pm 0.00013(0.0006)$ & 0.088 &0.51\\
\hline
$v$ &  $0.02\pm 0.0013(0.006)$ & $0.0198\pm 0.0009(0.0043)$ & 0.89 &0.042\\
\hline
$\theta^v$ &  $0.00385\pm 0.00029(0.0014)$ & $0.00418\pm 0.00023(0.0011)$ & 0.36 &0.27\\
\hline
$\delta^s$ &  $0.00915\pm 0.00056(0.0027)$ & $0.0116\pm 0.001(0.005)$ & 0.041 &0.62\\
\hline
$\delta^o$ &  $0.0117\pm 0.00095(0.0046)$ & $0.0135\pm 0.0014(0.0066)$ & 0.28 &0.33\\
\hline
$\phi^s$ &  $0.0101\pm 0.00074(0.0036)$ & $0.00959\pm 0.00096(0.0046)$ & 0.65 &0.14\\
\hline
$\phi^o$ &  $0.0111\pm 0.00093(0.0045)$ & $0.00992\pm 0.00045(0.0021)$ & 0.28 &0.33\\
\hline
\end{tabular}
\end{center}
\end{table}

\begin{table}[h!]
\begin{center}
\caption{Standard  metric comparison between Full Circular and No-short Circular models, $\rho_I=0.25$ ped/m$^2$ initial conditions.}
\label{Standard_4_low}
\begin{tabular}{|c|c|c|c|c|c|}
\hline
& Full $\langle \; \rangle \pm \varepsilon (\sigma)$ & No-short $\langle \;  \rangle \pm \varepsilon (\sigma)$ & $p$ & effect size\\
\hline
No body &  $3.51\pm 0.17(0.8)$ & $4.41\pm 0.22(1.1)$ & 0.0023 &0.95\\
\hline
$\rho$ &  $0.949\pm 0.025(0.12)$ & $0.908\pm 0.022(0.11)$ & 0.22 &0.36\\
\hline
$v$ &  $4.98\pm 0.49(2.3)$ & $5.3\pm 0.43(2.1)$ & 0.63 &0.14\\
\hline
$\theta^v$ &  $4.92\pm 0.4(1.9)$ & $5.74\pm 0.37(1.8)$ & 0.14 &0.45\\
\hline
$\delta^s$ &  $3.13\pm 0.3(1.4)$ & $5.6\pm 0.57(2.7)$ & 0.00037 &1.1\\
\hline
$\delta^o$ &  $3.57\pm 0.22(1.1)$ & $4.47\pm 0.31(1.5)$ & 0.023 &0.7\\
\hline
$\phi^s$ &  $1.36\pm 0.044(0.21)$ & $1.29\pm 0.035(0.17)$ & 0.25 &0.35\\
\hline
$\phi^o$ &  $2.14\pm 0.094(0.45)$ & $2.22\pm 0.058(0.28)$ & 0.49 &0.2\\
\hline
\end{tabular}
\end{center}
\end{table}

\begin{table}[h!]
\begin{center}
\caption{EMD metric comparison between Full Circular and No-short Circular models, $\rho_I=1.5$ ped/m$^2$ initial conditions.}
\label{EMD_4_medium}
\begin{tabular}{|c|c|c|c|c|c|}
\hline
& Full $\langle \;  \rangle \pm \varepsilon (\sigma)$ & No-short $\langle \;  \rangle \pm \varepsilon (\sigma)$ & $p$ & effect size\\
\hline
No body &  $0.015\pm 0.00067(0.0032)$ & $0.0199\pm 0.0012(0.0057)$ & 0.0009 &1\\
\hline
$\rho$ &  $0.0101\pm 0.00069(0.0033)$ & $0.0125\pm 0.0011(0.0052)$ & 0.072 &0.54\\
\hline
$v$ &  $0.0248\pm 0.0019(0.009)$ & $0.0377\pm 0.004(0.019)$ & 0.0059 &0.85\\
\hline
$\theta^v$ &  $0.00694\pm 0.0005(0.0024)$ & $0.00765\pm 0.00045(0.0022)$ & 0.3 &0.31\\
\hline
$\delta^s$ &  $0.0224\pm 0.0012(0.0058)$ & $0.0268\pm 0.00096(0.0046)$ & 0.006 &0.85\\
\hline
$\delta^o$ &  $0.0108\pm 0.0018(0.0088)$ & $0.0147\pm 0.0013(0.006)$ & 0.086 &0.52\\
\hline
$\phi^s$ &  $0.0192\pm 0.0023(0.011)$ & $0.0177\pm 0.0014(0.0066)$ & 0.58 &0.16\\
\hline
$\phi^o$ &  $0.0259\pm 0.0024(0.012)$ & $0.0254\pm 0.0024(0.012)$ & 0.89 &0.041\\
\hline
\end{tabular}
\end{center}
\end{table}

\begin{table}[h!]
\begin{center}
\caption{Standard  metric comparison between Full Circular and No-short Circular models, $\rho_I=1.5$ ped/m$^2$ initial conditions.}
\label{Standard_4_medium}
\begin{tabular}{|c|c|c|c|c|c|}
\hline
& Full $\langle \; \rangle \pm \varepsilon (\sigma)$ & No-short $\langle \;  \rangle \pm \varepsilon (\sigma)$ & $p$ & effect size\\
\hline
No body &  $6.28\pm 0.4(1.9)$ & $8.28\pm 0.48(2.3)$ & 0.0024 &0.95\\
\hline
$\rho$ &  $2.62\pm 0.23(1.1)$ & $2.77\pm 0.28(1.3)$ & 0.68 &0.12\\
\hline
$v$ &  $6.69\pm 0.86(4.1)$ & $12.3\pm 1.7(8.2)$ & 0.0054 &0.86\\
\hline
$\theta^v$ &  $8.96\pm 1.1(5.2)$ & $8.94\pm 0.88(4.2)$ & 0.99 &0.0047\\
\hline
$\delta^s$ &  $10.1\pm 1.6(7.5)$ & $12.5\pm 0.9(4.3)$ & 0.19 &0.39\\
\hline
$\delta^o$ &  $3.03\pm 0.3(1.5)$ & $4.95\pm 0.42(2)$ & 0.00062 &1.1\\
\hline
$\phi^s$ &  $1.91\pm 0.076(0.37)$ & $1.92\pm 0.081(0.39)$ & 0.95 &0.02\\
\hline
$\phi^o$ &  $3.84\pm 0.24(1.1)$ & $4.03\pm 0.28(1.3)$ & 0.6 &0.15\\
\hline
\end{tabular}
\end{center}
\end{table}

\newpage

\section{Model pdf comparison}
\label{graph_comp}
\begin{figure}[h!]
\begin{center}
  \includegraphics[width=0.8\textwidth]{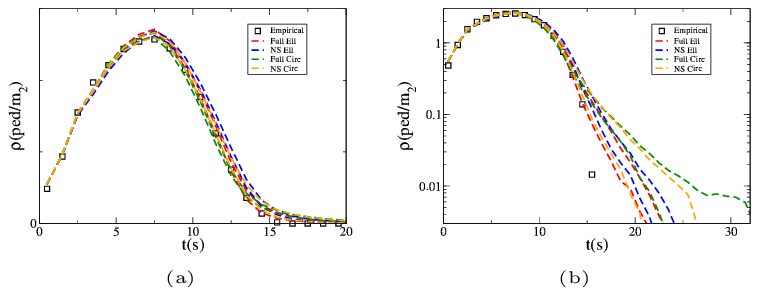}
  \caption{\label{rho_comp} $\rho$ standard error intervals in different models (dashed lines) compared to the experimental values (squares) for the $\rho_I=2.5$ ped/m$^2$ initial condition. (a): linear plot. (b): logarithmic plot.}
 \end{center}   
\end{figure}

\begin{figure}[h!]
\begin{center}
  \includegraphics[width=0.8\textwidth]{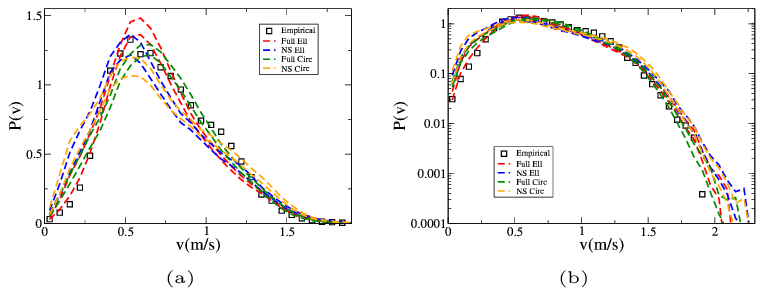}
  \caption{\label{v_comp} $v$ standard error intervals in different models (dashed lines) compared to the experimental values (squares) for the $\rho_I=2.5$ ped/m$^2$ initial condition.
 (a): linear plot. (b): logarithmic plot.
  }
 \end{center}   
\end{figure}

\begin{figure}[h!]
\begin{center}
  \includegraphics[width=0.8\textwidth]{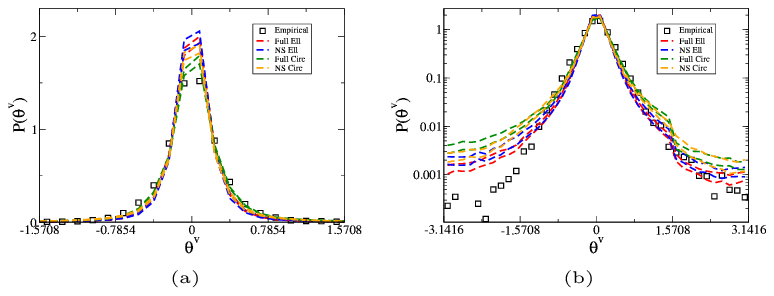}
  \caption{\label{theta_v_comp} $\theta^v$ standard error intervals in different models (dashed lines) compared to the experimental values (squares) for the $\rho_I=2.5$ ped/m$^2$ initial condition.
 (a): linear plot. (b): logarithmic plot.
  }
 \end{center}   
\end{figure}

\begin{figure}[h!]
\begin{center}
  \includegraphics[width=0.8\textwidth]{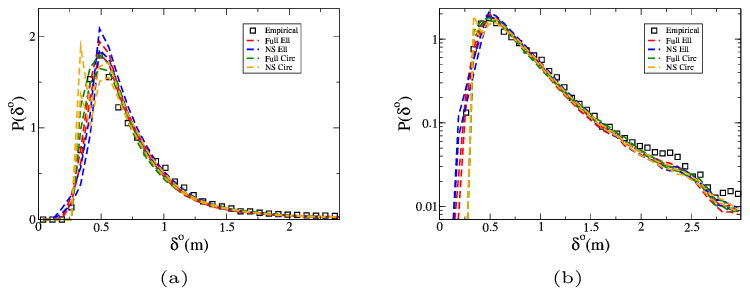}
  \caption{\label{do_comp} $\delta^o$ standard error intervals in different models (dashed lines) compared to the experimental values (squares) for the $\rho_I=2.5$ ped/m$^2$ initial condition.
 (a): linear plot. (b): logarithmic plot.
  }
 \end{center}   
\end{figure}

\begin{figure}[h!]
\begin{center}
  \includegraphics[width=0.8\textwidth]{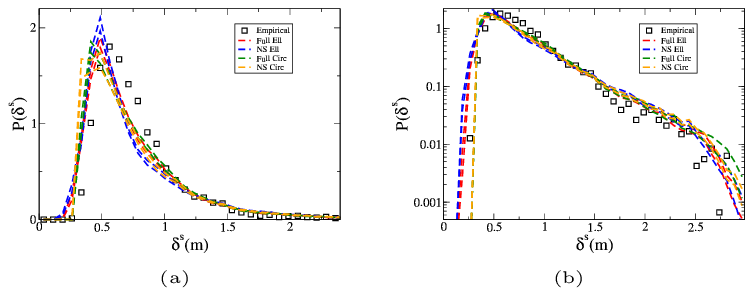} 
  \caption{\label{ds_comp} $\delta^s$ standard error intervals in different models (dashed lines) compared to the experimental values (squares) for the $\rho_I=2.5$ ped/m$^2$ initial condition.
 (a): linear plot. (b): logarithmic plot.
  }
 \end{center}   
\end{figure}

\begin{figure}[h!]
\begin{center}
  \includegraphics[width=0.8\textwidth]{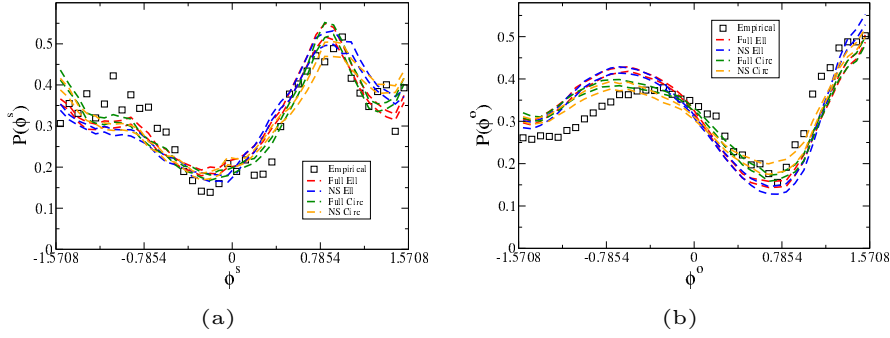} 
  \caption{\label{phi_comp}(a): $\phi^s$ standard error intervals in different models (dashed lines) compared to the experimental values (squares) for the $\rho_I=2.5$ ped/m$^2$ initial condition.
(b):  $\phi^o$ standard error intervals in different models (dashed lines) compared to the experimental values (squares).
  }
 \end{center}   
\end{figure}

\begin{figure}[h!]
\begin{center}
  \includegraphics[width=0.8\textwidth]{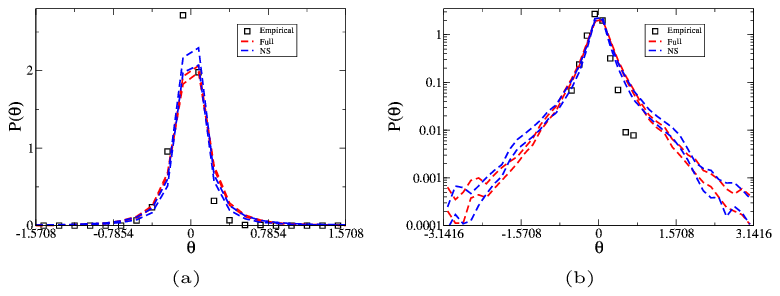} 
  \caption{\label{theta_comp} $\theta$ standard error intervals in different models (dashed lines) compared to the experimental values (squares) for the $\rho_I=2.5$ ped/m$^2$ initial condition.
 (a): linear plot. (b): logarithmic plot.
  }
 \end{center}   
\end{figure}

\begin{figure}[h!]
\begin{center}
  \includegraphics[width=0.8\textwidth]{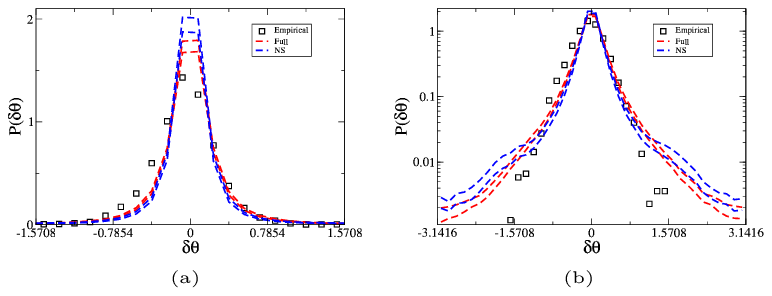}
  \caption{\label{delta_theta_comp} $\delta \theta$ standard error intervals in different models (dashed lines) compared to the experimental values (squares) for the $\rho_I=2.5$ ped/m$^2$ initial condition.
 (a): linear plot. (b): logarithmic plot.
  }
 \end{center}   
\end{figure}

In this section, we compare the observable pdf obtained by each model on the $\rho_I=2.5$ ped/m$^2$ initial condition with the experimental data (See Figs.
\ref{rho_comp}-\ref{delta_theta_comp}). Experimental data are reported as average over all experiment
repetitions, while the model standard errors are computed by averaging over all GA solutions (i.e., for each solution we first compute the average over experiment repetitions, and then use the
variation of such averages on GA solutions to compute the standard errors in the figures).

\clearpage

\bibliographystyle{unsrt}

\bibliography{cross}

\end{document}